\documentclass[a4paper,11pt]{article}
\pdfoutput=1 

\usepackage{jcappub} 

\usepackage[T1]{fontenc} 
\usepackage{graphicx}	
\usepackage{amsmath}	
\usepackage{float}
\usepackage{tikz}
\usepackage{graphicx} 
\usepackage{soul}
\usepackage{bm, framed, cleveref}
\usepackage{cleveref}
\usepackage{hyperref}
\usepackage{orcidlink}
\usepackage{jcappub}
\usepackage{aas_macros}

\title{Caustic Skeleton and the Local Cosmic Web:\\ the Coma Cluster node and the Pisces-Perseus ridge}

\author[a]{Amelie Read
\orcidlink{0000-0001-5388-8953},}
\author[b]{Job Feldbrugge \orcidlink{0000-0003-2414-8707},}
\author[a,b]{Celine Boehm \orcidlink{0000-0002-5074-9998},}
\author[c]{Rien van de Weygaert \orcidlink{0000-0001-8379-1263}}
\author[b]{and Benjamin Hertzsch \orcidlink{0009-0006-6231-8905}}

\affiliation[a]{School of Physics, The University of Sydney,\\ Sydney, Australia}
\affiliation[b]{Higgs Centre for Theoretical Physics, University of Edinburgh,\\ James Clerk Maxwell Building, Peter Guthrie Tait Road, Edinburgh, U.K.}
\affiliation[c]{Kapteyn Astronomical Institute, University of Groningen,\\ P.O. Box 800, 9700AV Groningen, The Netherlands}

\emailAdd{amelie.read@sydney.edu.au}

\abstract{Caustic skeleton theory is an analytical phase-space based formalism that states that the cosmic web is composed of a hierarchy of caustic singularities arising from the gravitational evolution of the dark matter cosmological fluid. Structurally these manifest themselves as walls, filaments and cluster nodes connecting in a pervasive complex spatial network. The present study seeks to identify the caustic skeleton singularities in the observational reality of the cosmic mass distribution in the surrounding Local Universe. To this end it bases itself on the \texttt{Manticore-Local} Bayesian constrained reconstructions of the Local Universe inferred from the 2M++ galaxy survey catalogue.
We extract the three-dimensional multi-scale caustic skeleton of two canonical weblike structures in our Local Universe, the Coma Cluster and the Pisces-Perseus ridge. They represent the most prominent cluster node and filamentary artery in the nearby Universe. We show that the Caustic Skeleton network of caustic singularities, including filaments and walls at this stage, accurately reproduces the observed large-scale organisation of galaxies in redshift space for one of the Manticore realisations.

One of the most interesting aspects of the theory is that it predicts two topologically distinct classes of filaments ($A_4$ swallowtail and $D_4$ umbilic caustics) that form through fundamentally different folding histories.
We find that the $A_4$ filaments form the most outstanding filamentary features of our local universe, whereas $D_4$ filaments appear to largely collect at cluster regions identified in the galaxy distribution. 
Thus, a preliminary galaxy classification scheme presented in this study finds indications of a higher proportion of galaxies in $D_4$/Node-dominated environments in the Pisces-Perseus Supercluster, which may be indicative of more clumpy substructure along its ridge, compared to the extended weblike structure around the Coma Cluster.
Moreover, because caustic skeleton theory also naturally encodes the formation time of caustic structures, it allows us to show that filaments are complex structures that emerge in disjoint segments and merge at different redshifts. In future work, we will explore the extensive potential of the full caustic skeleton framework including nodes, filaments and walls to unravel both the topology and full dynamical history of the structures we see in the sky.}

\begin{keywords}
{cosmology: large-scale structure of universe -- galaxies: clusters: general -- cosmology: theory}
\end{keywords}

\begin{document}
\maketitle
\flushbottom

\section{Introduction}
The present study assesses the skeletal spine - ridges, membranes and nodes - predicted by by the \textit{Caustic Skeleton}
formalism in and around two of the most outstanding and dominant structures in the local \textit{Cosmic Web},
the Pisces-Perseus chain and the weblike large-scale structure around the Coma cluster. The Caustic Skeleton
formalism is the phase-space based analytical theory for the formation and nonlinear hierarchical evolution of
the cosmic web and its composite structural features \citep{Feldbrugge+2018}.

The \textit{Cosmic Web} is the intricate multiscale network \citep{Zeldovich1970, Joeveer1977, BondKofmanPogosyan1996, WeygaertBond2008, Aragon2010, Cautun+2014, Libeskind+2018} defining and representing the fundamental spatial organisation of galaxies and matter on scales of a few up to a hundred megaparsecs. Galaxies, intergalactic gas and dark matter arrange themselves in a salient wispy pattern of dense compact clusters, elongated filaments, and tenuous, sheet-like walls surrounding near-empty void regions. The filaments are the most visually outstanding features of the Megaparsec Universe, in which around $50\%$ of the mass and galaxies in the Universe reside. 
On the other hand, almost 80\% of the cosmic volume belongs to the interior of voids \citep[see e.g.][]{Cautun+2014, Ganeshaiah+2018}. Together, they define a complex spatial pattern of intricately connected structures, displaying a rich geometry of various morphologies and shapes. 
Such filaments and structures have been found using topological methods such as the \texttt{DisPerSE} method \citep{Sousbie2011}, \texttt{SpineWeb} formalism \citep{AragonCalvo+2010} and the explicit multiscale geometric MMF-\texttt{NEXUS(+)} formalism \citep{AragonCalvo+2007, Cautun+2013}, stochastic methods such as the Bisous method \citep{Tempel+2016} and Graph and Percolation methods \citep{Alpaslan:2014}.

The cosmic web represents a key phase in the gravitationally driven dynamical development of structure in the Universe.
It emerges when the original long phase of linear evolution of the primordial density and velocity perturbations turns
into a more advanced nonlinear stage involving contraction and collapse of mass inhomogeneities. As such, it marks the transition from the primordial (Gaussian) random field into a mass distribution marked by highly nonlinear and fully collapsed features such as halos and galaxies. Structure at this transition stage assumes characteristic anisotropic morphologies, such as that of elongated filaments and flattened walls, as the result of the deformations induced by the anisotropic - tidal - gravitational force field. Hence, by implication, it is the tidal force field, induced by the inhomogeneous mass distribution, which is the key to modeling and understanding the structure of the cosmic web. Numerous studies recognized and studied  its role in shaping the cosmic web 
\citep[see e.g.][]{Zeldovich1970,Weyedb1996,BondKofmanPogosyan1996,
Hahn+2007,WeygaertBond2008,Paranjape2018,Paranjape2021,
Kugel+2026}. Amongst these, the Zeldovich formalism stands as the foundation on which our theoretical understanding of the cosmic web is based, and for its accurate description of structure formation in the (mildly) nonlinear regime. 

The weblike organisation of the emerging cosmic mass distribution is well reproduced by $N$-body simulations of structure formation, in which
structure emerges through gravitational growth and amplification of the tiny primordial Gaussian matter perturbations.
In this context, the large-scale organisation of dark matter, gas and galaxies is clearly an emergent property of the
tiny matter fluctuations in the simulations' initial conditions. The resulting multiscale structure is the product of the
interplay of a range of non-linear processes. $N$-body simulations, numerically simulating cosmological
structure formation, are very useful to obtain a good impression of the resulting structure. However, even though several up-to-date
projects provide a large number of simulations that offer a good representation of many aspects of the structure formation process \citep{millennium2005,magneticum2015,eagle2015,bahamas2017,illustris2017,flamingo2023,colibre2025}, they are often demanding, expensive, and usually focus on a single cosmological scenario. Most often this concerns the prevailing standard cosmological $\Lambda$CDM scenario, in which the main matter constituent is that of non-baryonic cold dark matter and the still mysterious dark energy component dominating the dynamical evolution of the Universe. In the past few years, more elaborate cosmological simulation pipelines have been developed that involve the assembly of a large sample of simulations for different cosmologies, such as the set of 82,000 Quijote N-body simulations  \citep[e.g.][]{quijote2020}, and that of
emulator techniques to enable exploration of varying cosmological parameters between a set of N-body simulations \citep[see e.g.][]{Wang+2026}.

Nonetheless, while the simulations may be seen as optimal observational windows into the entire cosmic history of
the emergence and buildup of cosmic structure, for a full physical understanding and appreciation of their results it is
crucial to be able to interpret these within the context of a complementary theoretical framework. The \textit{Caustic Skeleton} theory is the
first fully analytical framework that is able to follow the resulting buildup of structure far into the nonlinear regime
\citep{ArnoldShandarinZeldovich1982,Hidding+2014,Feldbrugge+2018,FeldbruggeWeygaert2024}. This is based on the recognition
of the intimate relationship between the formation of structural features in the cosmic matter distribution and the
emergence of multistream regions in the gravitationally induced flow field, and the key role of the anisotropic tidal and
deformation field in shaping the resulting spatial pattern in matter distribution,

It recognizes that clusters, filaments, and walls correspond to structural singularities that arise where streams of matter in the
cosmological fluid converge and fold into each other due to gravity. They thereby mark the locations where matter density spikes, manifesting as large-scale structures. Thus, the emerging geometric pattern of the large-scale cosmic matter distribution is the spatial manifestation of the folding of the dark matter distribution in 6D phase space \citep{Shandarin2011,AbelHahnKaehler2012,Falck:2012}. It implies the formation of a hierarchy of nonlinear structures following shell crossing in the corresponding multistream regions
\citep{Pogosyankofman1990,Sathyaprakash1996,Hidding+2016,Feldbrugge+2018,Hertzsch+2026}. The mathematical framework of \textit{Catastrophe Theory} \citep{Thom1972,Zeeman1977,Poston1978,Arnold1992} states that the stable singularities in a fluid can be classified in terms of a limited number of elementary catastrophes, and defines and classifies these stable folding configurations. Geometrically these assemble into the connecting framework of the caustic skeleton \citep{arnold1982b,Hidding+2014,Feldbrugge+2018,FeldbruggeWeygaert2024,Hertzsch+2026}. Arnold and collaborators were the first to apply this to the prediction of the structure of the cosmic web \cite{ArnoldShandarinZeldovich1982,
  Arnold1986}, pointing out that the different features of the cosmic web can be associated with mathematical catastrophes or caustics. While their work was limited to two dimensions, and therefore unable to capture the full richness of the cosmic web \citep{ArnoldShandarinZeldovich1982}, the \textit {Caustic Skeleton} theory of \cite{Feldbrugge+2018} opened the path towards its full treatment in any dimensional space \citep{ArnoldShandarinZeldovich1982,Hidding+2014,Feldbrugge+2018,FeldbruggeWeygaert2023,FeldbruggeWeygaert2024,Hertzsch+2026}. The resulting phase-space based \textit{Caustic Skeleton} description of the evolving weblike pattern in the cosmic matter distribution is codified in terms of a complete set of caustic conditions. \cite{Feldbrugge+2018} showed that a complete understanding of the cosmic web spinal structure is obtained through the spatial characteristics of the \textit{eigenvalue} and \textit{eigenvector} fields of the cosmic tidal force field. Underlining this is the realization that the embryonic outline of the cosmic web, in particular its filamentary network, can already be seen in the primordial tidal eigenvalue field \citep{FeldbruggeYanWeygaert2023,Wilding+2022,Ram+2024}.

The vast majority of studies of the cosmic matter and galaxy distribution resort to the inference of statistical measures involving averages over large cosmological volumes and samples. However, a more direct strategy is that of allowing a direct comparison between the structure and dynamics of specific individual observed patterns and objects and those predicted by a theoretical framework and/or related simulations. 
Following this strategy of observationally constrained predictions, the Pisces-Perseus ridge and the Coma cluster centered weblike network represent a compelling testing ground for the range of validity of the Caustic Skeleton theory as the significant structures it predicts can be directly compared with the features seen in the galaxy distribution in these two prominent regions. 
Thus, a motivation for our study is the ability to evaluate the connectivity of individual weblike features to known real and known structures in the local Universe. 
One of the most striking predictions of caustic skeleton theory is the existence of two distinct types of filaments, the A- vs. D-type filaments, involving different structures, mass distributions and formation histories. As such an additional motivation for this study is to assess whether these differences indeed manifest themselves in significantly different signatures in the Coma and Pisces-Perseus filamentary extensions. 
A third motivation is that the fine structure of filaments predicted by the theory - their branching into tenuous tendrils, their connectivity to membrane-like walls and dense compact cluster nodes, and their rendering of the low contrast infrastructure within vast underdense void regions - will provide a highly promising source for directed observational campaigns. These predictions may enable a detailed exploration of the weblike manifestations in the gaseous and dark matter distribution around such prominent large scale structures in the nearby Universe. 
Perhaps the most ambitious prospect for our program is that of expanding its theoretical context towards developing the instruments - finally -  for exploiting the vast information content of the intricate complex weblike network towards inferring the values of cosmological parameters, information on the nature of dark matter and dark energy, and other important cosmological constraints. The versatile analytical predictions of Caustic Skeleton theory, for a wide range of cosmological scenarios, will allow this once confronted with the highly resolved spatial galaxy distribution obtained by upcoming surveys such as 4MOST \citep[see e.g.][]{4MOST2019}.

Following the outlined motivation, the present study incorporates and combines two of the philosophies discussed above. It assesses the weblike (dark) matter distribution implied and constrained by the observed galaxy distribution in the local universe and confronts this with the implied spine of the cosmic web in these regions predicted by the Caustic Skeleton theory. To enable this, we need the primordial deformation tensor eigenvalue field in the local Universe. For this we use the constrained initial conditions inferred by the BORG (Bayesian Origin Reconstruction from Galaxies) Bayesian inference method \citep{BORG2013} from the 2M++ galaxy survey catalogue \citep{Huchra+2012}. We use three different initial density field realizations to assess both the significant (mean) implied structures and patterns as well as the corresponding spectral variations within these - although caustics are then only considered for one realisation in this study.

The corresponding Manticore library of constrained N-body simulations \citep{McAlpine:2025,McAlpine:2026} evolve the cosmic mass distribution from these constrained initial conditions and form the "observational" reference with respect to which we compare and interpret the predicted caustic features (as well as the physical character of the various structures seen in the simulated mass distribution). These results will have broad implications for our understanding of cosmic evolution and the nature of dark matter and could be particularly important in light of upcoming data releases from Euclid, Dark Energy Spectroscopic Instrument (DESI), 4MOST and the Square Kilometre Array (SKA), as well as other large-scale surveys expected to map the Universe with unprecedented precision.
The high-resolution Manticore constrained realizations enable a considerable improvement in scope, detail and dynamic range of the caustic spine of the cosmic web with respect to the earlier reconstruction of the caustic network in the local Universe by \cite{Bixerman+2022} on scales of the cluster distribution. While the latter allowed an overall identification of inferred caustics with the distribution of clusters in the local Universe, with the current initial conditions we are allowed to probe the cosmic web at much smaller scales such that we will be able to relate it to the details of the galaxy distribution.

In the present study we concentrate on a detailed assessment of the caustic structure in and around two of the most prominent components of the cosmic web in the local Universe. These are the large-scale structure around the Coma Cluster and that of the Pisces-Perseus chain. They allow a highly detailed rendering of the weblike mass distribution of which they form the central manifestation. At a distance of 103 $\rm{Mpc}$ and a total mass of around $\sim 7 \times 10^{14} M_\odot$, the Coma cluster, Abell 1656, is the most massive cluster node in the immediate vicinity in the Local Universe. This structure was of seminal importance as the site at which Zwicky uncovered the presence of a major share of invisible matter, ie. \textit{dark matter} \citep{Zwicky1933}. The Coma Cluster along with the Leo Cluster, Abell 1367, to which it is connected via a strong filamentary bridge, form the heart of the Coma supercluster \citep[see e.g.][]{EinastoM:2025,EinastoM+2025}. 
In fact, the surroundings of the Coma Cluster may rightfully be considered the region in which, for the first time, the reality and complexity of the weblike patterns in the galaxy distribution surfaced in the famous \textit{Stickman} map of de Lapparent and collaborators \citep{deLapparent1986}. The galaxy and mass distribution in and around the Coma cluster testifies of a complex pattern of branching filamentary bridges emanating from the central node at which the cluster resides \citep{EinastoM:2025,EinastoM+2025,Malavasi:2023,Mondelin2025,Pedratti+2026,Benisty2026}. They are embedded in one of largest known superstructures in the nearby universe, the flattened wall-like Coma Great Wall \citep{GellerHuchra1989,Platen2009,EinastoM:2025}. 

With the Coma cluster representing the canonical example of a massive cluster node in the cosmic web, the $130$ Mpc long Pisces-Perseus chain of galaxies, groups and clusters is without doubt the most outstanding and dynamically dominant filamentary artery in the local Universe \cite{HaynesGiovanelli+1986,Hidding+2016,EinastoM:2025}. With a distance of a mere 77 Mpc and extending along no less than 40 degrees along the northern sky, it represents an exceptionally prominent and detailed example of a filament, the dominant structural features of the cosmic web \citep{Cautun+2014,Kugel+2026}. 
Its relative vicinity and almost perfect transverse orientation on the sky, perpendicular to the line of sight, yields an almost ideal unbiased vista along the ridge  of the Pisces-Perseus chain. In addition to the main Pisces-Perseus chain, the entire complex is a  superb prototype for the complex spatial connectivity marking the cosmic web, as we see it branching into two major arms, together forming the surrounding boundary of a tenuous large wall \citep{HaynesGiovanelli+1986,Hidding+2016,Mondelin2025}. Cosmographic studies have also uncovered the enormous gravitational - force and tidal -  impact of the Pisces-Perseus supercluster complex in the Local Universe. Along with the Great Attractor region and the related Laniakea complex, and to a lesser extent the Local Void, it has been found to be the dominant influence in the motion of galaxies in the nearby universe \citep{Pomarede2017,Tully2019,Pomarede2026}.

In this paper, \Cref{sec:cst} reintroduces caustic skeleton theory and how it can be applied to simulations to extract cosmic web structures. \Cref{sec:data} provides details on the Manticore BORG (Bayesian Origin Reconstruction from Galaxies) \citep{Jasche:2010,Leclercq:2015,Leclercq:2017,BORG2013} constrained Local Universe simulations utilised in this study. With these data-constrained reconstructions, \cref{sec:skel} presents the first application of caustic skeleton theory to real, prominent structures in our Local Universe, including the Coma Supercluster (\cref{sec:coma}) and the Pisces-Perseus filamentary complex (\cref{sec:pp}). In \cref{sec:class}, we then use the extracted caustic skeletons to establish a simple classification scheme for the large-scale environments of 2M++ galaxies to illustrate a general idea of how the theory could be used in future. Finally, \cref{sec:history} exploits caustic skeleton theory's unique ability to unravel both structure and formation time to explore the formation histories of both prominent local cosmic web structures. In \cref{sec:con} we present a summary of our key results and conclusions.

\section{Caustic Skeleton Theory}
\label{sec:cst}
The caustic skeleton is a mathematical framework that describes the formation of cosmic structures as a hierarchy of singularities (caustics) arising in the dark matter field. In this picture, large-scale structure formation is understood as the progressive folding of the dark matter sheet. The first singularities to appear correspond to fold ($A_2$) caustics. These are associated with the process of shell-crossing. The further evolution of these folds gives rise to higher-order singularities known as the cusps ($A_3$), the swallowtails ($A_4$), the butterfly ($A_5$), the hyperbolic/elliptic umbilic ($D_4$) and the parabolic umbilic ($D_5$) caustics. The cusp caustic is associated with the walls in the Universe separating void regions. The swallowtail and hyperbolic/elliptic umbilic caustics correspond to the filaments connecting galaxy clusters and dark matter halos. The butterfly and parabolic umbilic caustics are associated with the dense clusters in the cosmic web.

Feldbrugge and collaborators \citep{FeldbruggeWeygaert2023, Hertzsch+2026} demonstrated that caustic skeleton theory successfully captures the formation of the largest structures in the Universe. When combined with the Zel'dovich approximation, the theory efficiently and accurately marks the regions in the initial conditions that form the walls and filaments in the present-day Universe as modelled by $N$-body simulations. In particular, they showed that filaments emerge at the foldings and joints of the walls. The ultimate challenge for the theory is the confrontation with real observed structures, for which the immediate surroundings in the Local Universe represents the region providing the full detailed inventory of cosmic web components. Following an earlier low resolution
study by \citep{Bixerman+2022}, in which we were able to identify caustics with known supercluster complexes in the
cluster distribution in the Local Universe, the present study addresses the majority of the inventory of caustic features (i.e. walls and filaments) in the detailed resolved galaxy distribution in the Local Universe.

An intriguing aspect of the identification of caustics with morphological features in the cosmic web is that following
Arnold \cite{Arnold1986} we may observe that there exist two channels to form filaments: the swallowtail ($A_4$) and the umbilic ($D_4$) filaments. This is a surprising result that may have important cosmological consequences, especially given the growing body of evidence supporting the influence of filaments on the properties of galaxies and other cosmic structures (see, for example,
\cite{HaynesGiovanelli+1986,Alpaslan:2014, Metuki+2015,Kraljic+2017, Kuutma+2017,Xu+2020, GalarragaEspinosa+2022, Kuchner+2022, Hasan+2023, Hasan+2025,Yu+2025, Todorache:2025, NandiPandeySarkar2026}). 
If these results also hold on smaller scales, this would suggest that the properties of the Local Group could be inferred from the structure of the local walls, thus also highlighting the importance of the void near us \citep[][]{Tully2019,Wempe2026}. This is particularly important as the local void is often invoked in the interpretation of anomalous observations \citep{Peebles2001,Tully2008,Tully2019,tully2026}, including those based on recent results from the DESI survey \citep{DESI1, DESI2}. Moreover, \cite{Gallagher2026} measured the strength of correlation between galaxy properties and large-scale environment in BORG constrained posteriors, similar to those utilised in this study. They found that correlations were strongest when the underlying Dark Matter density fields was smoothed to 2 Mpc/h, suggesting a scale relevant to galaxy formation that can be probed through constrained realisations of the Local Universe.

\subsection{Shell-Crossing}
\label{sec:caustic_def}

\begin{figure}[tbp]
    \centering
	\includegraphics[width=\textwidth]{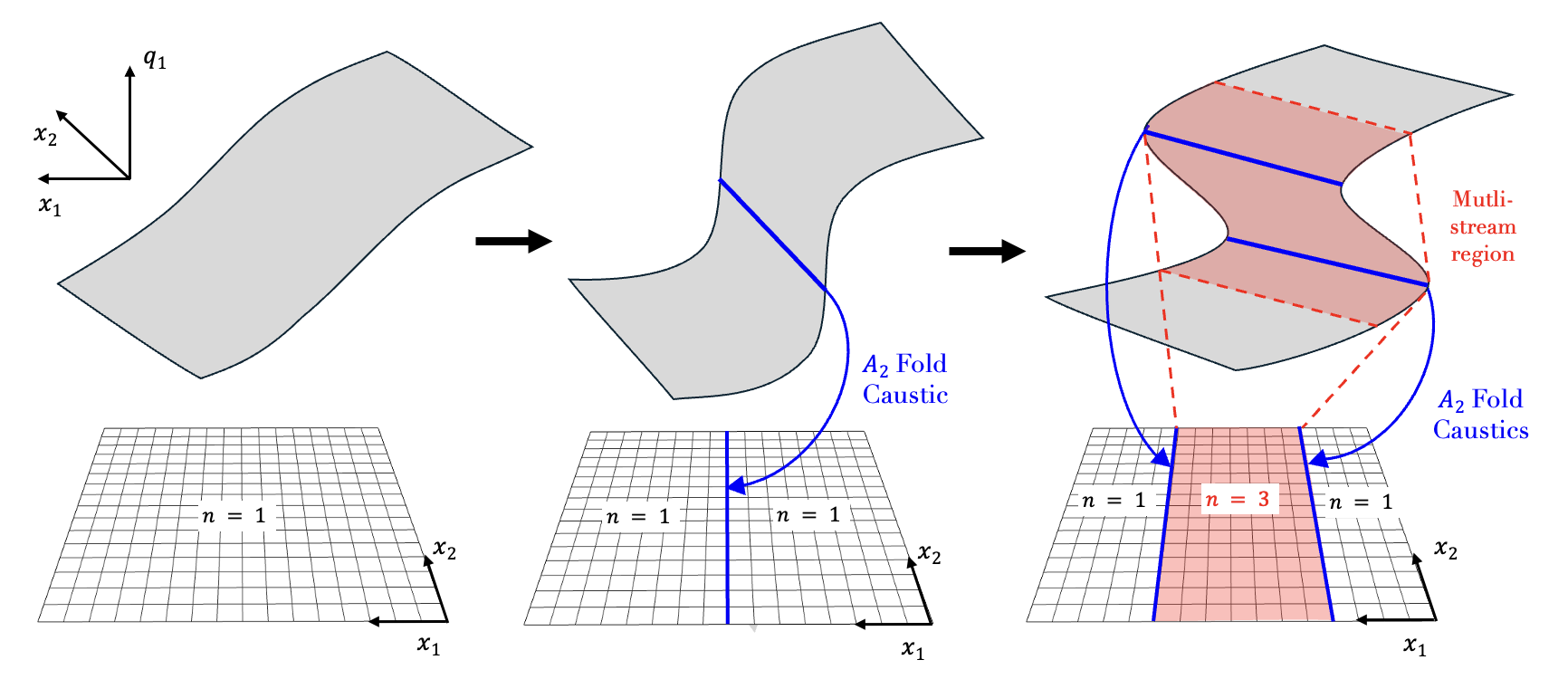}
    \caption{Example illustration of a Lagrangian mapping of a 2D continuous particle distribution. The vertical direction $q_1$ represents one of the initial (Lagrangian) positions of the distribution. The $x_1-x_2$ plane shown in the bottom panels represents the current/final (Eulerian) distribution at each stage. The top left image corresponds to an early time before the distribution has evolved, such that the mapping is still 1-1, and the distribution is single $n=1$ streaming in the Eulerian plane below.
    Subsequent panels towards the right show how the particle distribution evolves - since initial coordinates are fixed, the sheet (which is a slice of the Lagrangian mapping) only evolves in the planar direction.The blue line in the middle panel shows where a fold in the sheet first begins to form. This is the first instance of shell-crossing and as such is the first emergence of an $A_2$ fold caustic. In the last panel, the sheet has continued to fold over itself, such that the instantaneous shell-crossing is occurs in two separate parts of the sheet, manifesting as two separate $A_2$ caustic lines. These fold caustics bound a multistream region (red), separating a region with $n=3$ streams from regions with $n=1$ streams in the final Eulerian plane (bottom right).}
    \label{fig:fold}
\end{figure}


Caustics occur in the formation of multi-stream regions when mass elements move under gravity and eventually shell-cross. We will describe the formation of the cosmic web in Lagrangian fluid dynamics through the Lagrangian map 
\begin{align}
    \bm{x}_t(\bm{q}) = \bm{q} + \bm{s}_t(\bm{q})\,,
\end{align}
describing the position $\bm{x}_t(\bm{q})$ of a mass element at time $t$ in Eulerian space (the present-day cosmic web) that started at $\bm{q}$ in Lagrangian space (the initial conditions) through the displacement field $\bm{s}_t(\bm{q})$. When a point $\bm{x}$ is reached from $n$ points in the initial condition space, the point $\bm{x}$ is part of an $n$-stream region ($n$ being an odd number for an infinite sheet). At the boundary of an $n$-stream region, the dark matter sheet turns over in a fold caustic. 

An intuitive illustration is that of a thin sheet being folded (see \cref{{fig:fold}}). Initially, the sheet is unfolded (see the upper left panel in \cref{fig:fold}) such that it is in a single-streaming ($n=1$) configuration everywhere. 
As the sheet continues to fold, instantaneously in the middle panel there is a section of the sheet that is perpendicular to the Eulerian plane. As such, the blue line marks first shell-crossing, where the mapping from Lagrangian to Eulerian space first becomes no longer 1-1. This line represents the simplest fold caustic and is projected into the Eulerian plane below. Once the sheet has fully folded over itself, several points on the sheet correspond to the same position in $\bm{x}=(x_1,x_2)$ space (see the red regions shaded in the rightmost panels of \cref{fig:fold}) forming a multi-stream ($n=3$) region. The multi-stream region is bounded on either side and separated from the single-streaming regions by fold caustics, which, once again, correspond to the locations of critical points in the lagrangian mapping.

In cosmology, this “sheet” consists of the dark matter particles labelled by their initial coordinates $\bm{q}$. Starting from a close to homogeneous distribution of matter, the particles move, leading to the density field 
\begin{align}
    \rho_t(\bm{x}) &= \sum_{\bm{q} \in \bm{x}_t^{-1}(\bm{x})} \frac{\rho_0(\bm{q})}{|\det \ \nabla_{\bm{q}} \bm{x}_t(\bm{q}) |} \nonumber \\
    &= \sum_{\bm{q} \in \bm{x}_t^{-1}(\bm{x})} \frac{\rho_0(\bm{q})}{|\det \ (I + \nabla_{\bm{q}} \bm{s}_t(\bm{q}) )|}\,,
\end{align} 
where the sum ranges of the initial positions $\bm{q}$ that reach $\bm{x}$ in time $t$ and with the primordial density field $\rho_0(\bm{q})$. Initially, each point $\bm{x}$ in Eulerian space corresponds uniquely to a point $\bm{q}$ in Lagrangian space. As the initial overdensities attract their surrounding matter, the Jacobian $\det \nabla_{\bm{q}}\bm{x}_t$ decreases and eventually crosses zero. When this happens, the sheet folds over in a caustic, and the density spikes to infinity. This marks the creation of a multi-stream region. In terms of the eigenvalues $\mu_i$ of the deformation tensor $\nabla_{\bm{q}}\bm{s}$, defined by the eigen equation 
\begin{align}
    \left[\nabla_{\bm{q}}\bm{s}_t(\bm{q})\right]\bm{v}_i(\bm{q}, t) = \mu_i(\bm{q}, t)\bm{v}_i(\bm{q}, t)  
    \label{eq:general_deformation}
\end{align}
with the eigenvectors $\bm{v}_i$, the density assumes the form
\begin{align}
    \rho_t(\bm{x}) &= \sum_{\bm{q} \in \bm{x}_t^{-1}(\bm{x})} \frac{\rho_0(\bm{q})}{|1 + \mu_1(\bm{q},t) | \, | 1 + \mu_2(\bm{q},t) | \, | 1 + \mu_3(\bm{q},t) |}.  \label{eq:density}
\end{align}
Consequently, shell-crossing is neatly described in terms of the condition $1+\mu_i(\bm{q},t) = 0$ for some $i=1,2,3$ on the eigenvalue fields.

\subsection{Building the caustic skeleton}
\label{sec:caustic_class}
Catastrophe theory classifies the shell-crossing events in terms of the elementary catastrophes, also known as caustics. In the illustration in \cref{fig:fold}, the dark matter particles form simple fold caustics. In realistic scenarios, shell-crossing marks the onset of non-linear gravitational collapse and the formation of multi-stream regions that produce bound structures. For a clear introduction into catastrophe theory, see \cite{Saunders1980,Poston2014}. 

Although gravitational collapse can create intricate structures, only a handful of elementary types of caustics can stably occur in the Lagrangian flow. Classified by Arnold \citep{Thom1972, Arnold1986} these are: the \textit{fold} ($A_2$), the \textit{cusp} ($A_3$), the \textit{swallowtail} ($A_4$), the \textit{butterfly} ($A_5$), the \textit{elliptic} and \textit{hyperbolic umbilic} ($D_4$), and \textit{parabolic umbilic} caustics ($D_5$).


\begin{table*}
\centering
\begin{tabular}{l | l | l }
{Name} & {2D Cosmic Web} & {3D Cosmic Web} \\
\hline 
Fold, $A_2$& Shell-Crossing & Shell-Crossing   \\
Cusp, $A_3$ & Filament & Wall \\
Swallowtail, $A_4$ & Cluster & Filament \\
Butterfly, $A_5$ & Not stable & Cluster\\
Elliptic/Hyperbolic Umbilic, $D_4$ & Cluster & Filament  \\
Parabolic, $D_5$&Not stable & Cluster \\ 
\end{tabular}
\caption{The 7 types of caustics that can form in a 3-dimensional Lagrangian mapping with associated symbols. These caustics correspond to structures like walls, filaments and clusters in the 3D cosmic web. In the 2-dimensional version of the theory, a similar association can be made; the caustics effectively lose a dimension, and $D_5$ and $A_5$ points are not stable. } 
\label{tab:2d3d}
\end{table*}

\paragraph*{Caustic morphology}
Caustics fall into three families, $A$, $D$ and $E$, with the latter being virtually improbable to occur in realistic cosmological conditions. The key difference between the $A$ and $D$ singularities is that $A$-type caustics follow from conditions on a single eigenvalue field $\mu_j$ of the deformation tensor, whereas $D$-type caustics involve conditions on two eigenvalue fields, as shown by the explicit conditions given in Appendix \ref{appendixA}. As a consequence, the two families trace different formation histories, each with its own characteristic
formation time and morphology. $A_4$ swallowtail filaments, for instance, tend to be elongated and only mildly overdense, forming where walls meet or are creased, whereas $D_4$ umbilic filaments arise at the junctions of three walls and are correspondingly shorter and denser \citep{Feldbrugge+2018,FeldbruggeWeygaert2023,Hertzsch+2026}.
The $A_4$ filaments form at foldings in the wall sheets, while the $D_4$ filaments signal the junction of at least three walls. This analytically distinction is one of the more unique predictions of caustic skeleton theory: the observation that there should be two distinct types of filaments and two distinct types of cluster nodes (see \cref{tab:2d3d}).

The caustic conditions further imply a nested hierarchy, which is of seminal significance for understanding the connectivity
of the various structural features in the cosmic web. Each successive caustic lies on the manifold of the previous one, so that, at
the very least, filaments sit inside walls and clusters sit on filaments. Although less apparent from the conditions themselves, this nesting extends across the two families: $D$-type filaments sit on $A$-type walls, as they form where caustic walls of the first and
second eigenvalue fields meet, and $D_5$ clusters likewise form on $A$-type filaments. $D$-type caustics are therefore never fully isolated from an $A$-type caustic region. 
Classifying a structure as a particular caustic thus simultaneously characterises its geometry and encodes its dynamical formation history, with significant implications for the connectivity of the cosmic web (as we discuss in the following sections). 
The caustic formalism will allow a full general treatment of connectivity and extend what is known of the relation between cluster nodes and filaments \cite{Colberg+2005,Aragon2010,Codis+2018,Gouin+2026} and perhaps even more about the connection between walls, filaments and voids, as well as exploring the interscale connectivity between structural features and their substructures.

\bigskip

\paragraph*{Scale Space and the Multiscale Caustic Skeleton:} The cosmic web is a multiscale complex in which large-scale structures are built out of smaller-scale structures, the result of a hierarchical buildup of structure in which small structures emerged first and subsequently merged into ever larger structural features. In practice, to assess the multiscale structure of the caustic skeleton we focus on the caustics forming at a given length scale $\sigma$ by smoothing the displacement potential with a Gaussian filter 
\begin{equation}
    \Psi_{\sigma}(\bm{q}) = \int   \Psi(\bm{q} - \bm{q}^{\prime} ) W_{\sigma}(\bm{q}^{\prime} ) \mathrm{d}\bm{q}'
\end{equation}
with
\begin{equation}
    W_{\sigma}(\bm{q}) = \frac{1}{2 \pi \sigma^2} e^{-\frac{\bm{q}^2}{2 \sigma^2}}
\end{equation}
and evaluating the caustics associated with $\Psi_\sigma$ (following \cite{Hidding+2014,Feldbrugge+2018, FeldbruggeWeygaert2023, FeldbruggeYanWeygaert2023, FeldbruggeWeygaert2024, Hertzsch+2026}). By evaluating the caustic skeleton as a function of $\sigma$, we obtain a scale map of the multiscale structure of the caustic skeleton, the scale space caustic skeleton. On the basis of the dependence of growth rate on
the scale of features, the scale space structure of the displacement potential allows us to describe the multiscale structure of the
primordial field and the subsequent implied hierarchical development, and infer the spectrum of physical characteristics of the emerging objects and structural features \cite{PressSchechter1974,DoroshkevichShandarinSaar1978,BBKS,Cole1991,BondMyers1996,BondKofmanPogosyan1996,Shethwey2004,Shen2006}.

  \paragraph*{The Zel'dovich approximation:} While in general the caustic skeleton formalism can be based on the fully nonlinear displacement potential, inferring these even from eg. an N-body simulation, its power is particularly strong when it involves a fully analytical treatment. The Zel'dovich approximation \cite{Zeldovich1970,Shandarinzeldovich1989} provides a highly accurate first-order description of the displacement of mass elements up to the stage at which they enter evolved structures and their corresponding multistream regions. As the Zel'dovich approximation has proven to remain impressively accurate up to the regime at which mass elements enter the emerging structural features, it successfully and accurately identifies the location and timing of emergence of walls, filaments and even nodes in the cosmic web. Because caustic singularities arise at exactly these locations, for the computation of the various structural characteristics the Zel'dovich approximation provides the sufficient - and even ideal - framework  for the fully analytical formulation of the caustic skeleton. 

  Within the context of the Zel'dovich formalism, the displacement field factorises into a spatial and a temporal term 
\begin{align}
    \bm{s}_t(\bm{q}) = - D_+(t) \nabla_{\bm{q}} \Psi(\bm{q})\,.
    \label{eq:deformation}
\end{align}
with $\Psi(\bm{q})$ the displacement potential and where $D_+(t)$ is the linear perturbation growth factor (\cite{peebles1980}, see Appendix \ref{appendixB} for details).  Using \cref{eq:deformation}, we obtain \cref{eq:general_deformation} in terms of the Hessian of the displacement potential
\begin{align}
    - D_+(t) \ \left[\mathcal{H}\Psi(\bm{q})\right] \bm{v}_i(\bm{q}) = \mu_i(\bm{q},t) \bm{v}_i(\bm{q})\,,
\end{align}
or 
\begin{align}
\left[\mathcal{H}\Psi(\bm{q})\right] \bm{v}_i(\bm{q}) = \lambda_i(\bm{q}) \bm{v}_i(\bm{q})\,,
\end{align}
with $\mu_i(\textbf{q},t) = - \lambda_i(\textbf{q}) D_+(t)$. In the following, we will use the convention $\lambda_1 \geq \lambda_2 \geq \lambda_3$, leading to the simplified caustic conditions as in Appendix \ref{appendixB}. Hence, following this framework the eigenvalues $\lambda_i(\textbf{q})$ and eigenvectors $\textbf{v}_i(\textbf{q})$ of the deformation tensor determine the location and topology of the collapse. 

\bigskip

\subsection{Identification of caustics in the cosmic web: 2D illustration}
\begin{figure}
\centering
\includegraphics[width=\textwidth]{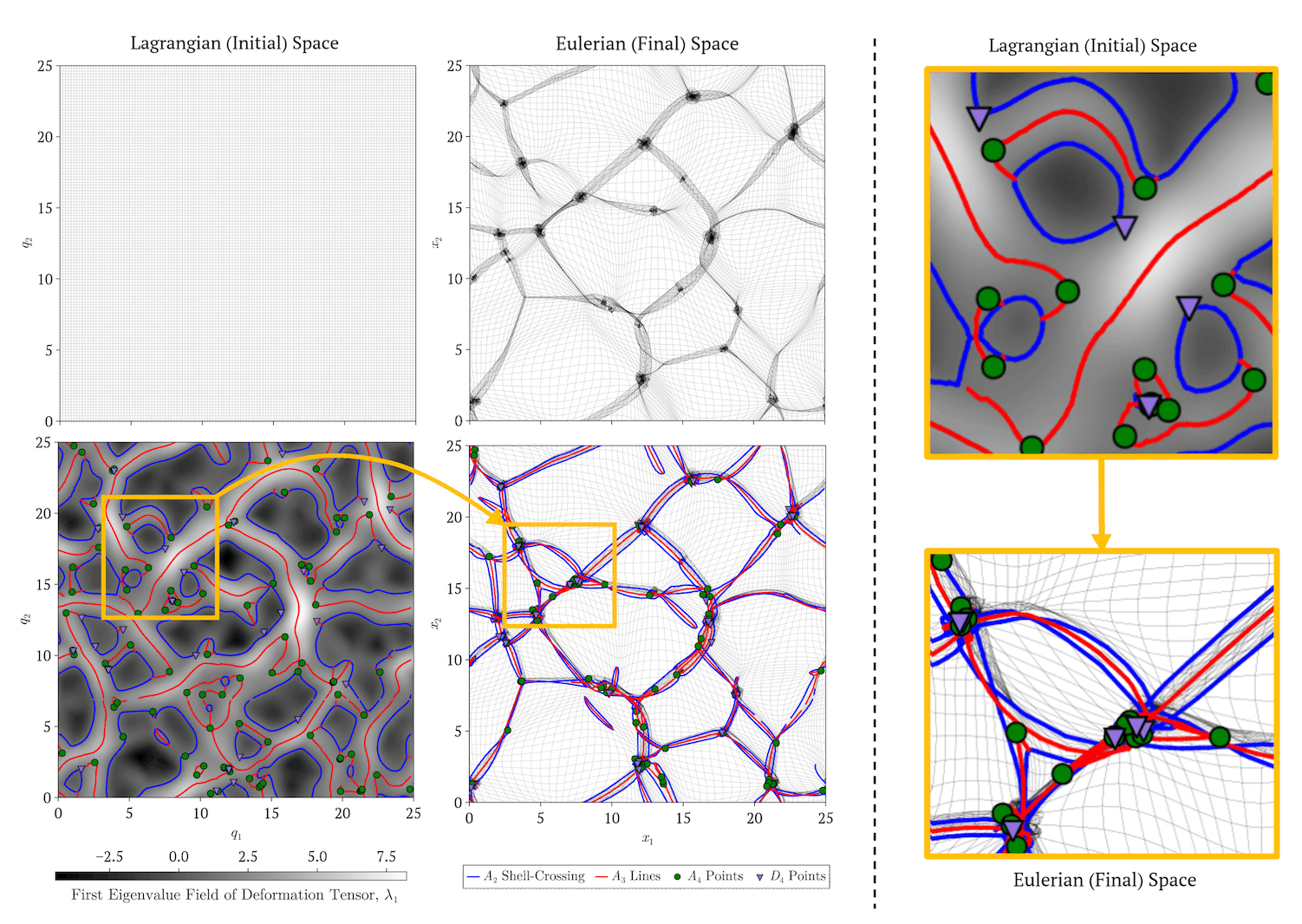}
\caption{Example of caustic skeleton theory applied to an unconstrained 2D $N$-body simulation. 
The Lagrangian-space initial particle displacements are shown in the top left panel, with the first eigenvalue field and caustics shown, also in Lagrangian space, in the panel below.
To the right, the final particle displacements are shown in the top panel with the Eulerian-space caustics overlaid in the panel below. 
On the right-hand side, the zoomed-in regions highlighted by the yellow boxes are displayed to demonstrate the evolution of caustics in the region from Lagrangian to Eulerian space.}
\label{fig:2dsim}
\end{figure}

We illustrate the caustic skeleton of the Zel'dovich approximation for a two-dimensional model of the cosmic web in \cref{fig:2dsim}. Starting with a homogeneous distribution of $N$-body particles (consisting of $128^2$ particles in a $25$ Mpc box, corresponding to a spatial resolution of $\sim 0.2$ Mpc), we perturb the particles slightly with the Zel'dovich approximation using the primordial gravitational potential modelled as a realisation of a Gaussian random field (see top-left panel of \cref{fig:2dsim}). For convenience, we smooth the displacement potential with a Gaussian filter at the scale $\sigma=0.7$ Mpc. Using a two-dimensional leapfrog $N$-body code \citep{nbody2d2020}, we evolve the initial conditions into the non-linear regime (see the top-right panel of \cref{fig:2dsim}). The pale grey single-stream regions mark the voids. The dark, thick lines due to the overlap of the particle mesh indicate multi-stream regions forming filaments and clusters.

The first eigenvalue field $\lambda_1$ and the associated caustics of the Zel'dovich approximation form a network marking the proto-filaments and clusters in the initial conditions (see the lower-left panel of \cref{fig:2dsim}). When mapped to Eulerian space with the displacement field $\bm{s}_t(\bm{q})$ of the $N$-body simulation, we successfully identify the multi-stream regions of the $N$-body simulation with the caustics of the Zel'dovich approximation (see the lower-right panel of \cref{fig:2dsim}).

We observe that valleys in the first eigenvalue field (where $\lambda_1$ is negative) correspond to voids.
The fold curves delineate the locations undergoing instantaneous shell crossing at the present time (in the Zel'dovich approximation). The red cusp curves, corresponding to filaments in the 2D cosmic web, follow the ridges of the eigenvalue landscape and are bounded by the folds. The swallowtail and hyperbolic/elliptic umbilic clusters form at locations where several cusp curves either intersect or change direction. This provides an intuitive demonstration of how these unique caustic structures depend on the eigenvalue field of the initial conditions' deformation tensor. It is worth noting that the $A_3$ caustics in two dimensions consist of a collection of points that collectively form a line, while the $A_4$ and $D_4$ caustic points remain isolated, as they are transient and disconnected in both space and formation time. 

We emphasise that the wealth of information contained within the bottom left panel of \cref{fig:2dsim} was solely extracted from the initial deformation tensor under the Zel'dovich approximation. Because caustics correspond to critical points of the Lagrangian map, the associated dense multi-stream regions are expected to give rise to structures that survive into the fully non-linear phase of gravitational evolution. The outcome is visible in Eulerian space, as shown in the bottom right panel of \cref{fig:2dsim}. As one can see, the $A_3$ lines perfectly align with the filaments predicted using our two-dimensional $N$-body simulation, while the $D_4$ and $A_4$ points correspond to the location of the clusters. It is also apparent that the $A_2$ fold lines, which delineate the multi-stream regions, end up quite closely hugging the $A_3$ lines as a consequence of the strong gravity around these high-density regions. Hence, remarkably, even though the linear Zel'dovich approximation is no longer valid in the non-linear regime, caustic skeleton theory applied to the Zel'dovich approximation accurately predicts the locations of these structures.

The yellow boxes in the bottom panels of \cref{fig:2dsim} highlight the same region in both the initial Lagrangian and final Eulerian distribution, and are shown zoomed in on the right-hand side. This demonstrates that the topology and connectivity of the structures present in the final distribution have already been captured by the caustics of the primordial field. In the Lagrangian-space panel at the top, we see three blue circles surrounding proto-voids. Although these voids change shape and grow or shrink, we can still see the correspondence of these structures between the Lagrangian and Eulerian spaces. 

\section{Data: Local Universe Constrained Conditions}
\label{sec:data}
The present study entails the \textbf{Caustic Skeleton} analysis in and around the Coma supercluster region and the
Pisces-Perseus ridge, extracted from the Manticore set of constrained reconstructions \citep{McAlpine:2025,McAlpine:2026}
based on the 2M++ (near-infrared) galaxy survey \citep{Huchra+2012}. Our analysis concerns the implied dark matter distribution,
while we also compare the observed galaxy distribution in the two cosmic web complexes with the implied and
evolved dark matter distribution of the corresponding Manticore simulations.

\subsection{Caustic Skeleton and Manticore reconstructions}
The caustic skeleton provides a parameter-free first-principles phase-space based dynamical framework for the formation and evolution of the cosmic web, and its structural components -- the cluster nodes, filaments, walls and voids -- and their classification.
Taking it to the realm of observational reality forms a major step and challenge for our understanding of the dynamics that shape the cosmic web and their impact on the formation and evolution of galaxies within its various structural components.
It is also crucial for studying the location and thermal state of the gaseous baryonic material that settles in and around the nodes, filaments and walls in our Local Universe.

Thus far, it has been almost exclusively concentrating on the ideal world of theoretical configurations
\citep{Feldbrugge+2018,FeldbruggeWeygaert2024,Hertzsch+2026} represented by a series of cosmological N-body simulations. 
In this paper, we bridge the theoretical and observational realms by evaluating the caustic structure of our Local Universe using the  reconstructed primordial conditions for the Local Universe yielded by the Manticore project. The Manticore project consists of $80$ constrained simulations based on the BORG (Bayesian Origin Reconstruction from Galaxies) method \citep{Jasche:2010,BORG2013,ata2015,Leclercq:2017}, providing access to the most plausible initial density field fluctuations of our Local Universe consistent with the 2M++ catalogue \citep{Huchra+2012}. These initial conditions are evolved to the present time using state-of-the-art $N$-body simulations, giving us unprecedented insight into the geometry and formation of our cosmological backyard.

The high-resolution Manticore simulations enable a considerable improvement in scope, detail and dynamic range of
the caustic spine of the cosmic web with respect to our earlier study of the caustic network in the
local Universe by \cite{Bixerman+2022}. The
study by \citep{Bixerman+2022} allowed a comparison of the inferred caustics with the distribution of clusters in
the local Universe, based on an earlier lower resolution BORG inferred set of initial conditions. The current
study allows an assessment of the cosmic web at much smaller scales, such that we will be able to relate it to
the details of the local galaxy distribution.

\subsection{Constrained Simulations}
The use of constrained cosmological simulations was initiated by Bertschinger \cite{Bertschinger1987}, who in a seminal publication developed the formalism for generating a field of Gaussian initial density and velocity perturbations that fulfilled one or more linear constraints \citep{Bertschinger1987,RybickiPress1992,Hofrib1991}. His formalism stated that any field realization $f(\bm x)$ can be written as the sum of a \textit{mean field} and a \textit{residual field}, 
  \begin{equation}
    f(\bm x)\,=\,{\bar f}(\bm x)+F(\bm x)\,,    
  \end{equation}
  in which the mean field ${\bar f}(\bm x)$ encapsulates the imposed constrained signal. Formally, it is the mean of
  all field realizations that fulfil the imposed constraints and the field that minimizes the root mean squared difference
  with all possible - constrained - field realizations. In the case of linear constraints, Bertschinger demonstrated it to be the
  weighted sum over the constrained (measured) values, the weight a direct expression of the correlation between
  the imposed constraints and the field through the field-constraint correlation functions. For a more extensive and detailed
  discussion of constrained simulations, see Appendix~\ref{app:constrsimul}. 
  
  The Manticore simulations used in the present study are based on a body of more advanced constrained reconstruction techniques that translate the nonlinear matter distribution, sampled by the observed galaxy distribution beset by incompleteness and various systematic effects, into representative primordial conditions with their inferred subsequent nonlinear gravitational evolution. 
  These statistical Bayesian inferece techniques involve the ability to translate the observed spatial distribution of galaxies within a given volume of the (local) Universe towards its implied primordial Gaussian density and velocity perturbations out of which these galaxies and structures emerged through gravitational growth. To this end, advanced and sophisticated Bayesian reconstruction schemes have been developed, which involve elaborate and computationally demanding  MCMC sampling procedures \cite{Kitaura:2008,Kitaura:2009, Jasche:2010,Leclercq:2015,Hess:2016, Bos:2016, Leclercq:2017, McAlpine:2022}. For large galaxy surveys in the local Universe, such as the 2MRS and SDSS surveys, impressive results have been obtained that allow us to study the properties and evolution of its detailed mass distribution. Telling examples are the 2MRS constrained simulations of the Local Universe by Kitaura and collaborators on the basis of their KIGEN formalism \cite{Hess:2013} (also see \cite{Hidding+2016}) and the recent detailed Sibelius simulation \citep{McAlpine:2022} and Manticore simulations \citep{McAlpine:2025,McAlpine:2026} on the basis of the BORG formalism. These simulations can be used to address a variety of problems,
  including BAO measurements and the removal of redshift space distortions.

\subsection{Manticore project: reconstruction of the Local Universe}
The primary goal of the Manticore project \citep{McAlpine:2025,McAlpine:2026} is to reconstruct the most plausible set of initial density fluctuations (in the form of the initial Gaussian white noise field $n_w({\bf q})$) that, under gravitational evolution, give rise to the present-day observed large-scale structure in the Local Universe. 

The displacement potential is then simply a rescaling of the white noise field by the $\Lambda$CDM potential power spectrum in fourier space:
\begin{equation}
\hat{\Psi}(\bm{k}) = \hat{n}_w(\bm{k})\sqrt{P(|\bm{k}|)}
\label{eq:wnf}
\end{equation}

The Manticore project improves upon previous BORG (Bayesian Origin Reconstruction from Galaxies) implementations \citep{Jasche:2010,Leclercq:2015,Leclercq:2017,BORG2013} by introducing a more flexible galaxy bias model, a generalised Poisson likelihood, and physics-informed priors that enforce statistical consistency with the standard $\Lambda$CDM cosmology (e.g., flat power spectrum for the white noise field, and appropriate moments for the final density field).

The pipeline uses the approximate gravity solver COLA to forward-model the density field during the inference process, utilising Hamiltonian Monte Carlo (HMC) sampling to explore the high-dimensional posterior of the initial conditions and bias parameters. After obtaining the posterior distribution of the initial white noise fields, the authors perform high-resolution dark-matter-only "re-simulations" using the \textsc{monofonIC} initial conditions generator~\cite{Hahn_2020} and the SWIFT $N$-body code~\cite{Schaller_2024} to obtain the final dark matter density field at $z=0$ that could plausibly host the observed galaxies.
 Since the resolution of the white noise field inference is only $3.9$ Mpc, smaller-scale Gaussian fluctuations consistent with $\Lambda$CDM are injected into the simulation by \textsc{monofonIC}. These re-simulations demonstrate an excellent agreement with observational data, accurately recovering the masses and positions of 14 prominent local galaxy clusters (e.g., Virgo, Coma, Perseus) and providing a highly accurate peculiar velocity field that surpasses previous models.

This suite of eighty Bayesian constrained white noise fields $n_w({\bf q})$ and corresponding posterior dark-matter only re-simulations are referred to as \texttt{Manticore-Local}. These are available through the Python \texttt{manticore$\_$data} package. In this project, we use the initial white noise fields and final configuration of the $N$-body particles of three of the \texttt{Manticore-Local} simulations. Specifically, we use the simulations marked by the flags  \texttt{mcmc = 0,1,2}. We refer to these simulations as  $M_0,$ $M_1$, and $M_2$.  

\subsection{Set-up before applying caustic skeleton theory}
The Manticore dark-matter only posterior re-simulations are performed in a periodic box of $L=1000$ Mpc with $N=1024^3$ particles giving a length scale resolution $\sim 0.98$ Mpc. However,  the region constrained by the 2M++ galaxy data only extends to $R < 200$ Mpc from the location of the Milky Way, which is equivalent to a redshift $z < 0.05$. The purpose of extending the simulation box far beyond the constrained region allows for periodic boundary conditions to be maintained without disturbing the density distribution towards the centre. For the analysis presented in this work, we do not use the full simulation volume. Instead, we extract a subbox of $L = 400$ Mpc centred on the location of the Solar System. This reduction significantly improves the computational efficiency of the caustic calculations, although it necessitates abandoning the periodic boundary conditions of the parent volume.

 As the \texttt{Manticore-Local} simulations do not provide the first order deformation tensor, we used \textsc{monofonIC} on the white noise fields (see \cref{eq:wnf}) to generate the linear displacement field $\bm{s}_{t}(\bm{q})\sim -\nabla_{\bm{q}} \Psi(\bm{q})$, under the Zel'dovich approximation. We then apply an 8th-order finite difference scheme on the components of the initial displacement field in the non-periodic $L=400$ Mpc sub-box to compute the deformation tensor  $\nabla_{\bm{q}} s_{t_0}(\bm{q}) = -D_+(t_0) \mathcal{H} \Psi(\bm{q})$ of the linear initial conditions scaled to today, $t_0$.
 
Due to the highly multi-scale nature of the cosmic web, directly calculating the caustic skeleton from the given simulation's initial displacement field would result in an overwhelming proliferation of small-scale structures. Thus, as mentioned in \cref{sec:caustic_class}, we first smooth the initial linear displacement potential  $\Psi_\sigma(\bm{q})$ using a Gaussian kernel with smoothing length $\sigma$. The caustic analysis can be tuned to probe structures from the largest scales (large smoothing) to the smallest scales resolved by the simulation grid (with minimal smoothing) by varying $\sigma$.

\subsection{Dark Matter Maps}
The dark matter density maps in our study are obtained from the Manticore simulation particle data by using the
Phase-Space Delaunay Tessellation Field estimator \citep{Feldbrugge+2024,FeldbruggeHertzsch2025}. It takes into account the
foldings in phase-space and associated caustics while approximating \cref{eq:density}. This density estimator generalises
and combines the Delaunay Tessellation Field Estimator introduced by \cite{SchaapWeygaert2000, WeygaertSchaap2009}, and
the earlier phase-space density field estimators by \cite{Shandarin2011, ShandarinSalmanHeitmann2012, AbelHahnKaehler2012, Hahn+2015}. 

\section{Skeleton of the Local Universe}
\label{sec:skel}
In our exploration of the intricate and complex weblike network in the Local Universe, we chose to study in detail the regions in and around two of the most prominent structures of the cosmic web. These include the large-scale structure around the Coma Cluster and the Pisces-Perseus chain, respectively the most massive cluster node and the most outstanding filamentary artery in our local neighbourhood. 
The surroundings of the Coma Cluster may rightfully be considered the region in which, for the first time, the reality and complexity of the weblike patterns in the galaxy distribution surfaced in the famous \textit{Stickman} map of de Lapparent and collaborators \citep{deLapparent1986}. 
The close vicinity of the two structures, the Coma Cluster and the Pisces-Perseus chain \citep{HaynesGiovanelli+1986,Hidding+2016}, augmented by the favourable orientation of the latter, will allow an optimal comparison of the predicted caustic features and the observed distribution of galaxies in and around them.

This will not only allow a quantitative scrutiny of caustic skeleton theory and its ability to predict, on the basis of phase-space dynamics, the intricacies of the multi-scale network around the Coma Cluster and Pisces-Perseus, but also facilitate a sharpening of the formalism and identification of possible caveats. 

\subsection{Coma Cluster and Coma Great Wall}
\label{sec:coma}

\begin{figure}
    \centering
    \includegraphics[width=0.8\textwidth]{ 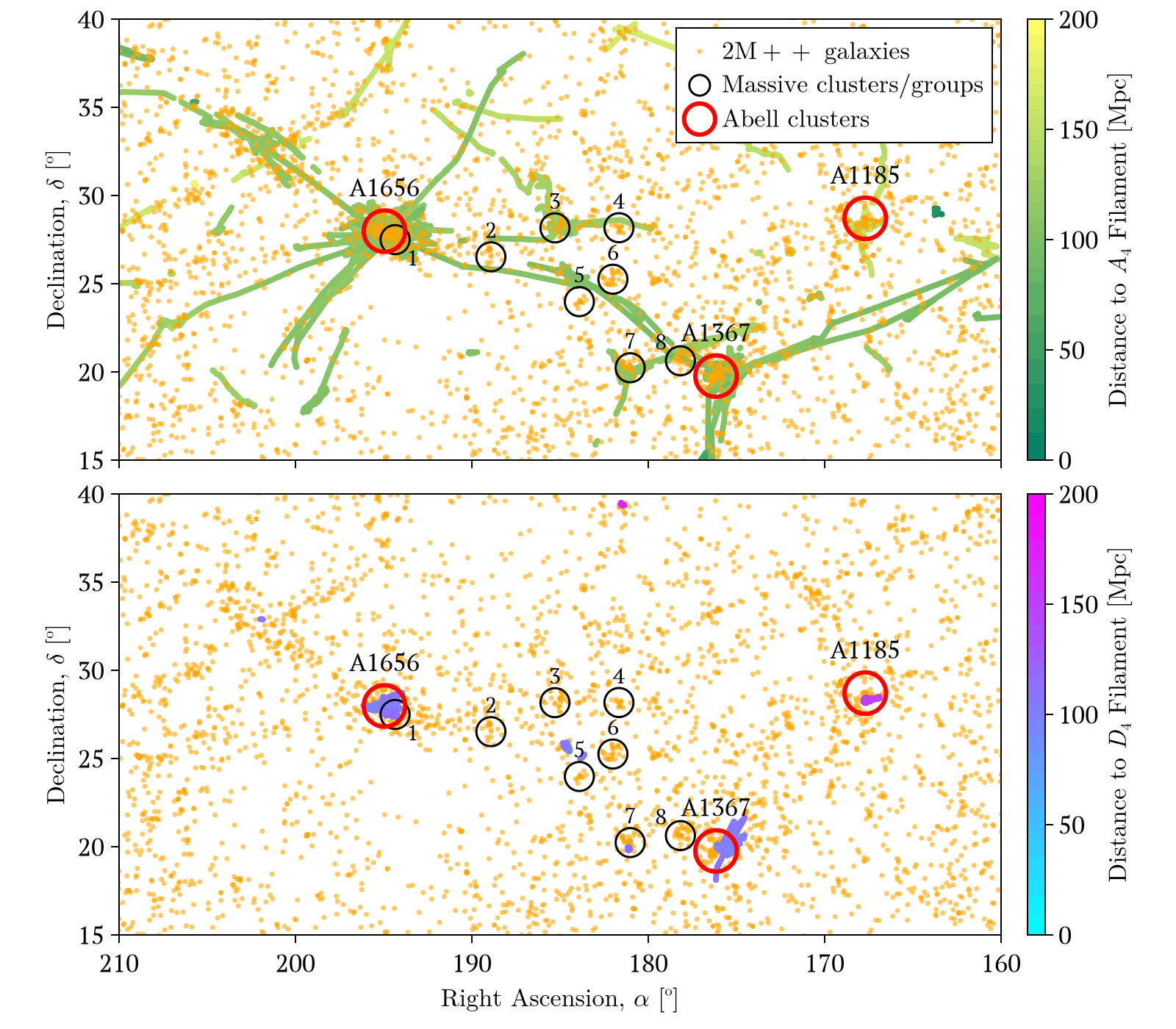}
    \caption{Projection of $\sigma = 3$ Mpc $A_4$ filaments (top panel) and $D_4$ filaments (bottom panel) within $R \leq 200$ Mpc onto sky in direction of Coma cluster (A1656). Filaments are calculated from $M_1$ realisation and are coloured according to distance along line of sight. All 2M++ galaxies within this section of the sky are shown. Other prominent Abell clusters in this region (A1185 and A1567) are also shown. Here, the A1656 and A1367 clusters are clearly bridged by the $A_4$ filaments. A few massive clusters/groups are circled in black: (1) NGC4839, (2) NGC4555, (3) NGC4295, (4) NGC4104, (5) NGC4213, (6) NRGb184, (7) NGC4065, (8) NGC3937. See Appendix \ref{appendixC} \cref{fig:coma_radec_slice} for additional materials showing slices of caustic skeleton in this region of the sky, including both walls and filaments, at various distances.}
    \label{fig:coma_radec}
\end{figure}

\begin{figure}
\centering
	\includegraphics[width=\textwidth]{ 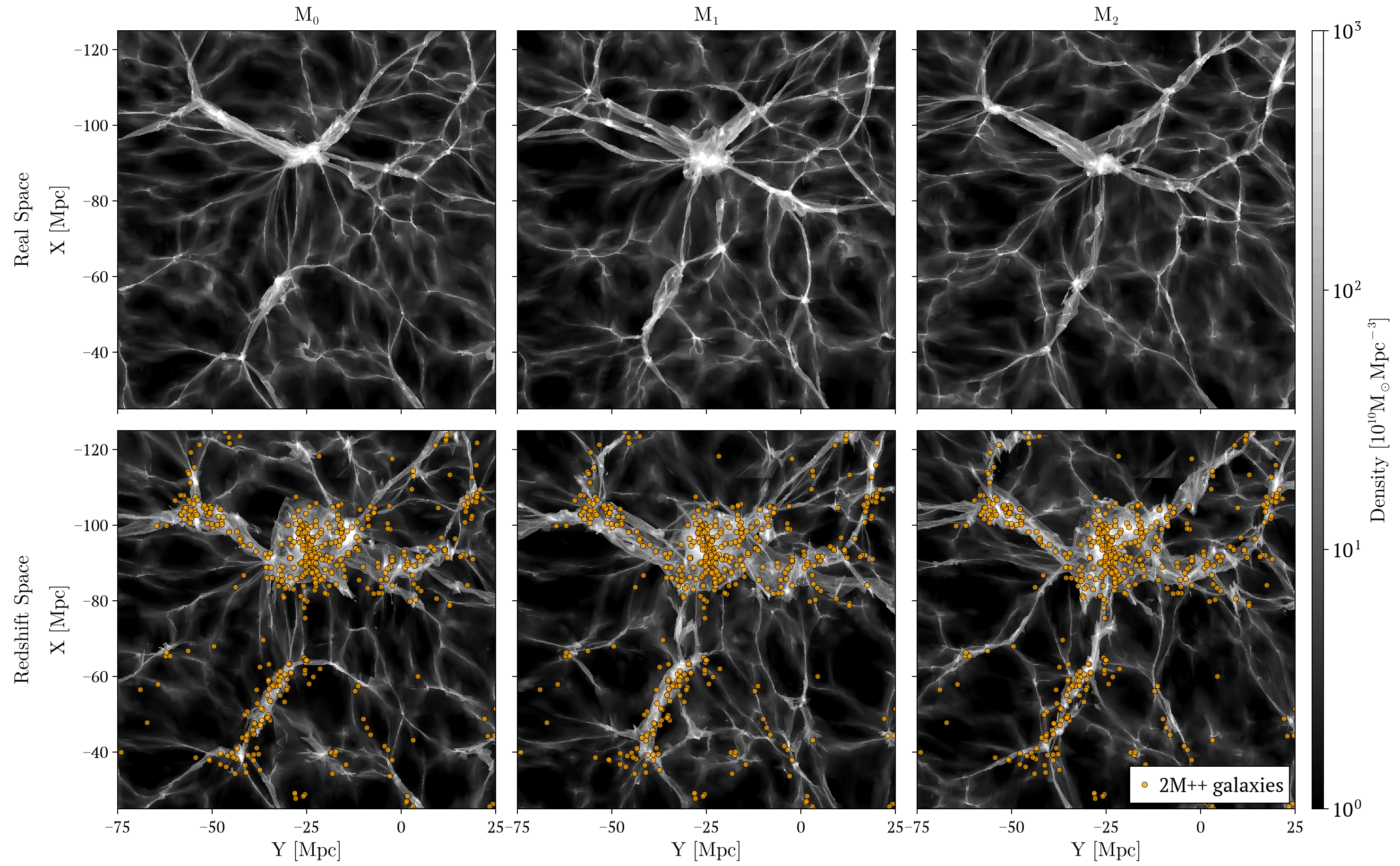}
    \caption{This figure shows the dark matter density field around the Coma Supercluster in a $100$ Mpc square region of the galactic $(X,Y)$ plane for three different \texttt{Manticore-Local} simulations (left to right: $M_0$, $M_1$, $M_2$). This is obtained by taking a 2D slice through the calculated 3D simulation density field at $Z_\mathrm{Coma} = 51.42$ Mpc, the observed galactic Z-coordinate of the Coma Cluster. Top panels: Dark Matter only density fields in real-space coordinates given by \texttt{SWIFT} final $z=0$ snapshot. Bottom panels: Dark Matter only density fields projected into redshift space with simulation peculiar velocity fields. Redshift space slices are taken at same Z-coordinate. Orange points show 2M++ galaxies -  which the \texttt{Manticore-Local} density fields were reconstructed from - within a $10$ Mpc thick slice around $Z_\mathrm{Coma}$. The density is estimated with the Phase-Space Delaunay Tessellation Field Estimator \citep{Feldbrugge+2024, FeldbruggeHertzsch2025}.}
    \label{fig:coma_den}
\end{figure}

\begin{figure}
\centering
	\includegraphics[width=\textwidth]{ 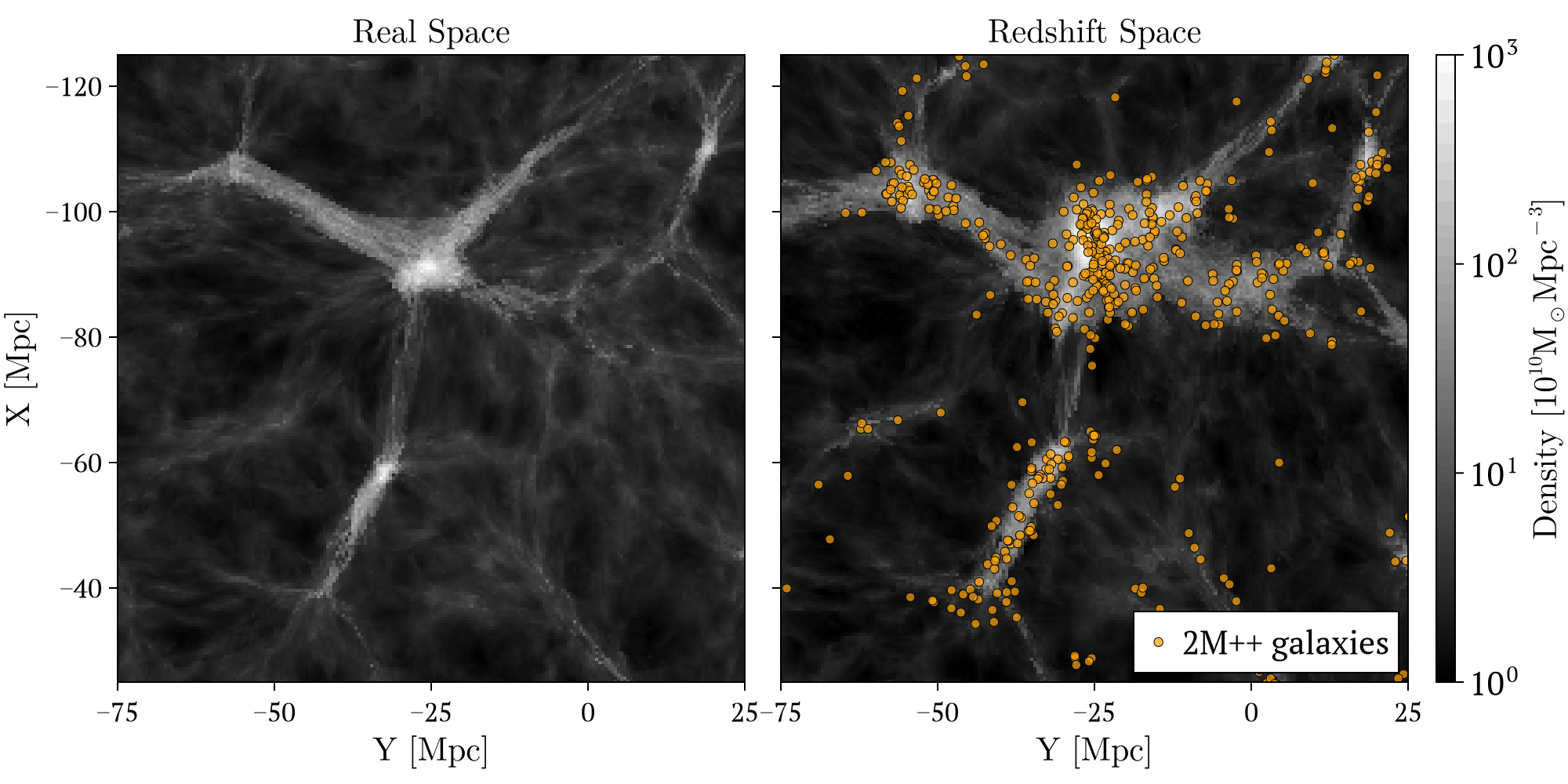}
    \caption{This figure shows the median density field around the Coma Cluster of 10 different \texttt{Manticore-Local} realisations in real space (left) and redshift space (right).
    The density in a 2d slice at $Z_\mathrm{Coma}$ is estimated for each realisation with the Phase-Space Delaunay Tessellation Field Estimator \citep{Feldbrugge+2024, FeldbruggeHertzsch2025} as shown in \cref{fig:coma_den}, and the median value of the density in each pixel is shown here.
    Orange points show 2M++ galaxies within a $10$ Mpc thick slice around $Z_\mathrm{Coma}$. }
    \label{fig:med_den}
\end{figure}

\begin{figure}
\centering
	\includegraphics[width=\textwidth]{ 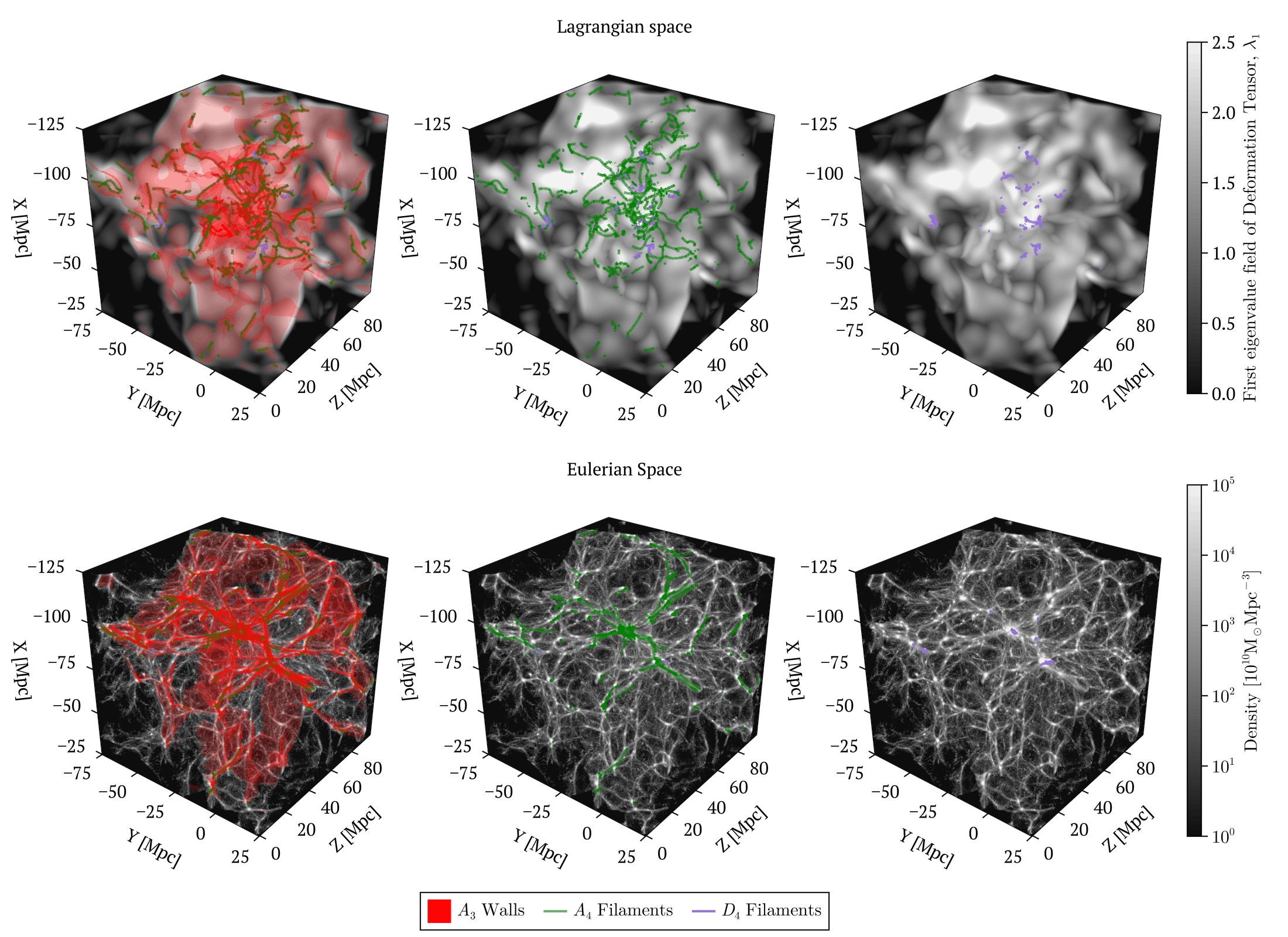}
    \caption{This figure shows the caustic skeleton around the Coma Cluster in both Lagrangian space (top panels) and Eulerian space (bottom panels) in a $L = 100 $ Mpc box extracted from one example of a \texttt{Manticore-Local} Universe re-simulation (refer to as $M_1$). We use the smoothing scale of $\sigma =3$ Mpc. Left: $A_3$ walls with $A_4$ and $D_4$ filaments. Middle: Both $A_4$ and $D_4$ filaments. Right: $D_4$ filaments alone. The 3-dimensional first eigenvalue field of the deformation tensor is also included in the top Lagrangian-space panels, and the density field in the bottom Eulerian-space panels. The density is estimated with the Phase-Space Delaunay Tessellation Field Estimator \citep{Feldbrugge+2024, FeldbruggeHertzsch2025}. A rotating and zoom-in animation of this figure is available on the additional video materials page at \href{https://jfeldbrugge.github.io/Caustics-Local-Universe/}{jfeldbrugge.github.io/Caustics-Local-Universe/}.}
    \label{fig:3d}
\end{figure}

\begin{figure}
\centering
\includegraphics[width=0.9\linewidth]{ 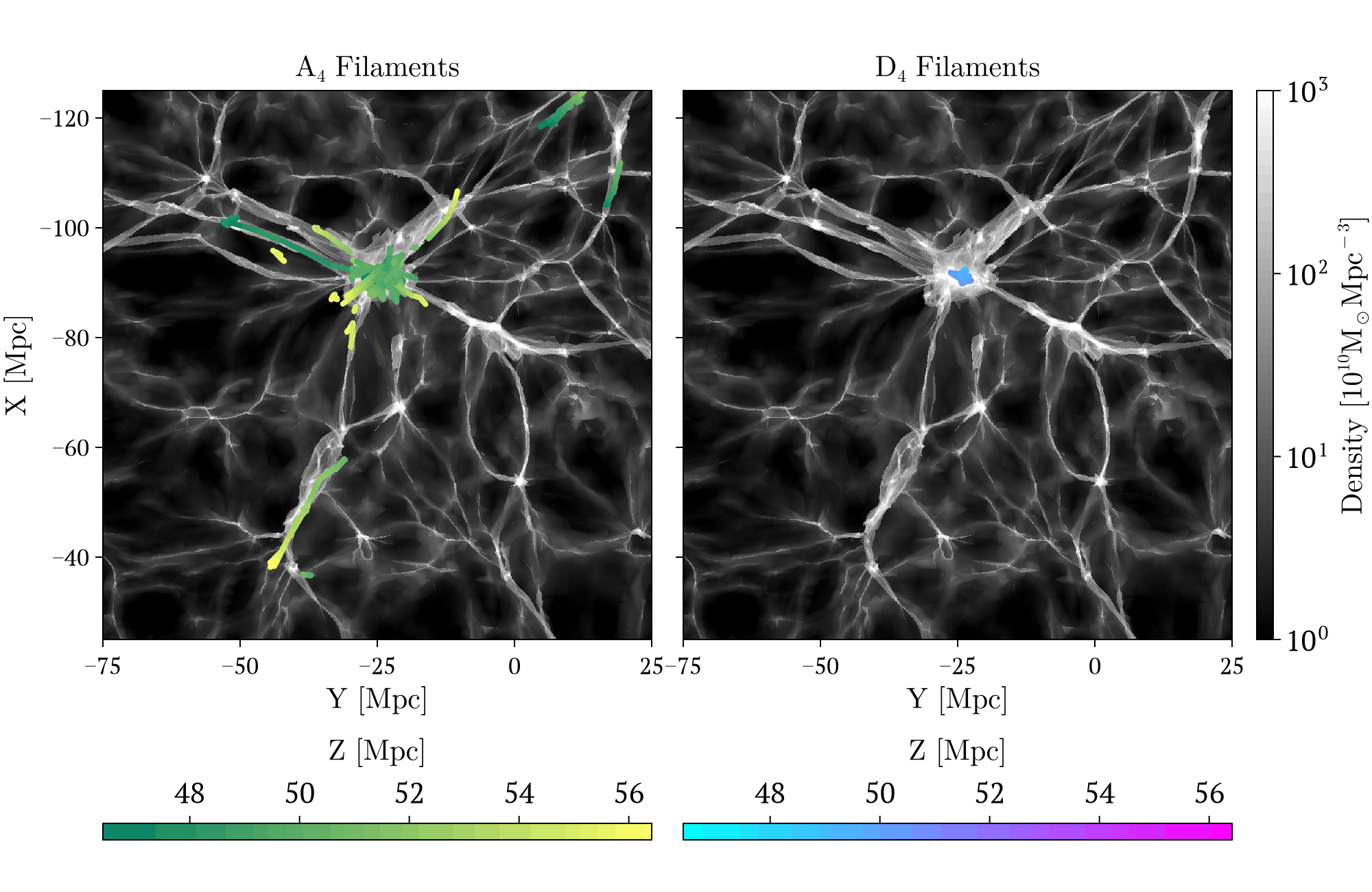}
\caption{Filamentary ($\sigma = 3$ Mpc) caustics of the Coma Supercluster for $M_1$ simulation within $10$ Mpc of $Z_\mathrm{Coma} = 51.42$ Mpc projected onto X-Y galactic plane in $100$ Mpc square region. Z-coordinates of filaments are shown for $A_4$ filaments and $D_4$ filaments on the bottom left and right, respectively. 2D real-space density field slice of $M_1$ simulations underlaid, same as upper middle panel of \cref{fig:coma_den}. The density is estimated with the Phase-Space Delaunay Tessellation Field Estimator \citep{Feldbrugge+2024, FeldbruggeHertzsch2025}.}
\label{fig:slice}
\end{figure}

\begin{figure}
\centering
	\includegraphics[width=\textwidth]{ 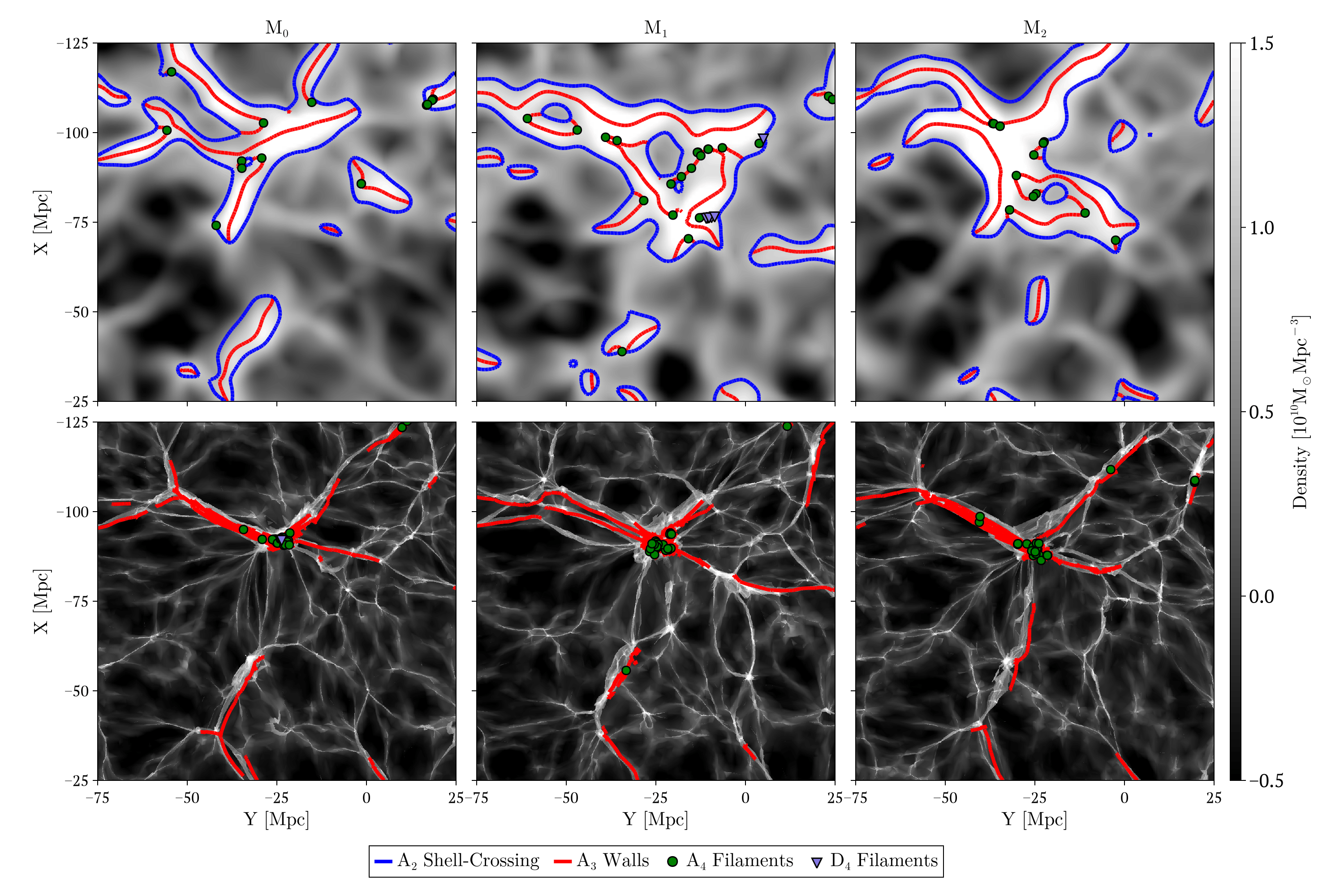}
    \caption{This figure shows the caustic skeleton at a $\sigma = 3$ Mpc smoothing scale of the Coma Supercluster in a $100$ Mpc square region of the galaxtic $(Y,X)$ plane in real space for three different \texttt{Manticore-Local} simulations ($M_0$, $M_1$, $M_2$). This is obtained by taking an infinitesimally 2D slice through the calculated 3D skeleton and simulation density field in real space at $Z_\mathrm{Coma} = 51.42$ Mpc, which is the observed Z-coordinate of the Coma Cluster.
    Blue lines correspond to instantaneous shell-crossing $A_2$ folds. Red lines show the intersection of 3D walls with a 2D slice. Intersection of $A_4$ and $D_4$ filaments with slice is shown by green points and purple triangles, respectively. 
    Top panels: Caustic skeleton in Lagrangian space with first eigenvalue field $\lambda_1$ of deformation tensor underlaid. Bottom panels: Caustic skeleton projected in Eulerian space with the simulation density field underlaid. The density is estimated with the Phase-Space Delaunay Tessellation Field Estimator \citep{Feldbrugge+2024, FeldbruggeHertzsch2025}.}
    \label{fig:coma_sigma_LE}
\end{figure}

\begin{figure}
    \centering
    \includegraphics[width=\textwidth]{ 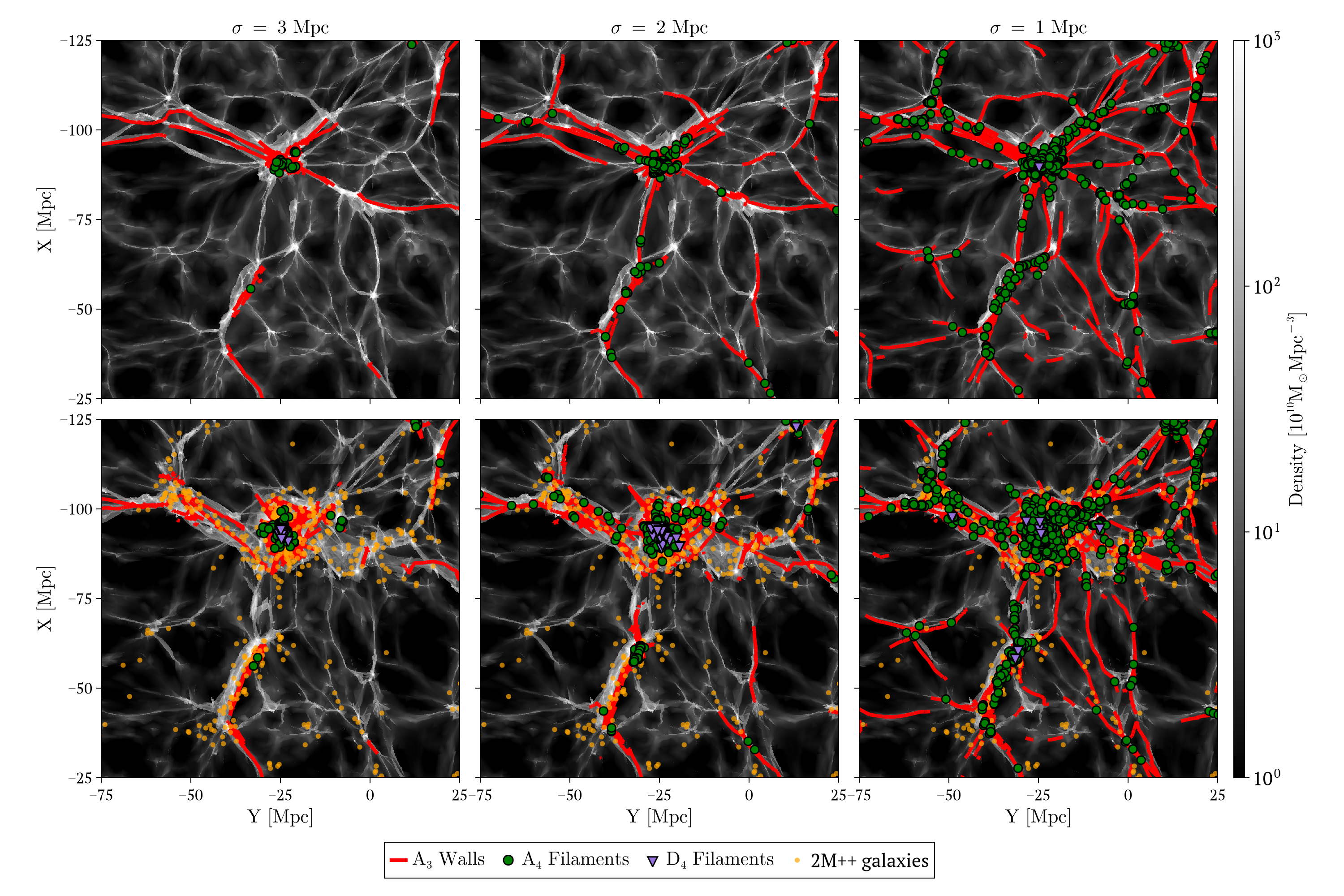}
    \caption{2D slice through caustic skeleton of Coma Cluster at $Z_\mathrm{Coma} = 51.42$ Mpc in $M_1$, one example of a Manticore Local Universe re-simulation. Red lines show the intersection of 3D walls with the thin 2D slice. Intersection of $A_4$ and $D_4$ filaments with slice is shown by green points and purple triangles, respectively. Left to right from larger scales at $\sigma = 3$ Mpc to progressively smaller scales at $\sigma = 2$ Mpc, then $\sigma = 1$ Mpc. Top panels: Caustic skeleton and simulation density field in real-space as given by the $M_1$ simulation's final $z=0$ snapshot. Bottom panels: Caustic skeleton and density field in redshift space with 2M++ galaxies in a $10$ Mpc slice around $Z_\mathrm{Coma}$ Mpc overlaid. The density is estimated with the Phase-Space Delaunay Tessellation Field Estimator \citep{Feldbrugge+2024, FeldbruggeHertzsch2025}.}
    \label{fig:coma_sigma_redshift}
\end{figure}

The Coma cluster, Abell 1656, with a total mass of around $\sim 7 \times 10^{14} M_\odot$, is the most
massive cluster node in the immediate vicinity of the Local Universe. Located at $\alpha = 12^\mathrm{h} 59^\mathrm{m} 48.7^\mathrm{s}$, $\delta = +27^\circ 58' 52''$, with a recessional velocity relative to the CMB of $v_\mathrm{CMB} = 7463$ km/s it has a distance of $\approx$ 103 $\rm{Mpc}$. Due to its prominence and relative vicinity, the Coma cluster has been the subject of a large number of studies. 
The Coma Cluster is connected through a strong filamentary bridge to the Leo Cluster, Abell 1367, and together they form the heart of the Coma supercluster \citep[see e.g.][]{EinastoM:2025}. The filamentary extension is even seen in the X-ray maps
of the region \citep{Mirakhor2020}.

The Coma Cluster has long been considered the canonical example of a relaxed, old, massive galaxy cluster. However, subsequent
studies revealed that it contains significant substructure \citep{Biviano+1996,Caretta+2023}. In particularly significant is the
realization that the Coma cluster resides at the node of an intricate network of filamentary extensions
\citep{deLapparent1986,Platen2009,Tempel+2017,Malavasi:2023,Mondelin2025,EinastoM:2025,EinastoM+2025,Pedratti+2026,Benisty2026}, of which most are embedded in one of the
largest known superstructures in the nearby universe, the flattened wall-like Coma Great Wall
\citep{GellerHuchra1989,Platen2009,EinastoM:2025}. 
Indeed, the surroundings of the Coma Cluster may rightfully be considered the region in which, for the first time, the reality and complexity of the weblike organization of the galaxy distribution was revealed. The famous \textit{Stickman} map of de Lapparent and collaborators \citep{deLapparent1986} unveiled the embedding of the Coma cluster in a surrounding network of walls and filaments, in
which it stood out as the dominant compact head with branching filaments stretching out to neighbouring structures, which in
conjuncion with the boundary of the foreground void represents a striking resemblance of a bodily form. 

While visual inspection of the Coma supercluster region testifies of it forming the node within an intricate weblike network,
it has been challenging to obtain an objective and unequivocal characterization of its detailed geometry. Different analysis
tools reveal substantial variations in the detected filamentary network, reflecting differences in the corresponding
definition for filaments, criteria for structural connectivity, influence of systematic effects such as redshift distortions and
a range of technical aspects. For example, the use of the Bisous formalism identifies up to six filaments connecting to the
Coma cluster, amongst which the prominent A1656-A1367 bridge \citep{Tempel+2017,EinastoM:2025}, while Malavasi et al.\cite{Malavasi:2023} find three prominent large-scale filamentary extensions with the DisPerSE algorithm
(also see \citep{Malavasi:2020}. The recent detailed kinematic analysis of DESI DR1 spectroscopic data by \citep{Pedratti+2026} is also interesting as they report evidence for a potential filamentary extension along the line of sight.

It is by invoking the phase-space dynamical framework of the caustic skeleton formalism, we may expect to
obtain a more solid, dynamically justified, view on the structural intricacy around the Coma cluster. 
The structure and connections identified here are dynamically justified as opposed to filaments identified in this region with
methods such as \texttt{DisPerSE} \cite{Malavasi:2020,Malavasi:2023}. In the present section we provide additional insight into the
extraction and reality of the filamentary skeleton around the Coma cluster.
\\
\paragraph{Coma Caustic Skeleton}
\Cref{fig:coma_radec} shows the presence of caustic filaments in the region of the Coma supercluster, as viewed in the sky in RA and DEC, specifically extracted
from the $M_1$ Manticore realization. The figure distinguishes between the two classes of caustic filaments, the $A_4$
swallowtail filaments (top panel) and the $D_4$ umbilic filaments (bottom panel). The two classes are clearly of an entirely
different nature.

The $A_4$ filaments represent the strong filamentary arteries of the cosmic web, arranged in an intricate branching network.
The most outstanding feature is the strong extended bridge between the Coma cluster, A1656, and the other massive cluster A1367 in the Coma Great Wall. Other massive clusters in this region are circled in black and mostly sit along this filamentary spine and its off-shoots. 
At a scale of $\sigma = 3$ Mpc, we find $D_4$ filaments around the Coma cluster and A1367, and near the massive clusters NGC4065 and NGC4213/NRGb184. They appear to remain more
compact and are mainly concentrated in and around the prominent Abell clusters in the region. It seems to suggest they are to be identified with the massive filamentary extensions in the outer regions of cluster nodes.

\paragraph{Coma Manticore realizations}
The three selected Manticore re-simulations used in our study - $M_0, M_1, M_2$ - do all succesfully reveal the weblike network
around the Coma cluster. \Cref{fig:coma_den} illustrates this identification, showing the an infinitesimally thin slice of the
dark matter density field for the $M_0$, $M_1$ and $M_2$ re-simulations within a $100 \times 100$ Mpc region (in galactrocentric cartesian coordinates) centred on the Coma (super)cluster.  The slice is taken at $Z_\mathrm{Coma} = 51.42$ Mpc for all simulations, which is the observed position of the cluster in redshift space.

The upper panels of \cref{fig:coma_den} show the Coma cluster region in real space as it appears in the simulations' final $z=0$ snapshot. All three realizations reveal characteristic filamentary arms emanating from the Coma cluster, in particular the long extensions
along the $Y$ direction, that historically were considered to form the "arms" of the Stickman in de Lapparent et al.'s original CfA2 Slice of the Universe \citep{deLapparent1986}.
Furthermore, it is interesting to notice the considerable variation in substructure of these realisations, comprising either massive thick arteries or consisting of an assembly of several more tenuous thin filaments. Towards the lower end of the each image we observe even more variation. The three Manticore realizations agree on the existence of a moderate short filament in between $X = -40\sim -60$ Mpc, surrounded by a more varied pattern of small voids and tenuous tendrils. Only for $M_3$ we see a stronger connection between the moderate filament and the Coma complex. As discussed in \cref{sec:data}, these variations are result of the probabilistic Bayesian inference process, and the random fluctuations inserted by \textsc{monofonIC} on scales below 3.9 Mpc, the resolution of the constraints.

Corresponding redshift maps are shown in the lower panels of \cref{fig:coma_den}. It is
obtained by including the peculiar velocity of mass elements along the radial direction, as seen from our perspective.
The transformation allows a direct morphological comparison with observational data. It allows an assessment of the
level of agreement between the reconstruction and the observed location, in redshift space, of the galaxies in the
2M++ survey. These galaxies -- the same dataset used by the BORG framework to constrain the Manticore initial conditions --
are superimposed as orange dots on the redshift map. The galaxies are plotted from a $10$ Mpc thick slice in the $Z$-direction,
centered on the thin two-dimensional simulation slice taken at $Z_\mathrm{Coma}$. It is good to observe the high level of agreement
between the reconstructed redshift maps and the the galaxy sample in redshift space. In particular in the regions that have been
densely sampled the agreement is almost perfect. On the other hand, in less sampled regions we may notice a larger variation in
structure, a direct reflection of the fluctuations implied by the power spectrum.

Also outstanding is the considerable structural distortion of the redshift map in and around the Coma cluster itself, the product of several redshift distortion effects (RSDs, see \cite{Hamilton1998} for an instructive review). Prominently visible is the radial stretching of the central core of the cluster along the redshift direction, the finger-of-God resulting from velocities of galaxies - exceeding more than 1000 $\rm{km} \rm{s}^{-1}$ - swirling around in the virialized core. Equally interesting is the extended trumpet-shaped area
surrounding the inner cluster, induced by the infall of galaxies on to the cluster within the turnaround radius. Because it is very
dense and massive, galaxies out to considerable distances from the Coma cluster galaxies are pulled in by its gravity and peculiar
infall velocities high enough to exceed the corresponding Hubble expansion. The resulting strongly distorting effects were first
pointed out by Kaiser \citep{Kaiser1987}.
In all, the quasi-linear RSDs affecting the Coma cluster, ranging from the
contraction of filaments and coherent cluster infall, appear to have been sufficiently accounted for by the Manticore reconstruction process. 


To appreciate the level to which the 2M++ galaxy sample constraints manage to fix the weblike pattern of the (dark) matter
distribution, we also assess the variation between the different realizations. To this end, we compute the mean (median) density
field of 10 of the \texttt{Manticore-Local} re-simulations (\texttt{mcmc = 0,1,2,10,11,12,13,14,15,16}) in both real and redshift space. 
The 2d slices in the panels of \cref{fig:med_den} show the resulting pattern in the X-Y plane. Here, structures smaller than the 3.9 Mpc scale of the constrained resolution are washed out, whereas the strongly constrained structure above this scale remains clearly visible. Some 3 to 4 filamentary structures appear to protrude from the Coma cluster in both real and redshift space. The agreement between the median density field and the 2M++ galaxy distribution is particularly striking, evidencing the high quality of the constrained
reconstructions. 

The median field reveals the statistically most significant structures determined by observations and is therefore present in
all or most individual reconstruction realizations. It is in these regions that the most definite conclusions can be drawn about
the large-scale structure surrounding the Coma cluster. On the other hand, the more tenuous features that appear in one of a minor
fraction of individual realisations, and remaining virtually invisible in the median field, are not strongly constrained by
the 2M++ galaxy dataset underlying \texttt{Manticore-Local}. This is mostly a function of the distance of a region from the
galaxies in the constraining
sample, given that the strength of the constraints depend on the corresponding spatial correlations \citep[see e.g][]{Bertschinger1987,RybickiPress1992,Hofrib1991,FeldbruggeWeygaert2023}. As mentioned in \cref{sec:data}, these more sporadic features are
statistically implied, a manifestation of the spectral perturbations of the cosmic matter distribution. 

From here on in this study we mainly restrict our analysis to the caustic skeleton of the $M_1$ realisation shown in the centre panels
of \cref{fig:coma_den}. This caveat should be kept in mind, as the caustic skeleton calculated down to the resolution of an individual realisation will inevitably include such unconstrained features. A full consideration of the caustic structures of the median constrained field that are real and properly constrained is left for future work.

\paragraph{The Coma Caustic Skeleton}
We apply caustic skeleton theory to the $M_1$ realisation to extract the caustic skeleton in three dimensions, shown in Lagrangian (top panels) and Eulerian (bottom panels) space in \cref{fig:3d}. Here we used a smoothing scale of $\sigma = 3$ Mpc. 
The left panels show the whole network of both walls and filaments, whereas the middle panels display the $A_4$ and $D_4$ filaments, and then only $D_4$ filaments in the right panels. 
We defer the identification of higher-order caustics (such as $A_5$ and $D_5$) to future studies. 
The top left panel shows the strong dependence of the $A_3$ walls in Lagrangian space on the first eigenvalue field of the deformation tensor. In comparison, the dependence of the $A_4$ and $D_4$ filaments is more complicated due to the involvement of higher-order directional derivatives and the second eigenvalue field, respectively, in the caustic conditions listed in \cref{tab:conditions}. Nonetheless, since both of these filaments are constrained to sit on the $A_3$ walls, there is still a clear connection between both filament types and the eigenvalue field.

As one can see in the bottom Eulerian space panels, the $A_4$ filaments follow the prominent filamentary structures of the density field and appear to be densely entangled at the location of the Coma Cluster, located towards the centre of the box, suggesting a complex formation history for this galaxy cluster.
Even though the $D_4$ filaments are also present in the bottom middle panel of \cref{fig:3d}, they are clearly much less abundant and occupy the regions where $A_4$ filaments are already present. Although these different types of filaments cannot be fully disentangled in Eulerian space at this smoothing scale, it is still important to identify regions that are influenced by both filaments since this indicates a distinct and complex formation history.

To highlight the quality of the filamentary reconstruction with caustic skeleton theory of the Coma supercluster region, we show a $10$ Mpc thick slice, around $Z_\mathrm{Coma}$, of filamentary caustics from the 3D skeleton reconstruction of the $M_1$ realisation at the same smoothing scale $\sigma = 3$ Mpc. In \cref{fig:slice}, these 3D filaments are projected onto the 2D slice of the density field at $Z_\mathrm{Coma}$ Mpc. $A_4$ and $D_4$ type filamentary caustics are separated into the left and right panels, and both are coloured according to their Z-coordinate to retain depth information in this 2D projection. At this smoothing scale, it is apparent that the main filamentary arms are composed of $A_4$ filaments, whereas the $D_4$ filaments are constrained to the densest central regions of the cluster. 

Furthermore, as opposed to a 2D projection of the skeleton, we show an infinitesimally thin 2D slice of the 3D skeleton reconstruction for the $M_0, M_1, M_2$ realisations in \cref{fig:coma_sigma_LE}. 
The top panels show the first eigenvalue field $\lambda_1$ of the deformation tensor in Lagrangian space for each realisation, to illustrate the formation of the caustics in the Lagrangian map and their nature.  The blue contours mark the shell-crossing regions. The red lines represent the $A_3$ wall structures (see the left panel of \cref{fig:3d}) that intersect the plane of this thin slice. The purple and green points represent sliced $D_4$ and $A_4$ filaments, respectively. From this figure, it is further evident that at this scale $D_4$ filaments are rare in Coma for all realisations. 
We note that even though $M_0, M_1, M_2$ realisations differ in Lagrangian space, they look similar in Eulerian space. 
This suggests that the \texttt{Manticore-Local} reconstructions arrive at the present-day constrained structures via slightly different formation histories. In future work, we will examine the caustic skeleton of the mean constrained field to characterise the average formation history captured by the \texttt{Manticore-Local} re-simulations. Although the Manticore project currently provides the state-of-the-art Local Universe reconstruction, the value of caustic skeleton theory will grow as data-constrained reconstructions improve and place tighter constraints on formation history.

Nonetheless, \cref{fig:coma_sigma_LE} shows the walls in this slice reaching outwards from the Coma Cluster, which itself becomes discernible as a dense bunch of filaments in all three simulations.
This is already a clear demonstration of the power of caustic skeleton theory. Starting from the tiny fluctuations in the initial conditions, this technique can predict the formation of filaments and clusters in the cosmic web with great accuracy.  Again, the success of caustic skeleton theory with the Zel'dovich approximation is underpinned by the idea that caustics outline the emergence of multi-stream regions that inevitably collapse into bound structures under gravity. Despite this being a known success prior from unconstrained simulations, the fact that caustic skeleton theory captures the features present in the \texttt{Manticore-Local} re-simulations so well is particularly significant, as it opens up a wealth of information about the topology and formation history of our Universe.

These conclusions are drawn using simulations relying on a smoothing scale of $\sigma = 3$ Mpc. At this length scale, many smaller structures remain unresolved. Indeed, $3$ Mpc is significantly larger than the $0.98$ Mpc mean inter-particle spacing of the simulation. However, since the resolution of the \texttt{COLA} forward model and the BORG-constrained white noise fields have a resolution of $\sim 3.9$ Mpc, we expect density fluctuations below this scale to be dominated by unconstrained, randomly generated small-scale power injected during the initial conditions generation. 
Nonetheless, it is worth exploring the potential of this approach at smaller scales in order to investigate the level of detail recoverable around a galaxy cluster, keeping in mind the limitations of the current reconstruction.

We have already analysed the structure of the greater Coma Cluster-Great Wall complex, so now we can focus on the smaller structure at the scale of the Coma Cluster. The challenge here is whether the application of caustic skeleton theory to the Zel'dovich approximation is able to deconstruct the cosmic web down to a small scale close to the resolution limit of the simulation. For this study, we consider three values of the smoothing scale, namely $\sigma = 1,2, 3$ Mpc, as shown in the upper panels of \cref{fig:coma_sigma_redshift} for the $M_1$ simulation in real space. In the lower panels, we compare the observed 2M++ galaxies to the caustic skeleton projected into redshift space for the different smoothing scales. 
At larger scales, the fewer but longer more coherent structures, mostly consisting of walls, are highlighted. Whereas towards the smaller scales, the number of filaments detected increases significantly and the voids are filled with finer structures to the point that nearly all the structure present in the simulation is captured.

Similarly to the density field, we use the velocity field to reintroduce the redshift space distortions (RSD) and draw the caustics into redshift space.  Although they inevitably appear messier due to the nature of RSDs, one can see that the observed galaxies mostly sit along the $A_3$ walls shown in this slice. Moreover, we observe that the higher concentration of galaxies coincides with a higher concentration of filamentary points in the 2D slices shown in \cref{fig:coma_sigma_redshift}. This demonstrates the remarkable potential of caustic skeleton theory in accurately describing the level of detail at the scale of a galaxy cluster and also highlights the influence of the largest caustic scales.
It is quite remarkable that the most outstanding filamentary and wall-like features of the simulation and observed galaxy distribution have been successfully captured by the $A_3$ walls and $A_4$ filaments with just one eigenvalue and eigenvector field. Moreover, the $D_4$ filaments, which collect at the densest parts of the simulation field, may serve as an indication of the transition to more complex node-like regions and dynamics.

\subsection{The Pisces-Perseus filament}
\label{sec:pp}

\begin{figure}
    \centering
    \includegraphics[width=0.8\textwidth]{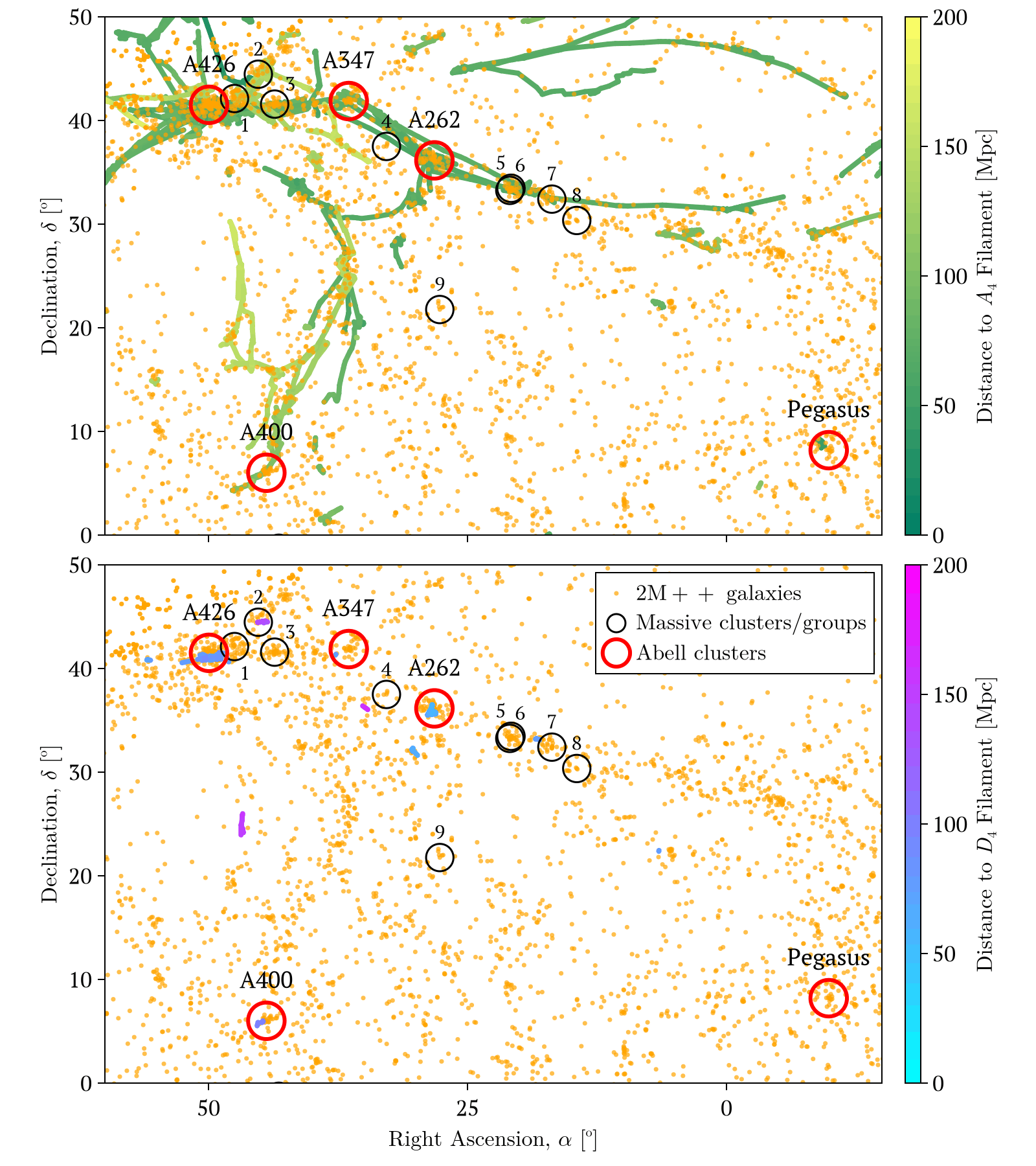}
    \caption{Projection of $\sigma = 3$ Mpc $A_4$ filaments (top panel) and $D_4$ filaments (bottom panel) within $R \leq 200$ Mpc onto sky in direction of Pisces-Perseus supercluster. Filaments are extracted from $M_1$ realisation and are coloured according to distance along line of sight. All 2M++ galaxies within this section of the sky are shown. Prominent Abell clusters in this region (A426, A347, A262, A400 and Pegasus) are shown. A few massive clusters/groups are circled in black: (1) UGC 2562, (2) CIZAJ0300.7+4427, (3) AWM7, (4) N841, (5) NGC499, (6) NGC507, (7) NGC383, (8) N315, (9) N691. See Appendix \ref{appendixC} \cref{fig:pp_radec_slice} for additional materials showing slices of caustic skeleton in this region of the sky, including both walls and filaments, at various distances.}
    \label{fig:pp_radec}
\end{figure}

\begin{figure}
\centering
	\includegraphics[width=\textwidth]{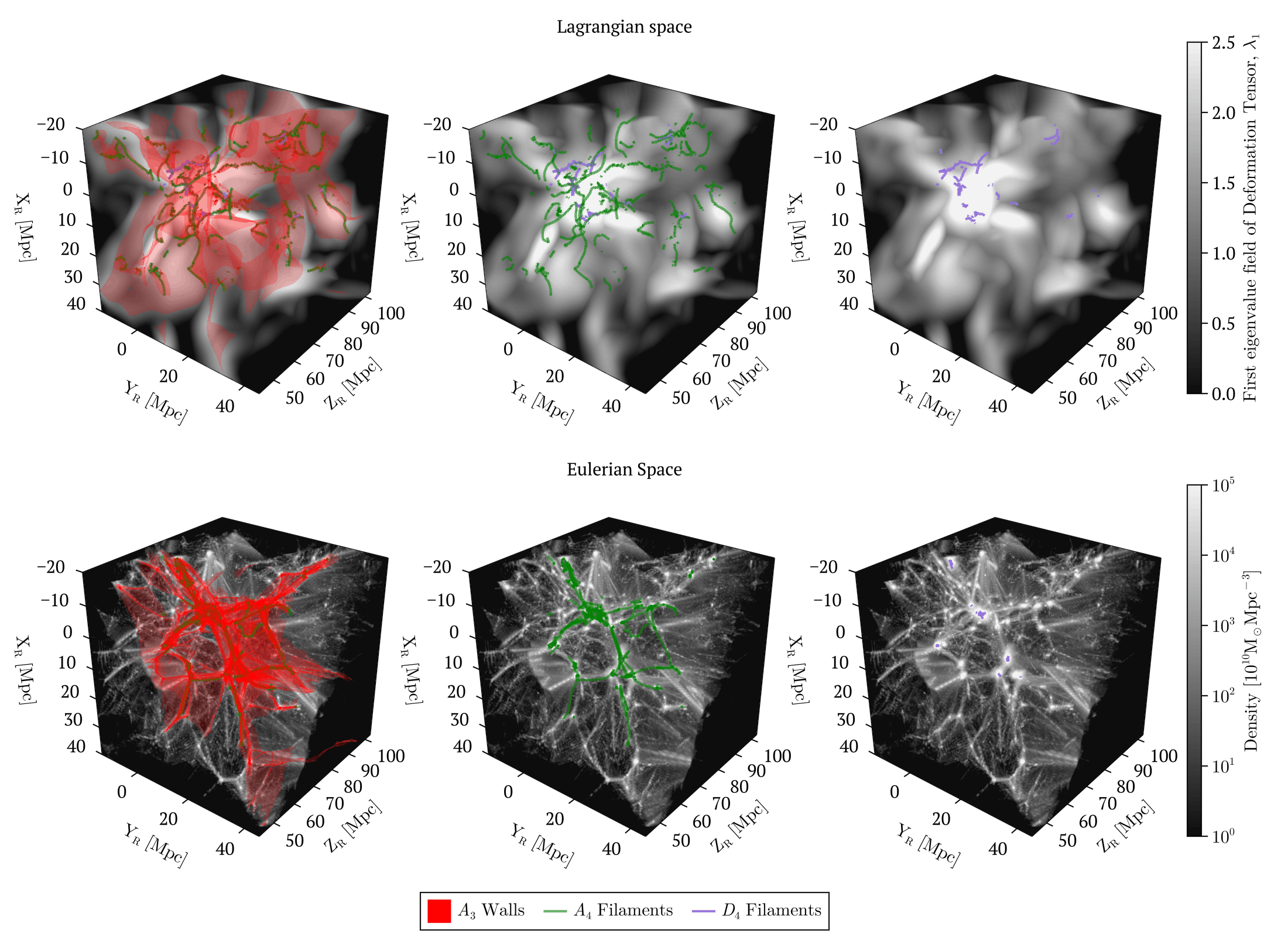}
    \caption{This figure shows the caustic skeleton of the Pisces-Perseus Supercluster in a $60$ Mpc box for $M_1$ Local Universe re-simulation. Coordinates are rotated so that the positive $Z_R$-direction is roughly along the line of sight and RA and DEC correspond to positive $Y_R$ and negative $X_R$ directions. Left: $A_3$ walls with $A_4$ and $D_4$ filaments, Middle: $A_4$ and $D_4$ filaments, Right: Only $D_4$ filaments. We use the smoothing scale of $\sigma =3$ Mpc. The density is estimated with the Phase-Space Delaunay Tessellation Field Estimator \citep{Feldbrugge+2024, FeldbruggeHertzsch2025}. A rotating and zoom-in animation of this figure is available on the additional video materials page at \href{https://jfeldbrugge.github.io/Caustics-Local-Universe/}{jfeldbrugge.github.io/Caustics-Local-Universe/}.}
    \label{fig:3dpp}
\end{figure}

\begin{figure}
\centering
	\includegraphics[width=\textwidth]{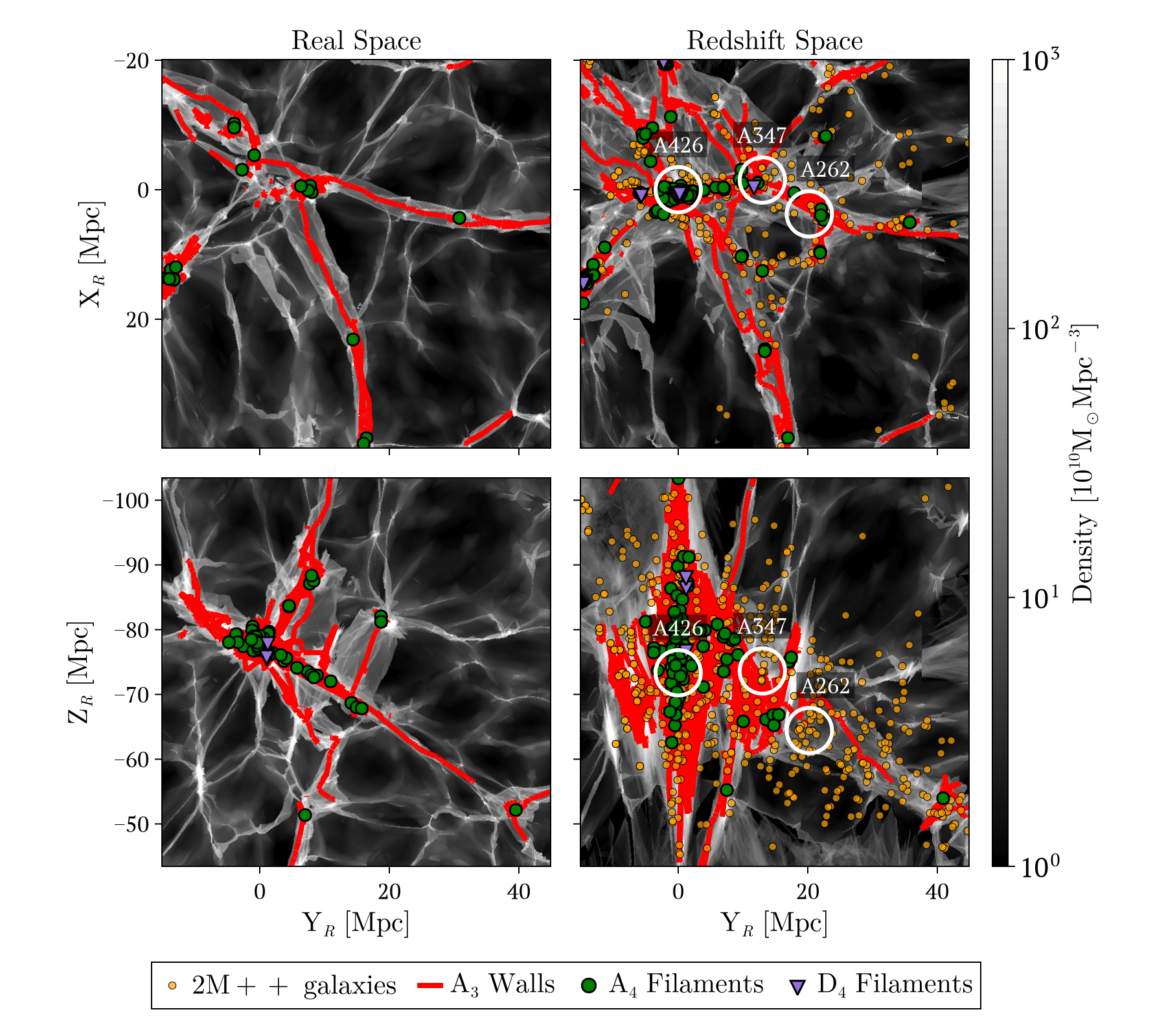}
    \caption{Infinitesimally thin slices of caustic skeleton of Pisces-Perseus cluster in $60$ Mpc square region for $M_1$ Local Universe simulation in real space on the left and redshift space on the right. Top panels: 2D $(Y_R,X_R)$ slice of density field and caustic skeleton at $Z_{R} = 73.3$ Mpc. Bottom panels: 2D $Y_R-Z_R$ slice at $X_{R} = 0$ Mpc. These slices are taken according to the observed location of the A426 cluster in the rotated coordinate system. The 2M++ galaxies are plotted as yellow dots and prominant clusters associated with Pisces-Perseus (A426, A347 and A262) are circled in white in the redshift space panels on the right. The density is estimated with the Phase-Space Delaunay Tessellation Field Estimator \citep{Feldbrugge+2024, FeldbruggeHertzsch2025}.}
    \label{fig:pp}
\end{figure}

\begin{figure}
\centering
	\includegraphics[width=\textwidth]{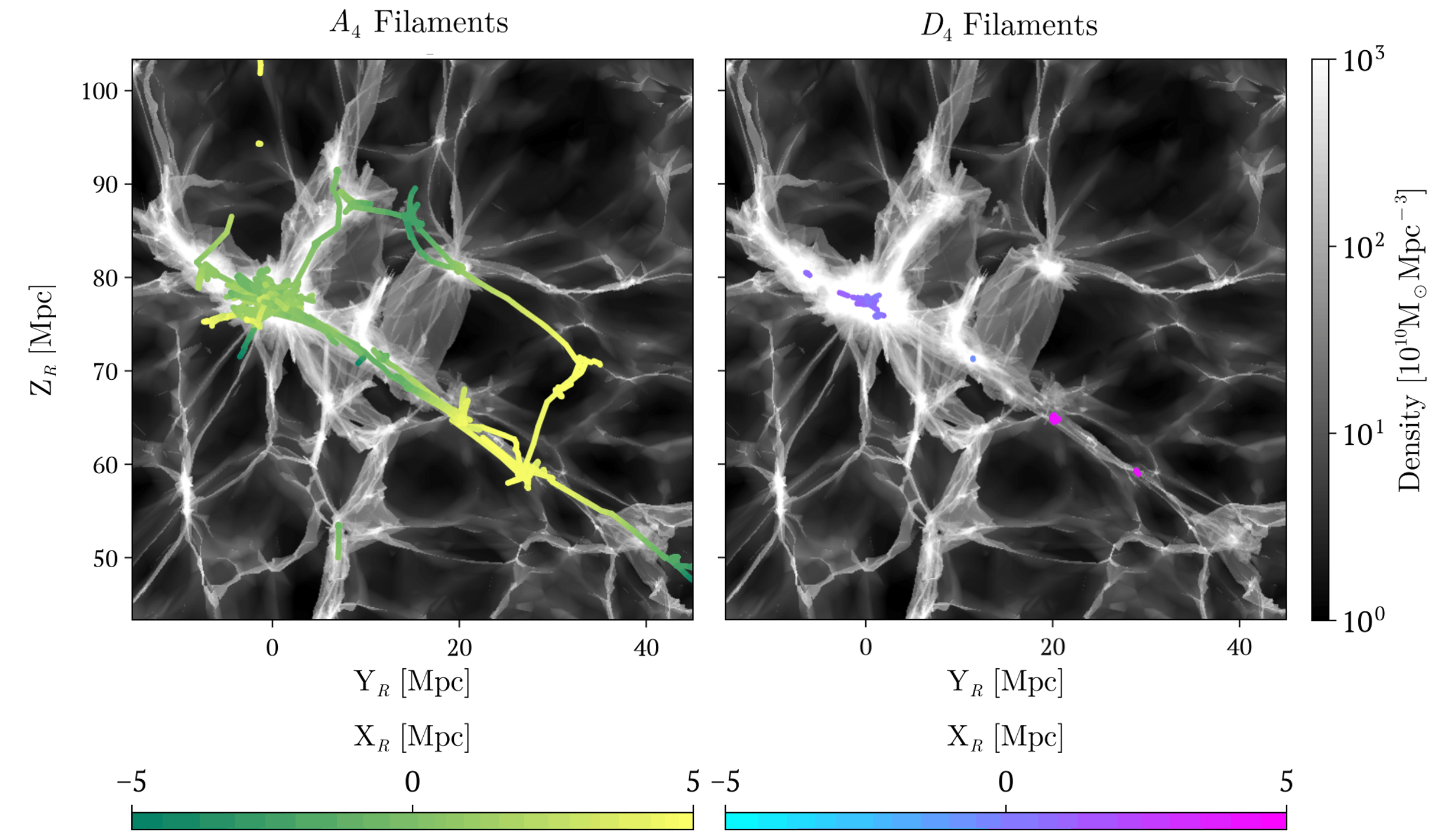}
    \caption{Filamentary ($\sigma = 3$ Mpc) caustics of Pisces-Perseus cluster for $M_1$ simulation within $10$ Mpc of $X_R = 0$ Mpc projected onto $(Y_R,Z_R)$ rotated coordinate plane in $60$ Mpc square region. $X_R$-coordinate of filaments are shown for $A_4$ filaments and $D_4$ filaments on the bottom left and right respectively. 2D real-space density field slice of $M_1$ simulations underlaid, same as the field underlaid in the bottom left panel of \cref{fig:pp}. The density is estimated with the Phase-Space Delaunay Tessellation Field Estimator \citep{Feldbrugge+2024, FeldbruggeHertzsch2025}.}
    \label{fig:pp_slice}
\end{figure}

The $130$ Mpc long Pisces-Perseus chain of galaxies, groups and clusters is without doubt the most outstanding and detailed example of a structurally and dynamically dominant filamentary artery of the local cosmic web\cite{HaynesGiovanelli+1986,Hidding+2016,EinastoM:2025}. The Pisces-Perseus Supercluster was first identified by Jõeveer, Einasto, and Tago in 1977 \citep{Joeveer1977} and is one of the most prominent and extensively studied large-scale structures in the nearby Universe \cite{GiovanelliHaynes1983}.  This supercluster complex forms a long, dense ridge of galaxies that hosts at three main Abell clusters: the Perseus cluster (Abell 426), Abell 347, and Abell 262. There are even indications it may be part of an ever more extended structure - the Perseus-Pegasus filament, identified by \cite{Batuski1985}, which stretches no less than one billion light-years ($>300\,  h^{-1}$ Mpc). This would make it one of the largest known coherent structures in the observable Universe.

Besides its relative vicinity, the  prominence of the Pisces-Perseus complex is also the consequence of two fortuitous circumstances. Its almost perfect transverse orientation on the sky, perpendicular to the line of sight, yields an almost ideal unbiased vista along the ridge  of the Pisces-Perseus chain, as testified by the meticulous redshift
maps by the HI surveys of the region by Haynes \& Giovanelli \cite[see e.g.][]{HaynesGiovanelli+1986}. Moreover, within the cosmographic arrangement of structure in our local Universe, we find the Pisces-Perseus on the opposite side of the Local Void with respect to our own location in the flattened wall of the Local (Virgo) supercluster \cite{Hidding+2016,EinastoM:2025}. With the Local supercluster wall forming part of the flattened boundary of the Local Void, it implies that we reside at a cosmic vintage point allowing us to look at the Pisces Perseus supercluster along the stretch of a largely empty void region, and hence providing  an optimal view uninterrupted by foreground structures \cite{Hidding+2016} (and related PhD thesis work by J. Hidding).

The detailed structural maps by \cite{Hidding+2016} of the implied dark matter distribution in this region, the result of the adhesion evolved web structure of the Bayesian inference KIGEN reconstruction of the primordial local dark matter distribution inferred from the 2MRS galaxy sample \cite{Hess:2013}, provide the best visual illustration of these intricate structural relations. In addition to the main Pisces-Perseus chain, the entire complex is a  superb prototype for the complex spatial connectivity marking the cosmic web, as we see it branching into two major arms, together forming the surrounding boundary of a tenuous large wall \citep{HaynesGiovanelli+1986,Hidding+2016,Mondelin2025}.  The detailed maps by Hidding even reveals the complex multi branching filamentary substructure along the main ridge \citep{Hidding+2016} (also see the recent \texttt{DisPerSE} map by \cite{Mondelin2025}). It makes the Pisces-Perseus region the archetype for the spatial organization of nodes, filaments and walls in the cosmic web and the Pisces-Perseus ridge the canonical example of a cosmic web filament and thus a natural target for our caustic analysis.

Besides the visual prominence, cosmographic studies have also uncovered the
enormous gravitational - force and tidal - impact of the Pisces-Perseus supercluster complex in the Local Universe. Along with the
Great Attractor region and the related Laniakea complex, and to a lesser extent the Local Void, it has been found to be
the dominant influence in the motion of galaxies in the nearby universe. The analysis of the cosmic velocity flows in the Local
Universe by the Cosmicflows peculiar velocity surveys \citep{Pomarede2017,Tully2019,Pomarede2026} clearly reveal the sizeable
imprint of Pisces-Perseus, inducing a strong migration flow along its ridge, evoking sizeable shearing motions along several
filaments and due to its vicinity influencing our own motion through the Universe. 

The Pisces-Perseus complex has also been the subject of numerous studies addressing the dependence of galaxy properties on environment, with the intention to establish the various physical and environmental processes affecting the formation and evolution of galaxies. As part of their extensive survey of the Pisces-Perseus region, the seminal study by Haynes \& Giovanelli \citep{HaynesGiovanelli+1986} produced highly informative and tantalizing maps of the spatial distribution of different galaxy morphologies, arguably the first significant evidence for morphological segregation on Megaparsec scales. They showed that early-type galaxies concentrated tightly around the ridge of the supercluster and in the clusters. Late-type galaxies revealed a far wider distribution, and also filled in the major share of the corresponding wall. Amongst the more recent studies on this aspect is the analysis by \cite{Mondelin2025}, who applied the \texttt{DisPerSE} algorithm to deep CFHT/MegaCam imaging. They also found clear evidence of galaxy morphology and stellar-mass segregation along
the filament spines: massive early-type galaxies being highly concentrated near the filament axes and dense nodes, whereas late-type and irregular systems seem more broadly dispersed. Interestingly, they also established that 10–13$\%$ of galaxies in the Pisces-Perseus complex exhibits signs of gravitational interaction, pointing to active pre-processing driven by the filamentary environment. 
\\
\paragraph{Pisces-Perseus Caustic Skeleton}
The filamentary structures extracted as a result of this caustic analysis are displayed in \cref{fig:pp_radec}. The $A_4$ filaments once again form the main bridge between the prominent Abell clusters along the strong filamentary back-bone of the Pisces-Perseus supercluster. An $A_4$ filament also extends downwards towards Abell 400. At the chosen smoothing length of $\sigma = 3$ Mpc the Pegasus cluster appears to be associated with a short and isolated $A_4$ filament. The $D_4$ filaments are concentrated at the centres of the massive clusters, in particular A426 and A262, appearing to form a more clumpy substructure along the ridge. The dynamical justification of these identified filaments and striking agreement with observations is once again indicative of the potential of caustic skeleton theory as a dependable tool for structure identification in future.

The results of our three-dimensional caustic analysis for the Pisces-Perseus region are shown in Lagrangian and Eulerian space in \cref{fig:3dpp} at smoothing scale $\sigma =3$ Mpc. It should be noted that the Cartesian coordinate system of the simulation has been rotated from galactocentric $(X,Y,Z)$ coordinates to align with the way the Pisces-Perseus cluster is viewed on the sky. The new coordinates $(X_R, Y_R, Z_R)$ roughly correspond to DEC, RA and distance, respectively, achieved through a simple rotation about the A426 cluster, a prominent member of the Pisces-Perseus Supercluster. As such, in the rotated $60$ Mpc box region depicted, the A426 cluster sits at $(X_R, Y_R, Z_R) = (0,0, 73.3)$ Mpc.
The left panels display the full network of both walls and filaments, and the middle panels show both types of filaments, while the right panels isolate the $D_4$ filaments in each space.
As in \cref{sec:coma}, we find that the caustics map the eigenvalue field in Lagrangian space and the density field in Eulerian space very well. In particular, the strong $A_4$ filamentary ridge of the Pisces-Perseus Supercluster is highlighted in the bottom middle panel.
The ability to distinguish between topologically distinct classes of filaments -- such as standard $A_4$ bridges versus complex $D_4$ umbilical structures -- provides novel structural information.
Although the $D_4$ filaments are obscured in the bottom middle panel by the $A_4$ filaments in the dense regions, their presence highlights topologically distinct regions around which we might anticipate more complex dynamics. These $D$-type structures are often marked by phase-mixing, forming immediate infall regions, depending on the scale probed, and as such warrant further investigation to properly understand the nature and characterise the differences between the two types of filaments predicted by this formalism. 

To investigate the detailed internal structure of this filamentary complex, we display an infinitesimally thin 2D slice taken from the 3D caustic analysis in \cref{fig:pp}, with different slices of real-space and redshift space displayed in the left and right panels. In this projection, one can see that the primary caustic wall is oriented approximately perpendicular to the line of sight, within the $(Y_R,X_R)$ plane of the rotated coordinate system. This is particularly clear in the $(Y_R,Z_R)$ redshift space panel on the bottom right, as the line-of-sight redshift space distortions are significant along the plane of the wall.
As expected, this structure exhibits an intricate network of $A_4$ and $D_4$ filaments embedded within the walls, with the vast majority of the observed galaxies strongly clustering along these caustic spines. 
The filamentary character of the Pisces-Perseus Supercluster is further emphasised in \cref{fig:pp_slice}, which shows a $10$ Mpc slice of $A_4$ and $D_4$ filaments projected onto the $(Y_R,Z_R)$ plane. Since the $Y_R$ direction roughly corresponds to RA, we find that the strong extended filamentary spine of Pisces-Perseus, which is seen across the sky, is captured remarkably well by the caustic skeleton theory. The right panel of \cref{fig:pp_slice} shows that the $D_4$ filaments are less extended and tend to be located in the denser regions, clumped along the ridge, whereas the $A_4$ filaments also appear to be closely gathered in the left panel.

\section{Galaxy affiliations in the Caustic Skeleton}
\label{sec:class}

\begin{figure}
\centering
	\includegraphics[width=0.8\textwidth]{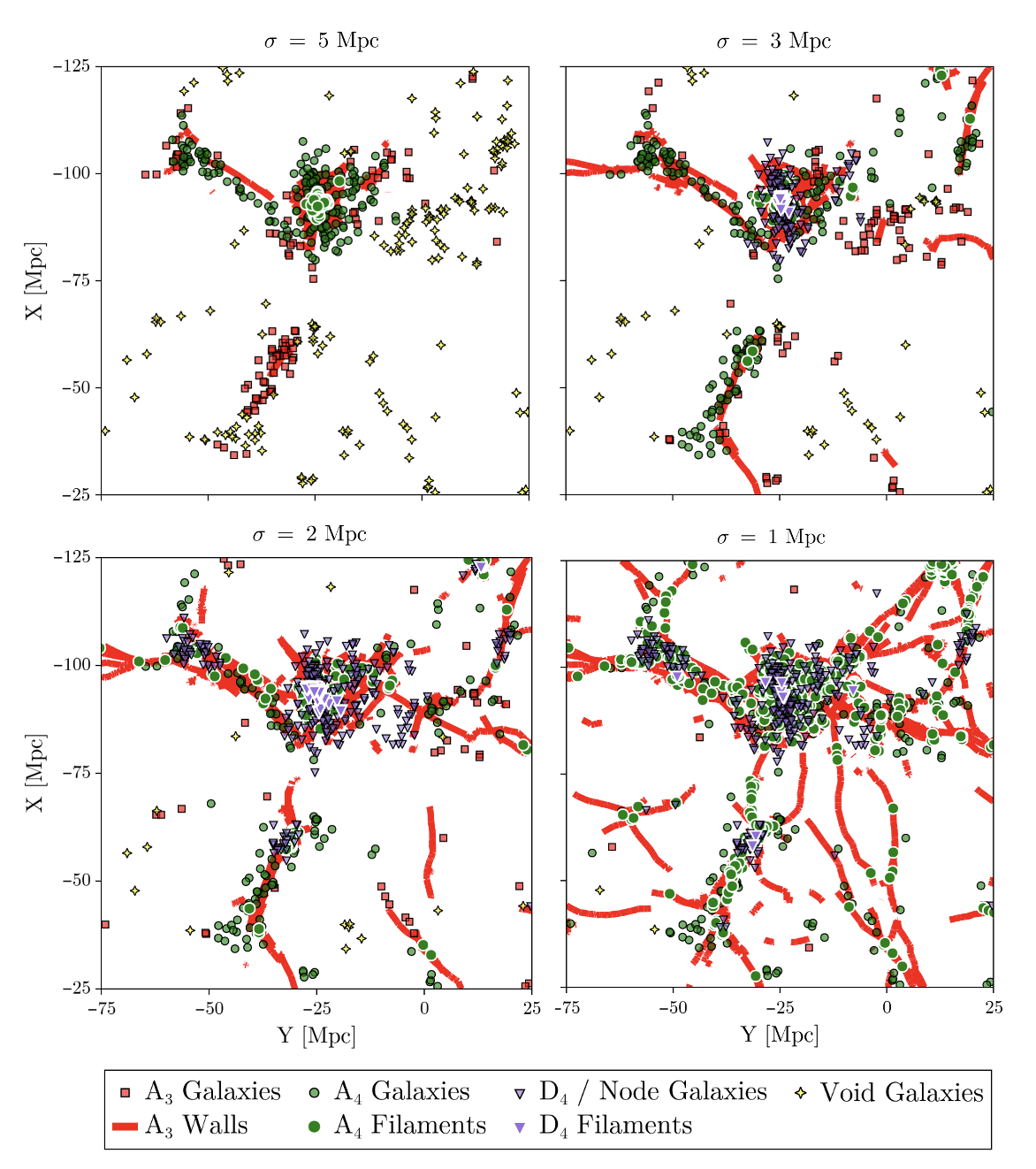}
    \caption{Classification scheme outlined in \cref{sec:class} applied to 2M++ galaxies in $10$ Mpc thick slice around $Z_\mathrm{Coma}$ (this is the same subset of galaxies shown in \cref{fig:coma_sigma_redshift}) using caustic skeleton extracted from the $M_1$ realisation. 
    Since the large-scale environment of a galaxy is scale dependent, the galaxy classification scheme is applied at smoothing scales $\sigma = 5, 3, 2, 1$, shown in the top left, top right, bottom left and bottom right, respectively.
    The large-scale environment of a galaxy at a given smoothing scale is classified as either dominated by $A_3$ wall caustics (red squares), $A_4$ filaments (green points) or $D_4$ filaments/Nodes (purple down triangles). In the case that no caustic structure is within $5$ Mpc of a galaxy, it is classified as a void galaxy (yellow 4-pointed stars)
    A slice of the corresponding redshift-space caustic skeleton from simulation $M_1$ for each smoothing scale is also included: $A_3$ walls appear as red lines, and $A_4$ and $D_4$ filaments appear as green points and purple down triangles, both with a white outline. The order of layering of elements in this figure reflects the hierarchy of the classification scheme - i.e. the $D_4$ filaments are likely associated with cluster nodes and thereby have the densest and have the strongest sphere of influence}, followed by $A_4$ filaments, then $A_3$ walls.
    \label{fig:class_coma}
\end{figure}

\begin{figure}
\centering
	\includegraphics[width=0.8\textwidth]{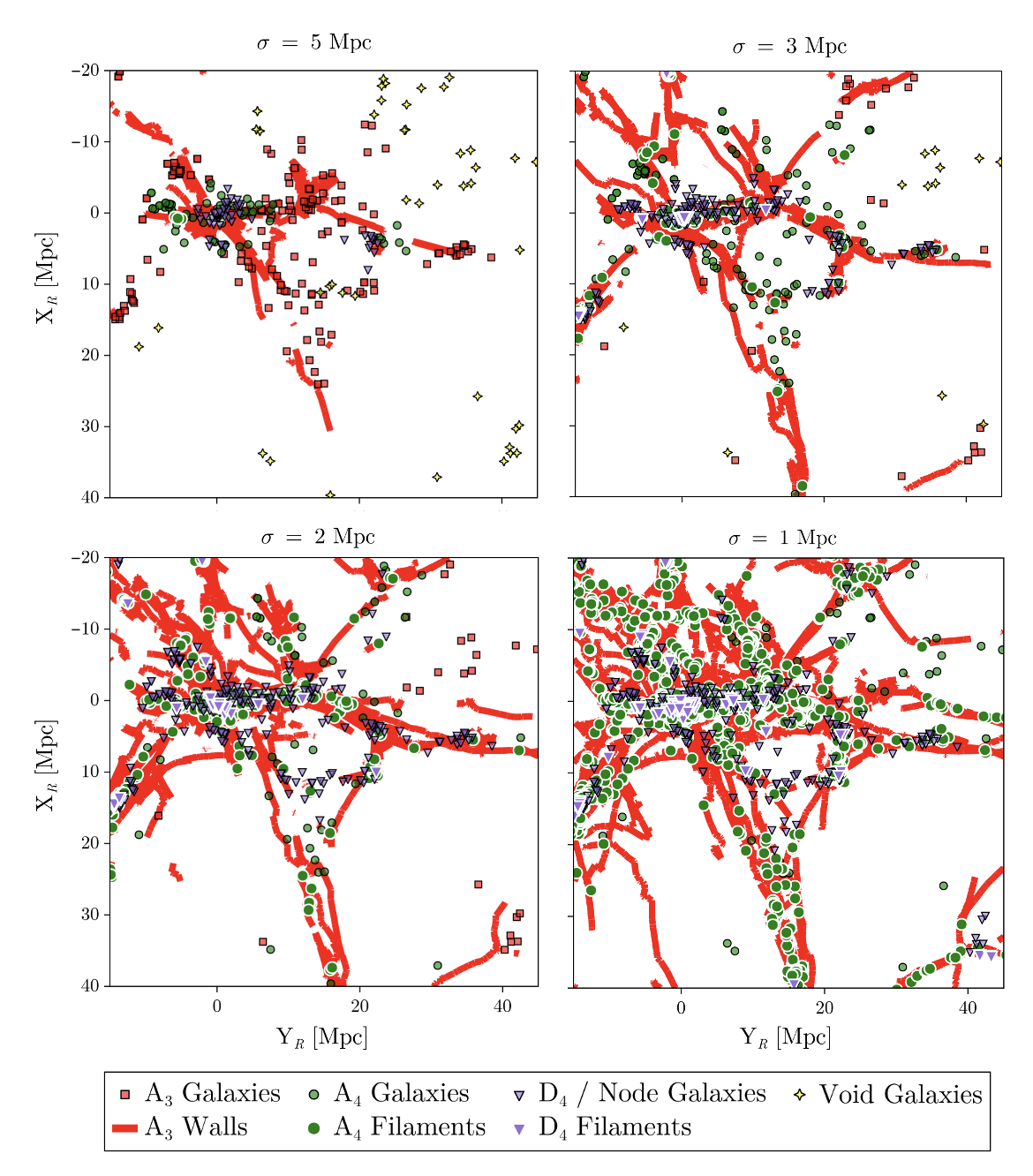}
    \caption{Same as \cref{fig:class_coma}, classification scheme applied to 2M++ galaxies in $10$ Mpc thick slice around $Z_{R} = 73.3$ Mpc in Pisces-Perseus Supercluster region (this is the same subset of galaxies shown in top right panel of \cref{fig:pp}) using the caustic skeleton extracted from the $M_1$ realisation.
    }
    \label{fig:class_pp}
\end{figure}

\begin{figure}
\centering
\includegraphics[width=\textwidth]{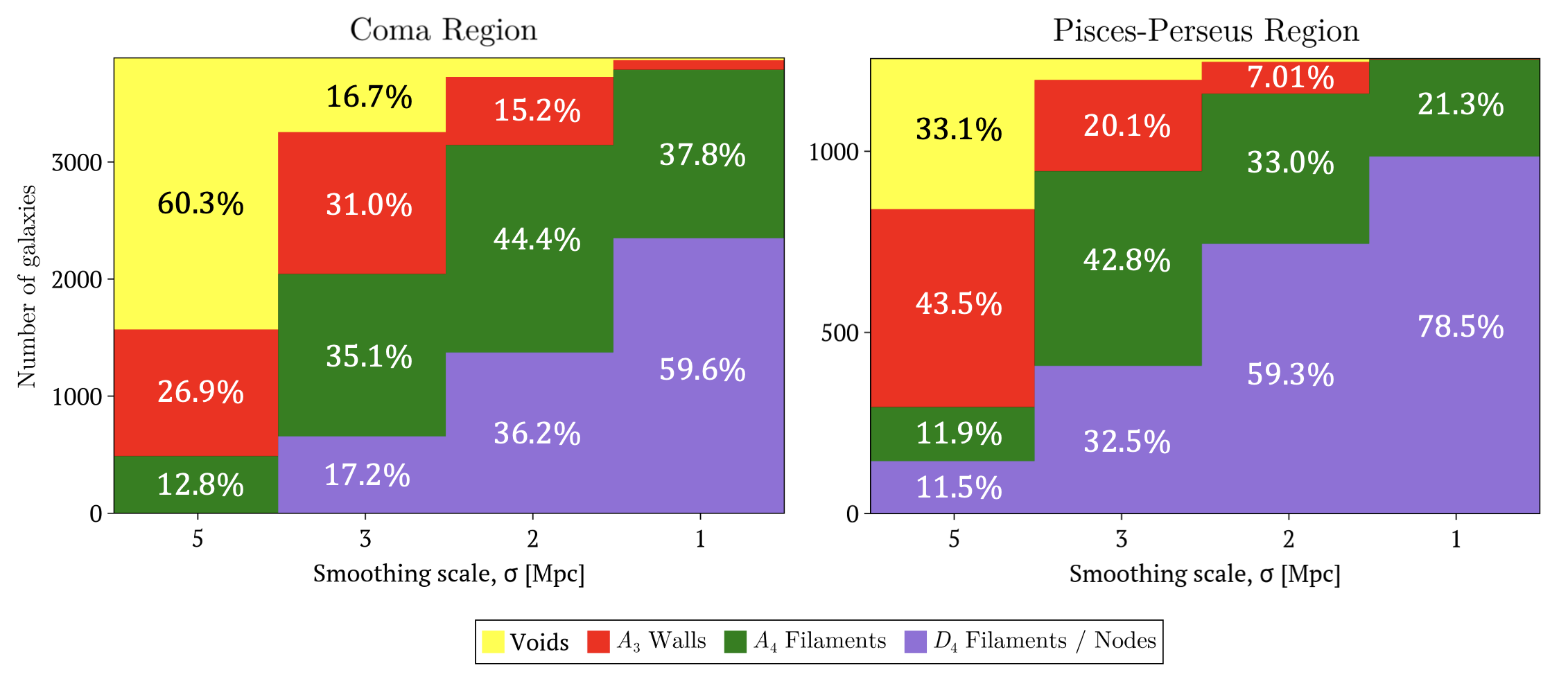};
    \caption{Stacked histograms displaying classification of dominant associated caustic structure for each 2M++ galaxy within a $100$ Mpc box around Coma supercluster (left panel) and a $60$ Mpc box (extracted in a rotated coordinate system) around Pisces-Perseus Supercluster structure (right panel) in realisation $M_1$. The classification scheme, outlined in \cref{sec:class}, is repeatedly applied with the caustic skeleton at a variety of smoothing scales - from large scales at $\sigma = 5$ Mpc to small scales at $\sigma = 1$ Mpc.  Total number of galaxies in box is fixed for both regions (3847 in Coma box and 1265 in Pisces-Perseus box). The proportion of galaxies associated with $A_3$ walls (red), $A_4$ filaments (green) and $D_4$ filaments/Nodes (purple)  and voids (yellow) are shown to strongly respond to scale probed with percentages overplotted.}
\label{fig:class_prop}
\end{figure}

In light of the observational evidence for morphological galaxy segregation along the various structures of the cosmic web, as
we also mentioned in some detail when discussing the Pisces-Perseus ridge, we also wish to investigate in how far caustic skeleton
theory is able to shed new light on this issue for our understanding of galaxy formation. 
Amongst the various relevant and outstanding manifestations, we also intend to in future address the specific issue of whether there is any indication or possibility that there are differential influences between those imparted by the regular large-scale $A_4$ caustic filaments and those of the $D_4$ filaments. 
Given the latter appear to be almost exclusively concentrated in or near the highest density concentrations of the cosmic web \citep{Arnold1986, Feldbrugge+2018} we would expect a noticeable and significant imprint of $D_4$ strands on the morphology of galaxies. Overall we know that the principal arteries of the cosmic web, which appear to be mainly the $A_4$ filaments are by far the most most outstanding component of the cosmic web, contain  more than $50\%$ of the dark
matter, gas and galaxy content of the universe \citep{Cautun+2014,Ganeshaiah:2019}. Moreover, they also dominate the dynamics
of the large scale matter distribution \cite{Kugel+2026}, and hence the transport of mass along the various large scale features
of the cosmic web. In how this translates into an effect on galaxy scale between $A_4$ and $D_4$ strands, given e.g. differences in mass and gas migration and accretion, remains an open question. 

Thus, in this section, we set the groundwork for a classification scheme which may be utilised in future studies to understand how the caustics could influence gas flow along filaments and subsequently impact galaxy properties such as spin alignment, morphology, and star formation rates.


\paragraph{Analysis of Galaxy-Caustic affiliations}
To obtain a rough impression of the level of affiliation of galaxies to the caustic singularities in the Local Universe, we invoke a rather simplistic analysis. It involves testing the sphere of influence of each caustic structure in the region, and examining the abundance and spatial distribution of galaxies associated with $A_3$ walls, $A_4$ filaments and $D_4$ strands. A more detailed study, also
involving morphology and other properties of galaxies, is left for future work.

We expect that if a galaxy is within proximity to both types of filaments (this is very likely, see \cref{sec:caustic_class}), the $D_4$ would dominate the large-scale cosmic environment of a galaxy since $D_4$ caustics have been established to be stronger and denser. Similarly, any filamentary caustic, which by definition is embedded within an $A_3$ wall, will gravitationally dominate the environment of a nearby galaxy over the influence of the wall itself. Therefore, the influence of galaxies' environment follows a clear topological hierarchy: $D_4$ filaments supersede $A_4$ filaments, which in turn supersede $A_3$ walls.

However it should be noted that a robust identification of $A_5$ and $D_5$ node-like caustics was not yet available for this current study, which would supersede the $D_4$ filaments in terms of density and influence.
Nonetheless, in \cref{fig:coma_radec} and \cref{fig:pp_radec}, even at a $3$ Mpc smoothing scale, the $D_4$ filaments appear to be mainly concentrated around nodal regions. In addition to the fact that they are found to be denser and thicker in simulations, these $D$-type filaments could be interpreted as the heavy filaments that stick out of clusters \citep{Arnold1986, Feldbrugge+2018}. 
Arguably galaxies in a dense clump or cluster that would, in a full caustic analysis, be assigned to a nodal caustic, are in this analysis instead associated with the $D_4$ filaments, which are by nature connected to the regions where galaxies are anticipated to concentrate. As such, we must be very careful when interpreting the preliminary results of this section: as the influence of $D_4$ filamentary caustics and $A_5$/$D_5$ nodes cannot yet be properly distinguished, from hereon galaxies associated with $D_4$ filaments will be classified as $D_4/$Node galaxies. 

The hierarchy of caustic influence allows us to establish a simple classification scheme for the large-scale environment in which a given galaxy resides for the purposes of this study. If a $D_4$ filament is present within a $5$ Mpc radius of a galaxy, the galaxy is classified as residing in a $D_4$/Node environment. In the absence of a nearby $D_4$ filament, we check for the presence of an $A_4$ filament within the same $5$ Mpc radius; if one is found, the galaxy is classified as $A_4$. If no filamentary structures are present, but an $A_3$ wall is located within $5$ Mpc, the galaxy is classified as $A_3$. Finally, galaxies that do not satisfy any of these proximity criteria are classified as residing in a Void. 
This choice of 5 Mpc is, of course, a rough approximation chosen for demonstration purposes in this study. There are many improvements that could be made in designing a robust galaxy classification scheme for future studies: i.e. modelling the density profiles of caustics to determine relative influence, adjusting chosen radius for different smoothing scales, considering caustic skeleton of the mean constrained field instead of the single realisation and most significantly identifying $A_5$ and $D_5$ node caustics as well.

\Cref{fig:class_coma} demonstrates the application of this classification technique to the $\sim 10$ Mpc thick Z-slice of 2M++ galaxies around the Coma Cluster (same subset shown in bottom panels of \cref{fig:coma_sigma_redshift}) using the 3D caustic skeleton obtained from the $M_1$ realisation for a range of smoothing scales from $\sigma = 5$ Mpc to $\sigma = 1$ Mpc. Infinitesimally thin slices of the caustic skeleton (similar to those displayed in upper panels of \cref{fig:coma_sigma_redshift}) are also displayed to show correspondance between caustics identified at each scale and galaxy classification. However, it should be noted the full 3D caustic skeleton was used to perform these classifications, so the caustic a galaxy has been associated with may not be visible in this slice.
\Cref{fig:class_pp} also shows the application of this classification scheme to a $\sim 10$ Mpc thick $Z_R$-slice of 2M++ galaxies around the Pisces-Persues region instead (same subset of galaxies shown in top right panel of \cref{fig:pp}) for the same range of smoothing scales.

Here, it is important to remember that within a single large-scale wall, there may be many finer, smaller-scale filaments permeating it. A large smoothing scale (e.g., $\sigma = 5$ Mpc) isolates the largest and most massive structures which form the most fundamental modes of the cosmic web. Conversely, decreasing $\sigma$ to smaller scales (e.g., $\sigma = 1$ Mpc) reveals the finer-grained, less massive structures.
Hence, the proportion of higher order caustics, and so proportion of galaxies associated with each, strongly depends on the scale $\sigma$ we probe with the Gaussian filter applied to the simulation. This is clear in the regions shown in \cref{fig:class_coma} and \cref{fig:class_pp}.

While the smoothing scale dependence prevents a single straightforward classification, tracking these changes across scales provides a wealth of dynamical information. Rather than treating the multi-scale nature of the cosmic web as a limitation, we can use it to define a galaxy's "nested" environment. For instance, \cref{fig:class_coma} highlights galaxies that are classified as residing within an $A_3$ wall at a larger scale ($\sigma = 3$ Mpc), but are revealed to be embedded within a dense $D_4$ filament (potentially a cluster) when resolved at a smaller scale ($\sigma = 1$ Mpc). The varying physical extents and densities of these nested structures may be reflected in the properties of the galaxies that reside within them. Moreover, galaxies that transition from an $A_4$ filament affiliation at large scales to a $D_4$/Node-dominated environment at small scales possess a distinct multi-scale environmental signature, which also may leave a distinct impression on their observable properties. The detailed connection between the multi-scale caustic environment of a galaxy and its specific physical properties is beyond the scope of this paper, but this will be explored in future work. 

The dependence of galaxies' classified large-scale environment on smoothing scale is further illustrated in \cref{fig:class_prop}. This figure shows the relative fraction of galaxies assigned to each caustic type as a function of the smoothing scale for the $100$ Mpc box around the extended Coma Cluster and Great Wall structure in the top panel and the rotated $60$ Mpc box around Pisces-Perseus in the bottom panel, both utilising the caustic skeletons extracted from the $M_1$ realisation. As noted in \cref{sec:data} and \cref{sec:coma}, the caustics identified at scales below the constraint resolution are less robust when considering a single realisation. Nevertheless, we proceed with this preliminary investigation to assess the potential of caustics as a galaxy classification tool. Robust galaxy classifications utilising the constrained median field will be presented in a subsequent study.

For the Coma supercluster region, we observe that at $\sigma = 5$ Mpc, galaxies are overwhelmingly classified as residing in voids, as the broader smoothing scale blurs out the finer structural details (as is visually evident in \cref{fig:class_coma} and \cref{fig:class_pp}). However, as we move towards smaller smoothing scales, the fine-grained structures permeating these voids and embedded within larger walls are captured, significantly altering the relative proportion of galaxies located in voids, walls, filaments and likely nodes as well.
\Cref{fig:class_prop} also shows that, at larger smoothing scales, the fraction of galaxies lying in voids is greater in the Coma region than in Pisces–Perseus. This reflects the fact that Pisces–Perseus more completely fills the $(60$ Mpc$)^3$ volume considered here, whereas the Coma Supercluster structure in the $100$ Mpc box consists of extended filamentary limbs enclosing large voids. As the smoothing scale decreases, however, small-scale structure within these voids is progressively resolved, and the void galaxy fraction in both regions drops to essentially zero by $\sigma = 1$ Mpc. This trend indicates the sizes of the structures that galaxies preferentially reside in, with most galaxies sitting in and around smaller structures rather than the few, heavy and large features.

Moreover, we observe that there is no $D_4$/Node-classified galaxy in the Coma Supercluster region obtained with the $\sigma = 5$ Mpc smoothing scale, as the fine-grained structures are smoothed away. However, as we move toward smaller scales, their relative proportion compared to $A_4$ galaxies steadily grows. 
Although we assume $D_4$ filaments gravitationally dominate their immediate local environment, their overall lower abundance in the Coma Supercluster region across the majority of smoothing scales means that there are more galaxies associated with $A_4$ filaments than $D_4$ filaments and nodes. However, when focusing specifically on $\sigma = 2$ Mpc for this region, the number of galaxies classified as residing in $A_4$ and $D_4$/Node environments becomes roughly comparable, and the $D_4$/Node environments then dominate at $\sigma = 1$ Mpc. This suggests that the dynamical impact of filaments and nodes around the Coma Cluster only becomes highly relevant on smaller scales.

However, as discussed under this simple classification scheme, most galaxies that appear to be affiliated with $D_4$ filaments, likely belong to the core of a clump or clusters. Nonetheless, there could also be a range of galaxies towards the outskirts of node, where the $D_4$ filaments connect, that would be dominated by a distinct $D$-type filamentary environment. 
Taking this into account, with caution, the trends in \cref{fig:class_prop} could potentially be interpreted as the transition in galaxy affiliations from walls, to filaments, to clusters as the scale of the structures probed decreases. With this interpretation, on large scales $D_4$/Node affiliation is low because galaxies preferentially do not sit in massive, heavy clusters, instead they concentrate into smaller groups on scales around 1 Mpc. This is a known phenomenon reflected in observations of our local neighbourhood, as even in the 18th century Messier's catalogue illustrated the clumping of galaxies in our small local group as opposed to the rich Virgo cluster \citep{Messier}.
Nonetheless, to properly understand these trends and the differences in $A$-type and $D$-type environments, the association of $D_4$ with nodes should be thoroughly investigated when the $A_5$ and $D_5$ nodal caustics can be robustly identified.

In comparison, we find a large proportion of $D_4$/Node-dominated galaxies across all smoothing scales for the Pisces-Perseus filament, including well above the constrained resolution at $\sigma = 5$ Mpc, as shown in the lower panel of \cref{fig:class_prop}. Even though \cref{fig:class_pp} shows an overwhelming number of $A_4$ filaments (similar to the Coma Supercluster region at $\sigma = 1$ Mpc in \cref{fig:class_coma}), the 2M++ galaxies within the Pisces-Perseus region tend to be positioned closer to $D_4$ filaments, resulting in a higher proportion of $D_4$/Node classified galaxies. Again, since at this stage we cannot yet disentangle the influence of $D_4$ filaments and nodes, if we interpret $D_4$ filaments as indicators of node-like regions this suggests that compared to the Coma Supercluster, Pisces-Perseus has much more substructure along its ridge. This is also highlighted in \cref{fig:pp_radec} as more galaxies concentrate in smaller clumpy groups.
This suggests a unique character and distinct formation histories of these well-known structures of our Local Universe.


\section{Formation History of the local cosmic web}
\label{sec:history}

\begin{figure}
\centering
	\includegraphics[width=\textwidth]{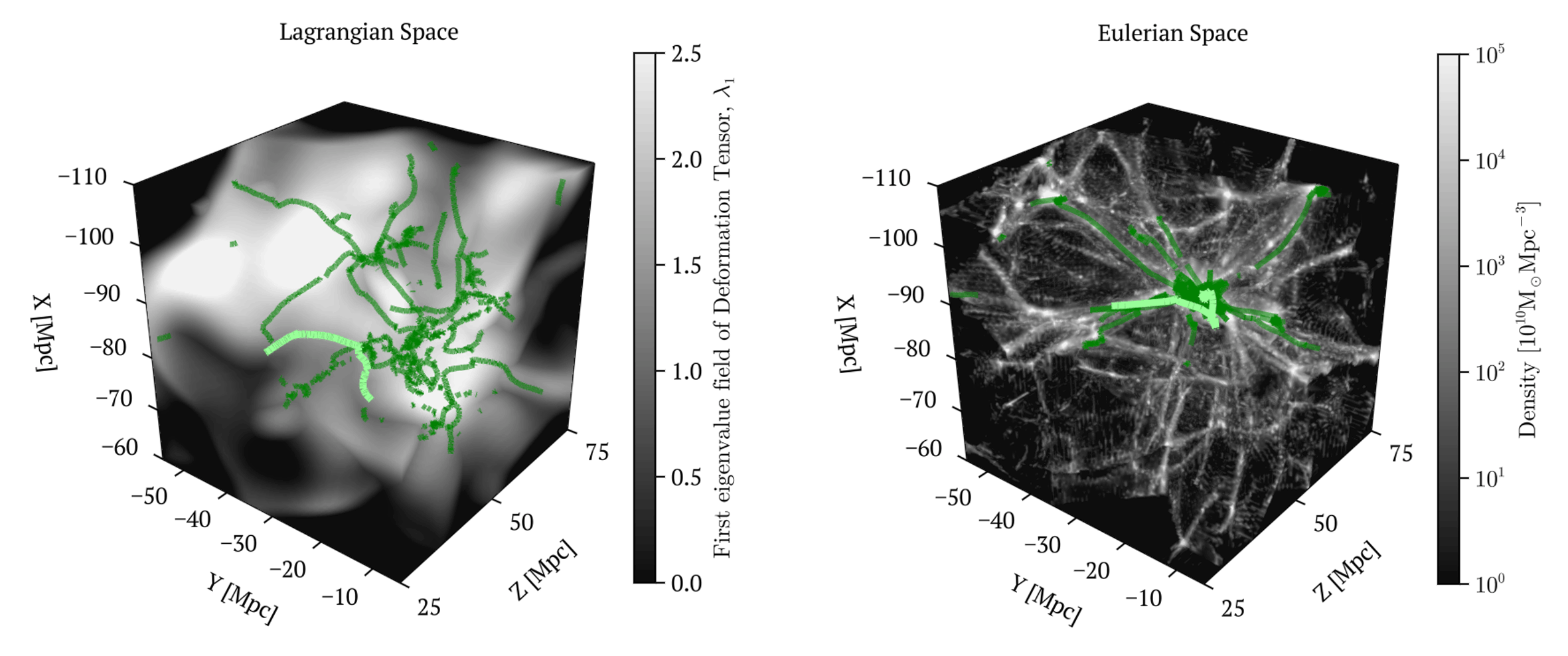}
    \caption{$A_4$ filamentary skeleton around Coma Cluster at smoothing scale $\sigma = 3$ Mpc in Lagrangian space (left) and Eulerian space (right). This is a zoomed in $(40 Mpc)^3$ region of the larger Coma Supercluster region shown in \cref{fig:3d}. Here, a single continuous section of a $\sim$30 Mpc long $A_4$ caustic filament is highlighted towards the box's centre. In Lagrangian space, the filament forms a relatively smooth curve. After the caustic skeleton is evolved into Eulerian space, one end of the filament appears to be drawn into the dense Coma Cluster region along with many other filaments, then extends outwards from the cluster mostly in the Z-Y direction. The density is estimated with the Phase-Space Delaunay Tessellation Field Estimator \citep{Feldbrugge+2024, FeldbruggeHertzsch2025}.}
    \label{fig:fil3d}
\end{figure}

\begin{figure}
\centering
	\includegraphics[width=\textwidth]{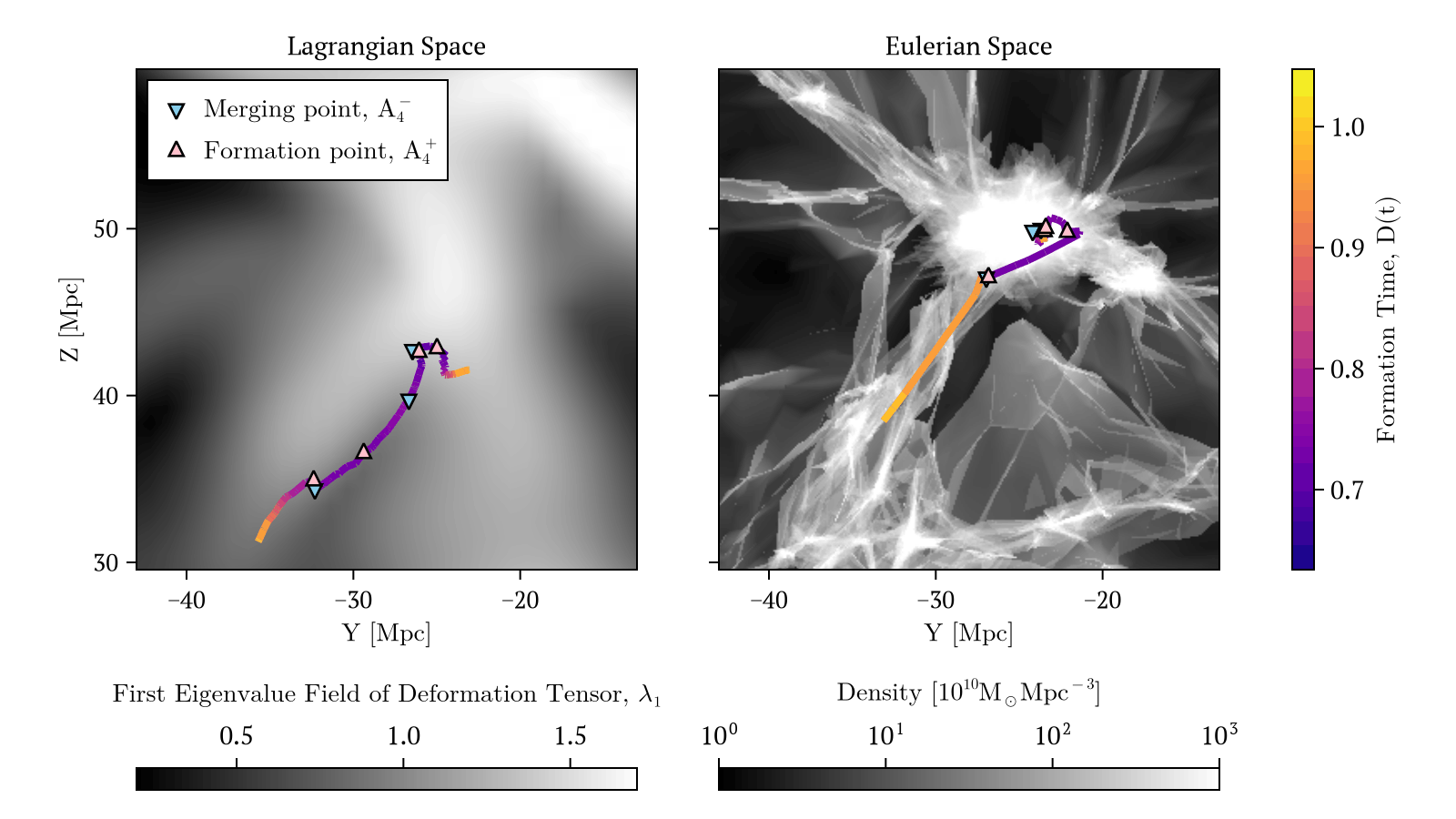}
    \caption{
   Right: 2D $(Y,Z)$ projection of 3D filament from \cref{fig:fil3d} in Eulerian space. Filament's extent in X direction is limited, so the slice of the simulation density field for comparison is taken around halfway through the $X$-extent at $X = -91$ Mpc.
    Left: Filament traced back to initial conditions (Lagrangian space) projected into the same $(Y,Z)$ plane. Following the same logic as in Eulerian space, a slice of the first eigenvalue field of the deformation tensor is taken at $X = -83$ Mpc, which is around halfway through the filament's $X$-extent in Lagrangian space. The density is estimated with the Phase-Space Delaunay Tessellation Field Estimator \citep{Feldbrugge+2024, FeldbruggeHertzsch2025}.
    In both panels, the filament is coloured according to formation time $D_{form}(
    q_{c})$ - i.e. the value of the growth factor at the time at which shell-crossing first occurred at the location of the caustic in Lagrangian space $q_c$. The $A_4^+$ points at which the filament first emerged are shown by pink triangles. The $A_4^-$ points where these separate growing sections of filament merged are shown by blue upside-down triangles. These are a few prominent examples of formation and merging points identified as local minima and maxima, respectively, in the formation time along the filament as shown in \cref{fig:fil_Dt}. }
    \label{fig:fil_2dproj}
\end{figure}

\begin{figure}
\centering
	\includegraphics[width=\textwidth]{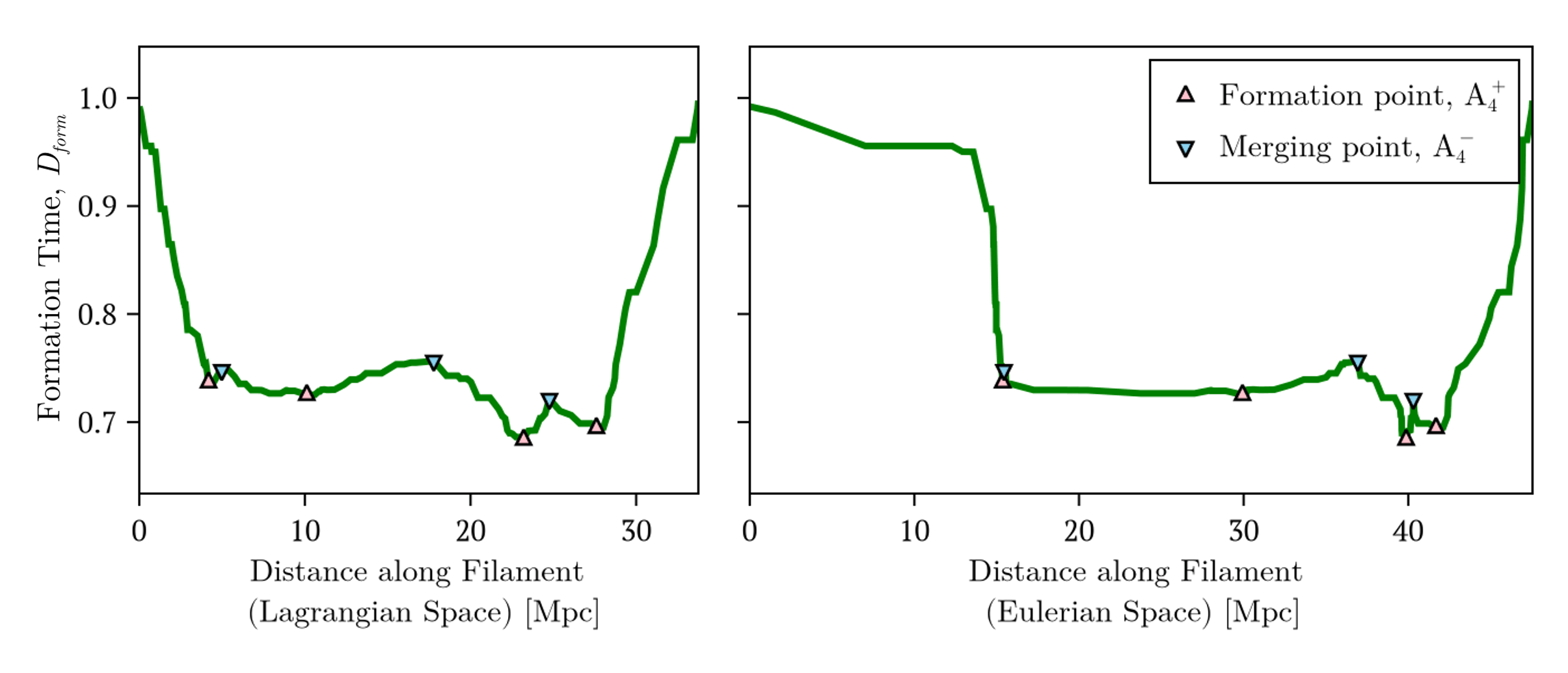}
    \caption{Formation time $(D_{form})$ as a function of distance along the length of example filament from \cref{fig:fil3d} in Lagrangian space (left) and Eulerian space (right). Distance along filament is defined from the bottom left to top right of the filament in \cref{fig:fil_2dproj}. Formation time is defined as the time at which shell-crossing first occurred and the relevant caustic condition is used to show this is equivalent to the statement $D_+(t_{form}(q_c)) = 1/\lambda_1(q_c)$. Thus, this graph is in fact the inverse of the first eigenvalue field of the deformation tensor at each point along the caustic filament in Lagrangian space.
    The filament emerges in initially disjoint segments at the local minima of formation time delineated by the pink triangles. Along an $A_4$ caustic, these local maxima of the eigenvalue field are referred to as $A_4^+$ points. The disjoint sections of the filament then merge at the local maxima of the formation time (local minima of the eigenvalue field) and are known as $A_4^-$ points along the caustic filament. The transformation of the filament from Lagrangian to Eulerian space clearly shows the sections of the filament that were stretched and compressed by the evolution of the cosmological fluid.}
    \label{fig:fil_Dt}
\end{figure}

\begin{figure}
\centering
	\includegraphics[width=\textwidth]{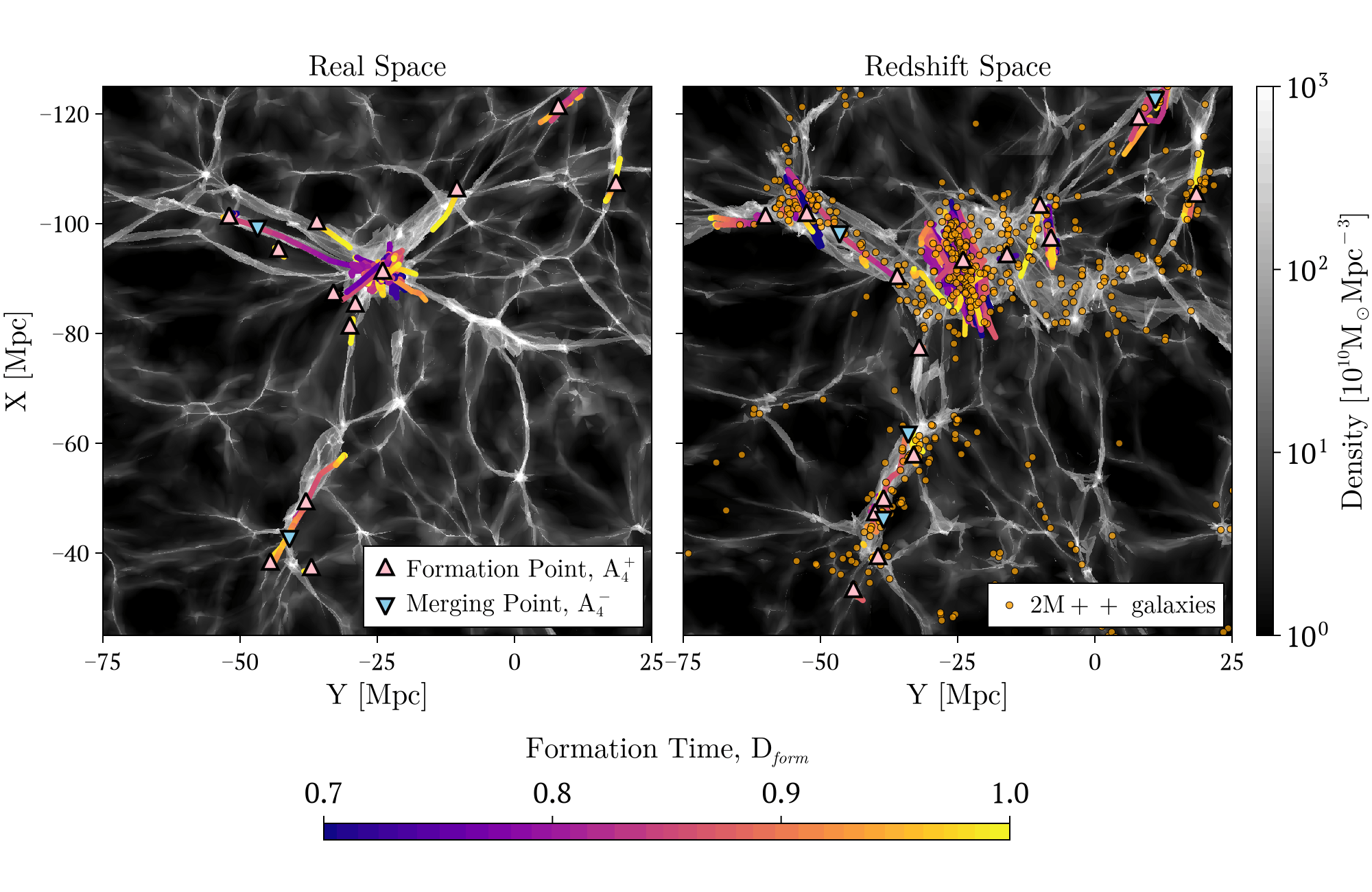}
    \caption{
    Formation times $(D_{form})$ mapped onto
    a $10$ Mpc thick slice of $A_4$ filaments around the Coma Cluster in a $100$ Mpc square region (this is the same slice of filaments at $\sigma = 3$ Mpc shown in the left panel of \cref{fig:slice}). The slice is taken in real space on the left and redshift space on the right. Due to the impact of RSDs, this is not exactly the same slice of the filamentary caustics around Coma, as some segments are shifted in or out of the slice. Significant merging ($A_4^-$: blue down triangles) and formation ($A_4^+$: pink triangles) are plotted in both panels. Due to the slice taken, some represent the points at which the filament first enters the slice, as the true formation point may sit outside the $10$ Mpc slice. 2M++ galaxies are plotted on the right in redshift space to demonstrate that the age of a galaxy's cosmic environment at a given smoothing scale can also be probed with caustic skeleton theory. The density is estimated with the Phase-Space Delaunay Tessellation Field Estimator \citep{Feldbrugge+2024, FeldbruggeHertzsch2025}.}
    \label{fig:coma_formation}
\end{figure}

\begin{figure}
\centering
	\includegraphics[width=\textwidth]{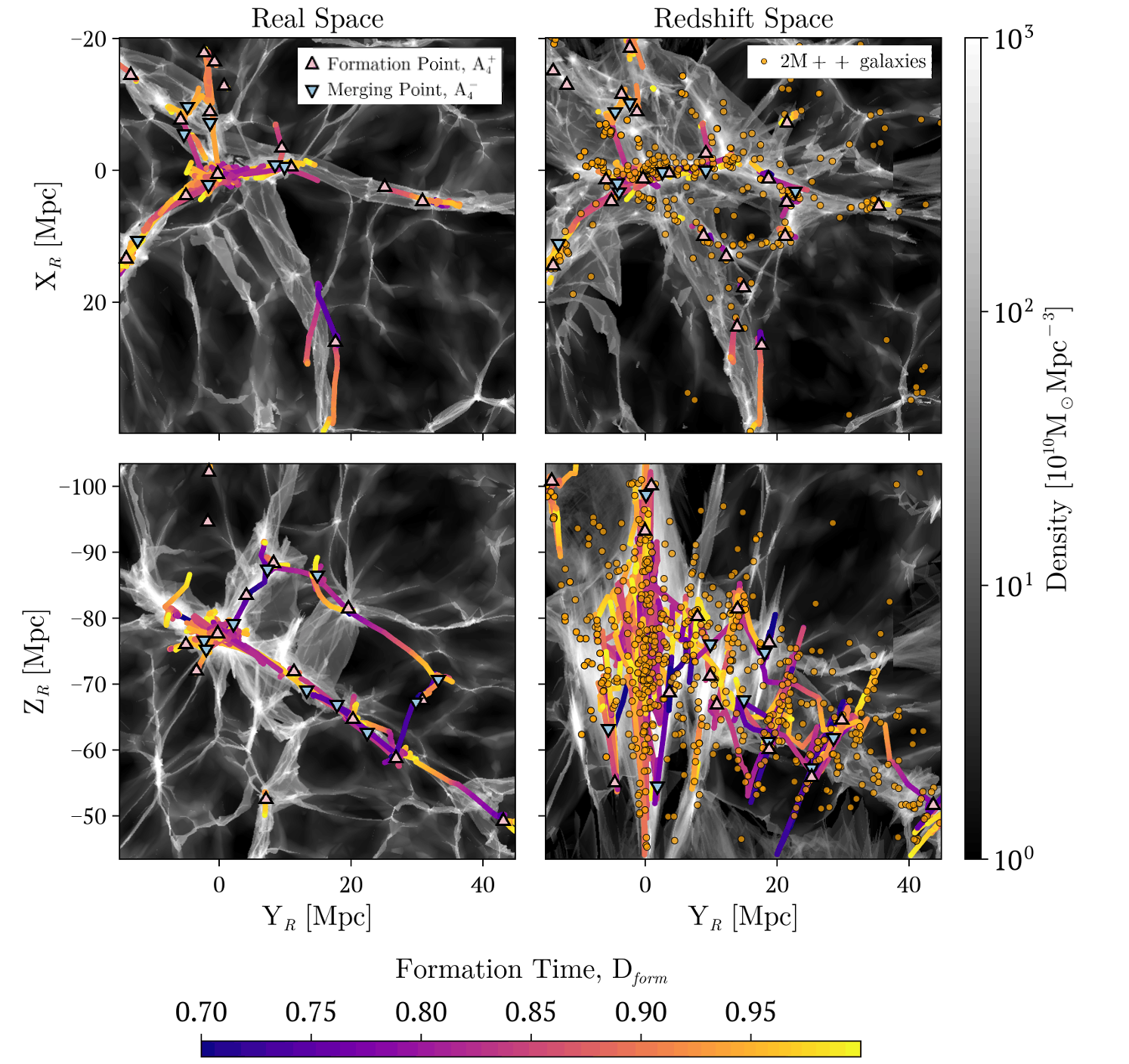}
    \caption{Formation times $(D_{form})$ of $A_4$ filaments in $10$ Mpc thick slices around Pisces-Perseus Supercluster at smoothing scale $\sigma = 3$ Mpc in a $60$ Mpc square region. Top panels: Slice taken about $Z_{R} = 73.3$ Mpc and filaments projected onto $(Y_R,X_R)$ plane. Bottom panels: Slice taken about $X_{R} = 0$ Mpc and projected onto $(Y_R,Z_R)$ plane, such that the bottom left panel is the same section of filaments shown in \cref{fig:pp_slice}. These slices are taken in real space on the left and redshift space with 2M++ galaxies included on the right. Significant merging ($A_4^-$: blue down triangles) and formation ($A_4^+$: pink triangles) are plotted in all panels. Same as \cref{fig:coma_formation}, some simply represent the points at which the filament first enters the slice. The density is estimated with the Phase-Space Delaunay Tessellation Field Estimator \citep{Feldbrugge+2024, FeldbruggeHertzsch2025}.}
    \label{fig:pp_formation}
\end{figure}

\begin{figure}
\centering
	\includegraphics[width=0.8\textwidth]{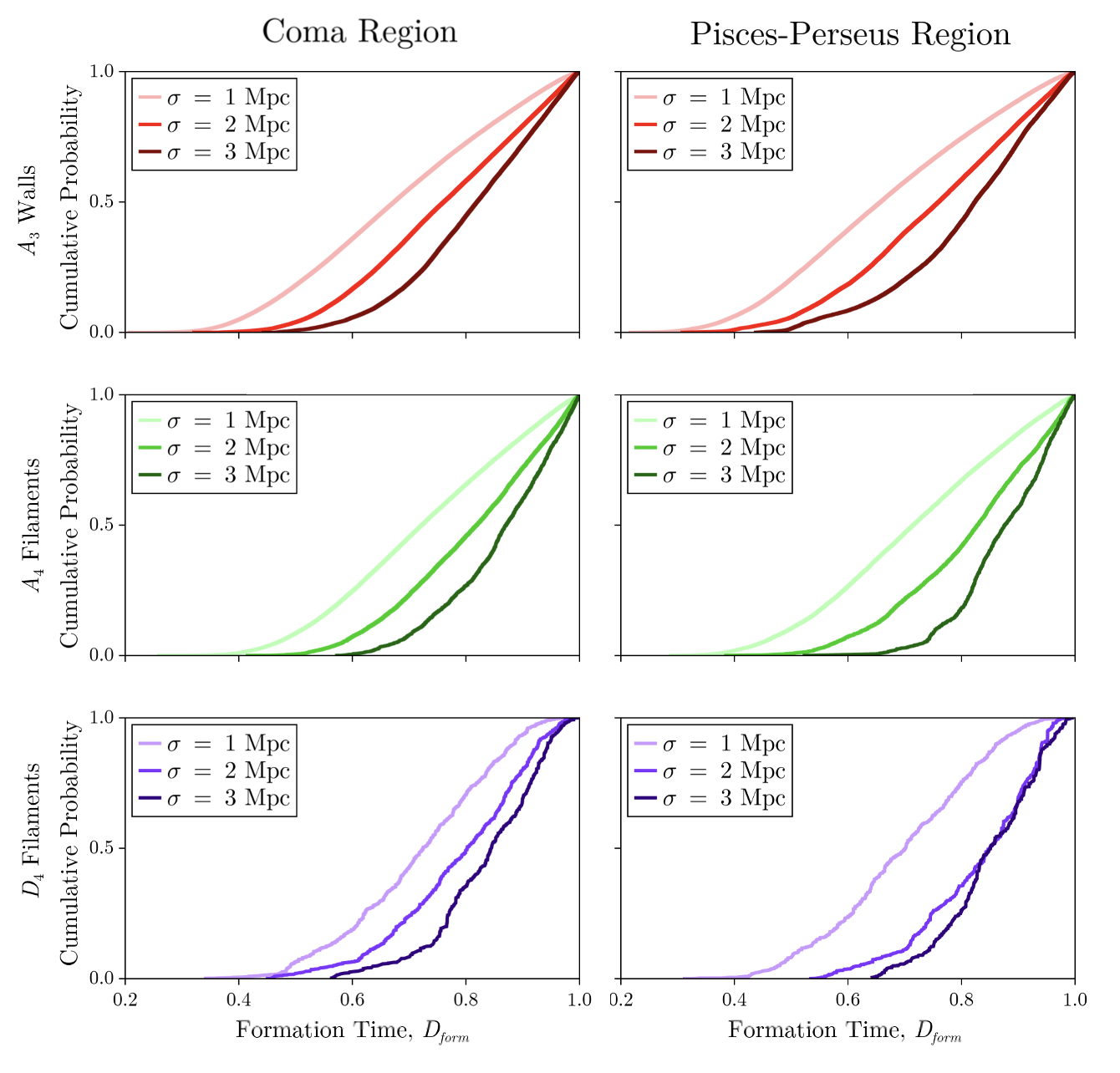}
    \caption{Cumulative probability distributions of formation times for Eulerian-space caustic skeleton structures in a $100$ Mpc region around Coma Cluster (left) and $60$ Mpc region around Pisces-Perseus Supercluster (right) for the $M_1$ realisation. $A_3$ walls, $A_4$ filaments and $D_4$ filaments are shown separately in the top, middle and bottom panels, respectively. The cumulative distribution for each caustic structure is given at three smoothing scales from $\sigma = 3$ Mpc to $\sigma = 1$ Mpc in progressively lighter colours. }
    \label{fig:hists}
\end{figure}

Since the cosmic sheet folds in different ways to create the different A- and D- type filaments, distinct formation histories are expected to leave an imprint on the galaxies living within them. In that respect, defining a formation time is also critical to understanding the type of caustics that form the filamentary structure of the Local Universe. The caustic skeleton theory is unique in that it simultaneously provides insight into the formation of the large-scale environment through its topological structure, while also offering an intuitive estimate of the formation times of the caustics that define it.

The time at which a caustic forms is formally defined as the moment when shell-crossing first occurs. As explained in \cref{sec:cst}, this initial moment of non-linear collapse corresponds to the formation of the first $A_2$ fold caustic, upon which all subsequent higher-order $A$-family and $D$-family caustics are necessarily located. Mathematically, this condition is met when the eigenvalue field of the deformation tensor satisfies $\lambda_1 =1/D_+(t_{form})$. By evaluating the eigenvalue field at the location of the caustic in Lagrangian space $\bm{q}_c$, we directly infer the formation time from the time of first shell-crossing in the Zel'dovich approximation: $D_{form}(\bm{q}_c)= D_+(t_{form}(\bm{q}_c))= 1/\lambda_1(\bm{q}_c)$. 

As an illustration, we consider a long $A_4$ filament of length exceeding $30$ Mpc, identified at smoothing length $\sigma = 3$ Mpc in final Eulerian-space of the $M_1$ realisation. 
This filament is highlighted in both Lagrangian and Eulerian space in \cref{fig:fil3d}, towards the centre of a zoomed-in $40$ Mpc box around the Coma Cluster with density field and $A_4$ caustics underlaid. The Eulerian space in particular shows that this filament originates in the dense bundle of filaments corresponding to the Coma Cluster and extends outwards mostly in the $(Y,Z)$ plane.

This filament is shown in \cref{fig:fil_2dproj} as a 2D projection onto the $(Y,Z)$ plane. The left panel corresponds to Lagrangian space, while the right panel shows the projection in Eulerian space. In both cases, the slice of the eigenvalue field and the final density field is taken at approximately halfway along the filament’s extent in the $X$-direction, corresponding to $X = -83$ Mpc in Lagrangian space and $X = -91$ Mpc in Eulerian space. In Eulerian space, the association of the filament with the 2D density slice is clear: one end is tightly wound within the dense Coma Cluster, while the other extends outward along the prominent filamentary structure visible in the density field. 
However, since the filament is significantly more extended in the $X$-direction in Lagrangian space (as seen in \cref{fig:fil3d}), its alignment with the eigenvalue field in a single slice is less apparent.
Nonetheless, even when we consider the full 3D Lagrangian $A_4$ caustic structure the correspondence between filaments and the eigenvalue field is complex and not immediately evident without the $A_3$ walls present, which was also the case in the upper middle panel of \cref{fig:3d}. This is due to the involvement of directional derivatives with the eigenvector field.

The formation time, $D_{form}$ along the filament is displayed according to the colour map in \cref{fig:fil_2dproj}.
This filament formed around $D_{form}\sim 0.7$ and then grew outwards from the cluster towards where it ends now, which corresponds to a portion that formed at later times.
Since both the caustic structure and formation time are computed prior to projection into Eulerian space, they provide additional insight into the dynamical evolution of the filament within the cosmological fluid. The relatively short, young tail toward the bottom left in Lagrangian space becomes significantly more extended in Eulerian space, reflecting the stretching of this region of the filament. Conversely, there is evidence of compression where the formation time changes rapidly at the edge of the Coma Cluster. Within the cluster itself, the opposite and also young end of the filament appears tightly wound, resembling a knot in a string. This is likely the region where higher-order caustics, such as $A_5$ or $D_5$, emerge. A more detailed investigation of this will be pursued in future work.

The pink and blue triangles in \cref{fig:fil_2dproj} mark the points along the filament where the $A_4$ caustic first formed in disjoint segments, and later merged into one long filament, referred to as $A_4^+$ and $A_4^-$ points respectively.
To understand how these formation and merging points were identified, one can imagine a horizontal line (isochrone) overlaid on \cref{fig:fil_Dt} slowly moving up with time. This figure shows the formation time along the length of the filament in Lagrangian space (left) and Eulerian space (right).
The segments of the filament that lie below this time threshold represent the portions that have already shell-crossed and collapsed by that epoch. Consequently, local minima along the filament represent the exact spatial locations where the filamentary structure emerges; these birth sites are formally classified as $A_4^+$ points. As time progresses, isolated filamentary segments eventually merge. The locations of these mergers correspond to local maxima in the formation time of the final filament and are classified as $A_4^-$ points. 

The formation time map along the filament in \cref{fig:fil_Dt} shows many local minima and maxima. We thus selected a subset of the most significant examples of $A_4^+$ birth points and $A_4^-$ merger points in Lagrangian space, mapped them to Eulerian space, and displayed them in \cref{fig:fil_2dproj} as well. 
As one can see, it is clear in both figures that the points move relative to one another along the filament as the system evolves from the initial Lagrangian space to the present-day Eulerian space and the filament is stretched, compressed, and wound up in different sections.
This evolution, as segments emerge and merge, could be studied simultaneously within the overall simulated Dark Matter flow to get a full picture of this filament's evolutionary history.

Instead of a single filament, in \cref{fig:coma_formation} this approach is applied to the $10$ Mpc thick slice of $\sigma = 3$ Mpc $A_4$ filaments around the Coma region (from \cref{fig:slice}) to illustrate how the formation history of the Coma Cluster and Great Wall complex can be unravelled through caustic skeleton theory.
The $A_4^+$ and $A_4^-$ points indicate where the $A_4$ filamentary caustics first emerge and merge in this slice (note: points where a pre-existing filament first enters this slice are also labelled as formation points in this figure). The caustic skeleton of the Coma Supercluster is also shown in redshift space in the right panel, with the real 2M++ galaxies plotted on top. Hence, the age of the cosmic environment of a galaxy can be probed at any chosen smoothing scale.

\Cref{fig:pp_formation} shows the formation times of $A_4$ filaments from two different viewpoints of the Pisces-Persues region - a $10$ Mpc thick $Z_R$-slice projected into the $(Y_R,X_R)$ plane, and a $10$ Mpc thick $X_R$-slice projected into $(Y_R,Z_R)$ plane. The filaments are also transformed into redshift space for comparison to 2M++ galaxies in the right panels. Formation points tend to be located in dense regions, as filaments emerge and draw the surrounding dark matter sheet inward.
At this preliminary stage, we already find indications that these results are broadly consistent with the findings of \citep{Mondelin2025}, discussed in \cref{sec:pp}, which claim older, early-type galaxies preferentially populate the dense regions along filamentary spines. However, this requires a full investigation incorporating the early- and late-type classification of 2M++ galaxies, which is deferred to a future study.
All in all, \cref{fig:coma_formation} and \cref{fig:pp_formation} illustrate how we can dissect the history and structure of the local cosmic web simultaneously. These results show that one can infer the age of the cosmic environment of real galaxies using caustic skeleton theory. In a future study, we intend to systematically investigate the impact of the ages of these structures on the age of galaxies and their properties.

\Cref{fig:hists} compares the cumulative distribution of formation times along walls and A- and D- type filaments at $\sigma = 1,2,3$ Mpc smoothing scales within the $100$ Mpc box surrounding the Coma Supercluster structure and the $60$ Mpc box around Pisces-Perseus, respectively. 
For the $M_1$ realisation, it is apparent that, overall, walls form at earlier times even though smaller-scale structures also form before the larger ones. This is consistent with hierarchical models of structure formation for $\Lambda$CDM, which was enforced during the bayesian inference of the \texttt{Manticore-Local} simulations' initial conditions \citep{McAlpine:2025}.
However, the differences in the shape of the distributions in the Coma and Pisces-Perseus regions are indicative of the unique ways the cosmic sheet underlying these cosmic web structures folded at each point in its history. For instance, the larger-scale $D_4$ filaments around Pisces-Perseus appear to form significantly later than the D-type filaments around Coma. This may be related to the apparent $D_4$ dominance of the Pisces-Perseus region we observe in \cref{sec:class}, which is perhaps suggestive of a more clumpy substructure that developed more recently. 

It is known that, overall, $D$-family caustics tend to continue forming at later times compared to standard $A$-family caustics from unconstrained simulations \citep{Hertzsch2026b}. This delayed formation is a natural consequence of the complex, multi-flow dynamics required to generate $D$-type structures, and it provides a promising indication that there will indeed be measurable statistical differences in the properties of the galaxies associated with these distinct environments. This is demonstrated in the Coma Supercluster region in the left panels of \cref{fig:hists}: the slope of the $D_4$ cumulative distribution is initially flatter then transitions to a steeper slope than the $A_4$ distribution, as the $D_4$ filaments experience comparatively greater growth towards late times. On the other hand, the cumulative distributions for the filaments in the Pisces-Perseus region (right panels) do not clearly follow this expectation. This is once again perhaps a consequence of the unique formation history of this structure captured in the $M_1$ realisation.
However, these remarks are in reference to only one realisation - a full analysis of the caustic character and the formation history of Pisces-Pisces in all \texttt{Manticore-local} realisations will be investigated in future. Moreover, the scales probed here are smaller than the resolution of the constraints.
Nonetheless, this is a powerful demonstration of the ability of the caustic skeleton theory to probe the multi-scale history of the cosmic web.

\section{Conclusions}
\label{sec:con}
In this paper, we present the first application of caustic skeleton theory to data-constrained simulations of the Local Universe. Using the \texttt{Manticore-Local} re-simulations, constrained by the 2M++ galaxy catalogue, we extract the caustic skeleton of two iconic structures of our Local Universe: the Coma Cluster and Great Wall complex, and the Pisces-Perseus Supercluster.
The caustic skeletons of these structures projected into redshift space show excellent agreement with the observed 2M++ galaxies, demonstrating that caustic skeleton theory can accurately describe the real cosmic web.

We explore the multi-scale nature of the cosmic web by varying the smoothing scale from $\sigma = 5$ Mpc down to $\sigma = 1$ Mpc, approaching the resolution of the simulations. Since the resolution of the BORG constraints $3.9$ Mpc, structures below this scale are shaped by unconstrained fluctuations injected by \textsc{monofonIC}. A comprehensive and systematic treatment of what caustic skeleton theory can reveal about the local cosmic web from the \texttt{Manticore-Local} simulations on constrained scales will be pursued in a future study.

A key prediction of caustic skeleton theory is the existence of two topologically distinct types of filaments and clusters. Both $A_4$ and $D_4$ filaments were successfully identified in the Coma and Pisces-Perseus regions. 
But as they formed through a fundamentally different folding of the dark matter sheet, these filaments possess unique local environments that may imprint distinct features on the galaxies within them that have yet to be uncovered.
Caustic skeleton theory achieves this novel classification by encoding the full folding history of the collisionless cosmological fluid, enabling it to distinguish the unique ways in which the dark matter sheet collapsed to produce these structures.
$A_4$ caustics appear to form the most outstanding filamentary-like features of the cosmic web.
However, much further investigation into the predictions of the caustic skeleton theory and to understand the nature of the $D_4$ filaments in particular due to their potential association with turnaround and infall regions around nodes. An investigation of the local cosmic web with the full caustic formalism, including higher order $A_5$ and $D_5$ node-like caustics, will be presented in an upcoming study.

The hierarchy of caustic density -- whereby $D_4$ filaments gravitationally dominate their local cosmic environment over $A_4$ filaments, which in turn dominate over $A_3$ filaments -- enables an intuitive and novel approach to classifying the large-scale environment of a galaxy. Moreover, by applying this classification across multiple smoothing scales, we demonstrate how a description of the nested, multi-scale environment in which each galaxy resides can be obtained.
In the regions around both prominent local structures, the observed transition from void and wall galaxy affiliations at large scales to an overwhelming association with filaments, in particular $D_4$/Node environments, at small scales, reflects the preference of galaxies to reside in smaller scale structures and groups.
In contrast to the Coma supercluster region, the Pisces-Perseus supercluster is revealed to be a significantly more $D_4$/Node-dominated structure in the one realisation considered. 
This is perhaps indicative of a significantly more clumpy substructure, although a thorough characterisation of $D_4$ filaments in conjunction with $A_5$ and $D_5$ node caustics is required to draw proper conclusions.
Nonetheless, this study shows promise that, in near future, the caustic skeleton theory could provide a novel topological characterisation of one of the most studied filamentary complexes in the Local Universe.

Caustic skeleton theory not only provides the spatial location at which topologically distinct caustics form, but also their formation times. The moment of first shell-crossing provides a natural and unambiguous definition of when structure formed in terms of the linear growth factor $D_+(t_{form}(\bm{q}_c)) = 1/\lambda_1(\bm{q}_c)$.
We find that walls and structures resolved at smaller smoothing scales systematically exhibit earlier formation times in both regions, consistent with the standard paradigm of hierarchical $\Lambda$CDM structure formation enforced in the \texttt{Manticore-Local} simulations. 
We find that for the Coma Supercluster region, $D_4$ filaments tend to form at later times than $A_4$ filament. This reflects expectations of the complex multi-stream dynamics required to generate these structures involving two eigenvalue fields of the deformation tensor. On the other hand, the Pisces-Perseus region displays unique and potentially contrary signatures that could be tied to its $D_4$/Node-dominated and clumpy nature and thereby unique formation history in the $M_1$ re-simulation considered.
Moreover, by identifying the points at which filaments first emerge ($A_4^+$) and merge ($A_4^-$), we demonstrate how the full dynamical history of filament -- including how it has been stretched, compressed and wound up over cosmic time -- can be uncovered.
Applied to the Coma and Pisces-Perseus regions, caustic skeleton theory reveals the intricate multi-scale topological structure and dynamical history of these well-known features of the local cosmic web.

The distinct caustic environments of galaxies can now be characterised by caustic type, scale and formation time. 
This opens new avenues to understand how the large-scale environment influences galaxy properties such as morphology, spin alignment and star formation rates. A detailed investigation of these connections will be pursued in future work, along with the extension of the caustic skeleton framework to the higher $A_5$ and $D_5$ caustics for the identification of clusters. These efforts are particularly timely given the upcoming data releases from Eulcid, DESI and the SKA surveys, which will map the cosmic web with unprecedented precision and volume.

\section*{Data Availability}
The \texttt{Manticore-Local} re-simulations \citep{McAlpine:2025} (white noise fields and final $z=0$ snapshots) are available at \href{https://cosmictwin.org}{cosmictwin.org}. The density fields in this paper were estimated with the Phase-Space Delaunay Tessellation Field Estimator \citep{Feldbrugge+2024, FeldbruggeHertzsch2025}. The \texttt{PhaseSpaceDTFE} package can be installed with the Julia package manager. The caustic skeletons of the Coma cluster and Pisces-Perseus regions presented in this paper are available upon request.

\acknowledgments
We thank the Aquila Consortium for providing early access to one of the \texttt{Manticore-local} simulations and for helpful discussions. Thanks to Dr Ciaran O'Hare for helpful advice on the visualisation of some results. 
RvdW acknowledges funding from EU Horizon Europe (EXCOSM, grant nr. 101159513).

\appendix

\section{Caustic skeleton theory}
\label{appendixA}

In \cref{sec:caustic_def}, we identified the shell-crossing events with the condition 
\begin{align}
    A_2:1+\mu_i(\bm{q},t) = 0\,,
\end{align}
for $i=1,2,3$. This condition is identified with the fold caustic ($A_2$). The fold caustic is itself folded into a cusp caustic ($A_3$) when, in addition, the directional derivative of the eigenvalue field in the associated eigenvector direction vanishes, \textit{i.e.},
\begin{align}
    A_3: \bm{v}_i(\bm{q},t) \cdot \nabla_{\bm{q}} \mu_i(\bm{q},t) = 0\,.
\end{align}
As we will see, these cusps mark where the dark matter sheet folds into the walls of the cosmic web. The cusp gets folded into a swallowtail caustic ($A_4$) when, in addition, a second-order directional derivative vanishes, given by 
\begin{align}
    A_4: \bm{v}_i(\bm{q},t) \cdot \nabla_{\bm{q}}\left(\bm{v}_i(\bm{q},t) \cdot \nabla_{\bm{q}} \mu_i(\bm{q},t)\right) = 0\,.
\end{align}
The swallowtail caustic marks the dark matter sheet folding into filaments of the cosmic web. Continuing this process, the swallowtail is folded into the butterfly caustic ($A_5$) when 
\begin{align}
    A_5: \bm{v}_i(\bm{q},t) \cdot \nabla_{\bm{q}}\left(\bm{v}_i(\bm{q},t) \cdot \nabla_{\bm{q}}\left(\bm{v}_i(\bm{q},t) \cdot \nabla_{\bm{q}} \mu_i(\bm{q},t)\right)\right) = 0\,.
\end{align}
The butterfly caustics mark the folding into clusters.
In addition to these $A$-type caustics, catastrophe theory includes the elliptic/hyperbolic caustic ($D_4$) given by 
\begin{align}
    D_4: 1+ \mu_i(\bm{q},t) = 1+\mu_j(\bm{q},t) = 0\,,
\end{align}
for $i,j \in \{1,2,3\}$ with $i \neq j$, corresponding to the formation of a second type of filaments. Finally, catastrophe theory predicts the existence of the parabolic caustic ($D_5$), where in addition 
\begin{align}
    D_5: \det (\mathcal{H}_{\bm{v}_i\bm{v}_j}(\det \nabla \bm{x}_t))  = 0\,,
\end{align}
with $\mathcal{H}_{\bm{v}_i\bm{v}_j}$ the Hessian in the $\bm{v}_i\bm{v}_j$-plane. The parabolic umbilic caustic forms a second formation channel for clusters in the cosmic web. For a detailed derivation of these conditions, we refer to \cite{Feldbrugge+2018}, extending the work by \cite{ArnoldShandarinZeldovich1982} and \cite{Arnold1986} to three-dimensional structure formation.

\begin{table*}
\centering
\begin{tabular}{l | l }
{Name} & {Caustic Conditions under Zel'dovich Approximation} \\
\hline 
Fold, $A_2$& $\lambda_{1}(\bm{q}) = 1/D_+(t)$  \\
Cusp, $A_3$ &  $\bm{q} \in A_2, \bm{v}_1(\bm{q})\cdot\nabla \lambda_{1}(\bm{q})=0$  \\
Swallowtail, $A_4$ &   $\bm{q} \in A_3, \bm{v}_1(\bm{q})\cdot\nabla (\bm{v}_1(\bm{q})\cdot\nabla \lambda_{1}(\bm{q}))=0$\\
Butterfly, $A_5$ &  $\bm{q} \in A_4, \bm{v}_1(\bm{q})\cdot\nabla(\bm{v}_1(\bm{q})\cdot\nabla (\bm{v}_1(\bm{q})\cdot\nabla \lambda_{1}(\bm{q})))=0$\\
Elliptic/Hyperbolic Umbilic, $D_4$  &  $\lambda_{1}(\bm{q}) = \lambda_{2}(\bm{q}) = 1/D_+(t)$ \\
Parabolic, $D_5$&  $\bm{q}\in D_4, \det (\mathcal{H}_{\bm{v}_1\bm{v}_2}\left(\det ( I - D_+(t) \mathcal{H} \Psi(\bm{q}))\right)  = 0$\\ 
\end{tabular}
\caption{The 7 types of caustics that can form in a 3-dimensional Lagrangian mapping with associated symbols. Regardless of the dimension of the problem, the caustic conditions under the Zel'dovich approximation the eigenvalue and eigenvector fields of the deformation tensor can be used to identify each type of caustic in Lagrangian space $q$. The key distinction between A and D type caustics is that the former enforces a condition on only the first eigenvalue field, whereas the latter must involve two eigenvalue fields.} 
\label{tab:conditions}
\end{table*}

\section{The Zel'dovich approximation:} 
\label{appendixB}
The caustic conditions displayed in Appendix \ref{appendixA} apply to a general Lagrangian fluid. In this paper, we assume the Zel'dovich approximation, which approximates the motion of the mass elements by ballistic trajectories following the primordial gravitational force field \citep{Zeldovich1970}. This approximation -- also known as first-order Lagrangian perturbation theory -- provides a good description of the growth of cosmological perturbations in the linear regime and captures the formation of the largest scales in the Universe and its filamentary structure. The caustics of the Zel'dovich approximation mark the places and ways in which the Zel'dovich approximation breaks down, and non-linear gravitational collapse takes over.

In the Zel'dovich approximation, the displacement field factorises into a spatial and a temporal term 
\begin{align}
    \bm{s}_t(\bm{q}) = - D_+(t) \nabla_{\bm{q}} \Psi(\bm{q})\,.
\end{align}
Here, the \textit{primordial displacement potential} $\Psi(\bm{q})$ is proportional to the primordial gravitational potential perturbation $\phi(\bm{q})$ 
\begin{equation}
    \Psi(\bm{q}) = \frac{2}{3 H_0^2 \Omega_m} \phi(\bm{q}) = \frac{2}{3 D_+(t) a(t)^2 H_0^2 \Omega_m} \phi_{\textrm{lin}}(\bm{q}) \,,
    \label{eq:displacement_potential}
\end{equation}
with the Hubble parameter $H(t)$, the scale factor $a(t)$, the matter density parameter $\Omega_m$ and the linearly extrapolated potential perturbation $\phi_{\textrm{lin}}$ at time $t$.
The temporal dependence is absorbed by the linear growth factor  $D_+(t)$, which gives the solution for the linear growth of perturbations through the defining equation
\begin{equation}
    \ddot{D}_+(t) + 2 H(t) \dot{D}_+(t) - 4 \pi G \rho_0 D_+(t)=0 \,,
\end{equation}
with the boundary conditions $D_+(0)=0$ and $D_+(t_0)=1$, the primordial mean density $\rho_0$ and the gravitational constant  $G$ (see \cite{Peebles1994}).

Working in terms of the deformation tensor -- now given by the Hessian of the displacement potential $\mathcal{H} \Psi$ -- and its eigenvalues $\lambda_i$ and eigenvectors $\bm{v}_i$ defined by 
\begin{align}
    \left[\mathcal{H}\Psi(\bm{q})\right] \bm{v}_i(\bm{q}) = \lambda_i(\bm{q}) \bm{v}_i(\bm{q})\,,
\end{align}
with the convention $\lambda_1 \geq \lambda_2 \geq \lambda_3$, the caustic conditions simplify, relating the present-day cosmic web to specific conditions on the primordial gravitational potential (see \cref{tab:conditions}). The caustic conditions in \cref{tab:conditions} apply to the caustics at first shell-crossing in the Zel'dovich approximation. Remarkably, the caustic skeleton of the Zel'dovich approximation turns out to capture the intricate structure of the cosmic web not only in the linear regime but also at mildly non-linear scales.

\section{Constrained Simulations}
\label{app:constrsimul}
    
The linear constrained Gaussian random field theory technique, first introduced by Bertschinger \citep{Bertschinger1987}, can also be rephrased as an optimization problem. With
  this perspective, linear constraints can be implemented for a large number of constraints resembling a local patch of
  Lagrangian space \cite[see for example the splice method][]{Cadiou:2021}. In an observational context involving data with
  measurement errors, it immediately implies the Wiener filtered processed data constraints as the estimate of the mean field $\bar{f}(x)$
  \citep{RybickiPress1992,Zaroubi1995c}. 
  
  The residual field $F(\bm x)$ represents the typical spectral fluctuations in the field. These are no longer entirely
  random but tuned and adjusted to the influence of, ie. correlation with, the imposed constraints: while the fluctuations
  at the location of the constraints will be zero, far removed from these locations they may assume again the average
  fluctuation level dictated by the cosmological power spectrum. It renders the generation of the permitted residual field
  values challenging, involving the constraint induced correlations. \cite{Bertschinger1987} based the sampling on a rather
  elaborate and complex simulated annealing technique involving the Metropolis algorithm. In a remarkably illuminating
  contribution, \cite{Hofrib1991} revealed that for constrained Gaussian random fields the residual field can be
  generated in a straightforward single operation involving the generation of only one corresponding Gaussian random
  field realization.   

  The Bertschinger-Hoffman-Ribak formalism provides a good basis for exploring - in a laboratory style approach - the impact
  of cosmologically interesting conditions. The first constrained simulations along these lines concerned that of
  voids emerging from troughs in the primordial density field \citep{Weykamp1993} and of filaments constrained via
  the primordial tidal field \citep{Haarlemwey1993}. In an observational context, as to a good approximation linearity may still
  be presumed to hold on large scales, the formalism allows the inference of the mass distribution from the mapping of the velocity
  flows in our Local Universe \cite{Bertschinger+1989,Dekel:1999,Courtois2012,Hoffman:2015}, and to investigate the formation and
  evolution of the cosmic web in our Local Universe. Well-known in this context are the CLUES simulations \citep[e.g.][]{Sorce+2016}. An
  important aspect of these procedures is to incorporate the strength/reliability of the observational data in the significance of
  the obtained reconstruction through Wiener filtering \cite{Zaroubi1995,Erdogdu2004}.  

   A range of avenues have been taken to extend the ability to impose linear constraints to that of the far more ubiquitous situation
   of constraints that pertain to the non-linear cosmic matter distribution, or that are a non-linear function of observable quantities. One option that has many applications in e.g. geophysics is that of (lognormal) Kriging, in which one exploits the information available at irregularly placed locations by taking into account the spatial correlations \cite{Platen+2011}. There have also been attempts to augment the linear Gaussian field formalism by invoking a (first order) model for the displacement of mass elements in order to follow these along their path backward in time \citep{Doumler2013}. Recently, \cite{FeldbruggeWeygaert2023} extended the linear constrained Gaussian random field formalism to the specific situation of nonlinear constraints that are explicit functional expressions of Gaussian random fields. This enables, amongst others, constraints within the Caustic Skeleton formalism. The latter are functions of the tidal/deformation field eigenvalues, distinctly non-Gaussian in nature \citep{Doroshkevich1970}. It enables the extensive exploratory study of walls in the cosmic web, such as described in \cite{Hertzsch+2026}, and of filaments. 
  
  Because of the restricted validity of the linear constraint formalism to regions that are not (yet) multistream, it cannot be applied
  for our purpose of modeling the mass distribution around the intrinsically multistream filamentary features. Given the fact that
  the majority of matter, halos and galaxies reside in filaments \citep{Cautun+2014,Ganeshaiah:2019}, more elaborate higher-order schemes
  will be necessary. One option is to incorporate higher-order Lagrangian schemes for modeling the orbits of galaxies and/or paths of
  mass elements. The Least Action Principle algorithm introduced by \cite{Peebles:1989}, and its further developments \citep{NusserBranchini2000},  is one such approach. It has recently provided a highly interesting assessment of the dynamical evolution of the galaxy
  distribution in the nearby universe, out to even a region of 80 Mpc probed by Cosmicflows-3 \citep{Shaya2022}. Other sophisticated
  higher-order Lagrangian formalisms have been proposed by \cite{Mohayaee2006, Lavaux:2010, Hada:2018, Shi:2018, Zhu:2018}. For example,
  \cite{Mohayaee2006} and \cite{Lavaux:2010} used the Monge-Amp\`ere-Kantorovitch algorithm to follow the orbits of objects back
  in time, culminating in a compelling reconstruction of the local velocity field corresponding to the 2MRS survey. Over recent years, 

  An approach that has become an important cosmological tool over the past 15 years concerns statistical Bayesian inferece
  techniques. They involve the ability to translate the observed spatial distribution of galaxies within a given volume of the (local) Universe towards its implied primordial Gaussian density and velocity perturbations out of which these galaxies and structures emerged through gravitational growth. To this end, advanced and sophisticated Bayesian reconstruction schemes have been developed, which involve elaborate and computationally demanding  MCMC sampling procedures \cite{Kitaura:2008,Kitaura:2009, Jasche:2010,Leclercq:2015,Hess:2016, Bos:2016, Leclercq:2017, McAlpine:2022}. The corresponding constrained simulations can be used to address a variety of problems,
  including BAO measurements and the removal of redshift space distortions. For large galaxy surveys in the local Universe, such as the 2MRS and SDSS surveys, impressive results have been obtained that allow us to study the properties and evolution of its detailed mass distribution. Telling examples are the 2MRS constrained simulations of the Local Universe by Kitaura and collaborators on the basis of their KIGEN formalism \cite{Hess:2013} (also see \cite{Hidding+2016}) and the recent detailed Sibelius simulation \citep{McAlpine:2022} and Manticore simulations \citep{McAlpine:2025,McAlpine:2026} on the basis of the BORG formalism. Bayesian Hamiltonian MCMC methods have also been succesfully adapted for significant improvements of density and velocity field reconstructions on the basis of peculiar velocity surveys such as Cosmicflows \citep{Valade2022,Valade2026} and, within the context of BORG, for high-resolution studies of the formation
  and evolution of the Local Group \citep{Wempe2024}.

  Finally, with the proliferation of Machine Learning and AI based tools new avenues for cosmological galaxy and matter distribution
  reconstructions have opened up. Recent techniques like developed by \cite{Ganeshaiah+2023,Lilow+2024}, producing convincing
  density and velocity field reconstructions based on the 2MRS survey, have demonstrated their high potential. 

\section{Slices of the Caustic Skeleton in the sky}
\label{appendixC}
\begin{figure}[h!]
    \centering
    \includegraphics[width=\textwidth]{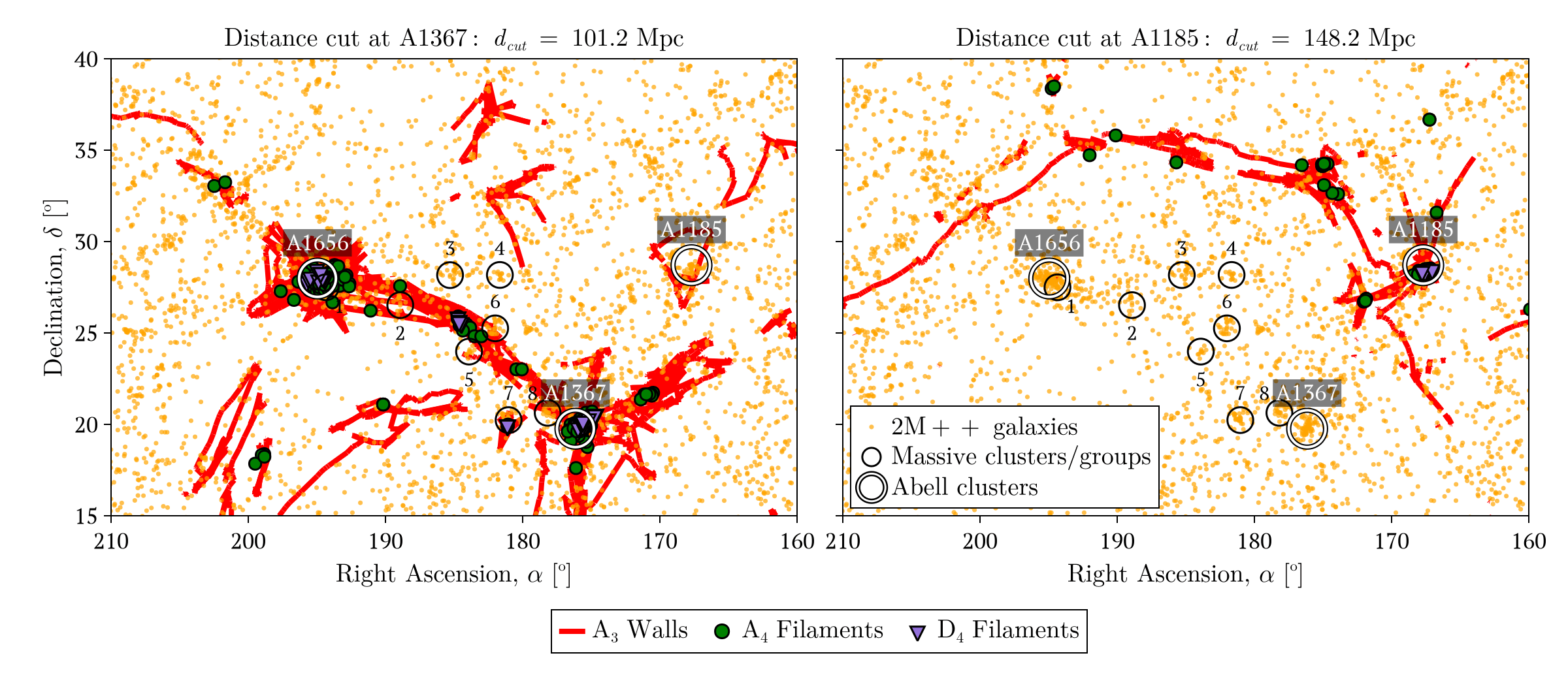}
    \caption{Slices of $\sigma = 3$ Mpc caustic skeleton from $M_1$ realisation at two different radial distances in vicinity of Coma cluster (A1656) viewed in equatorial coordinates. 
    In left panel, slice is taken at $d_{cut} = 101.2$ Mpc, corresponding to the distance of A1367 from observer position. The slice is taken further away at $d_{cut} = 148.2$ Mpc in right panel, corresponding to the distance of A1185. All 2M++ galaxies within this section of the sky are shown. A few massive clusters/groups are circled in black: (1) NGC4839, (2) NGC4555, (3) NGC4295, (4) NGC4104, (5) NGC4213, (6) NRGb184, (7) NGC4065, (8) NGC3937.}
    \label{fig:coma_radec_slice}
\end{figure}

\Cref{sec:coma} \cref{fig:coma_radec} showed a projection of $A_4$ and $D_4$ filaments onto the sky in equatorial coordinates for the $M_1$ realisation with a smoothing scale of $\sigma = 3$ Mpc. To provide additional insight into the distribution of structures, including both filaments and walls, with depth \cref{fig:coma_radec_slice} shows a slice of the same caustic skeleton at two distances, corresponding to two of the prominent Abell clusters, A1367 and A1185, in the vicinity of Coma. In the left panel, the cut is made at the location of the A1367 cluster, which sits at a similar radial distance to the Coma cluster (A1656). As such, the dense bunch of $A_4$ and $D_4$ filaments at the centre of both clusters are picked up in this slice. Moreover, the filamentary bridge between these two clusters observed in \cref{fig:coma_radec}, is also evident here as the $A_3$ walls intersect along this ridge. The $D_4$ filaments near NGC4065 and NGC4213/NRGb184 also intersect this slice. In comparison, A1185 sits around $\sim 47$ Mpc further away, so the $D_4$ and $A_4$ filaments at the clusters center are only picked up in the right panel.

\begin{figure}
    \centering
    \includegraphics[width=\textwidth]{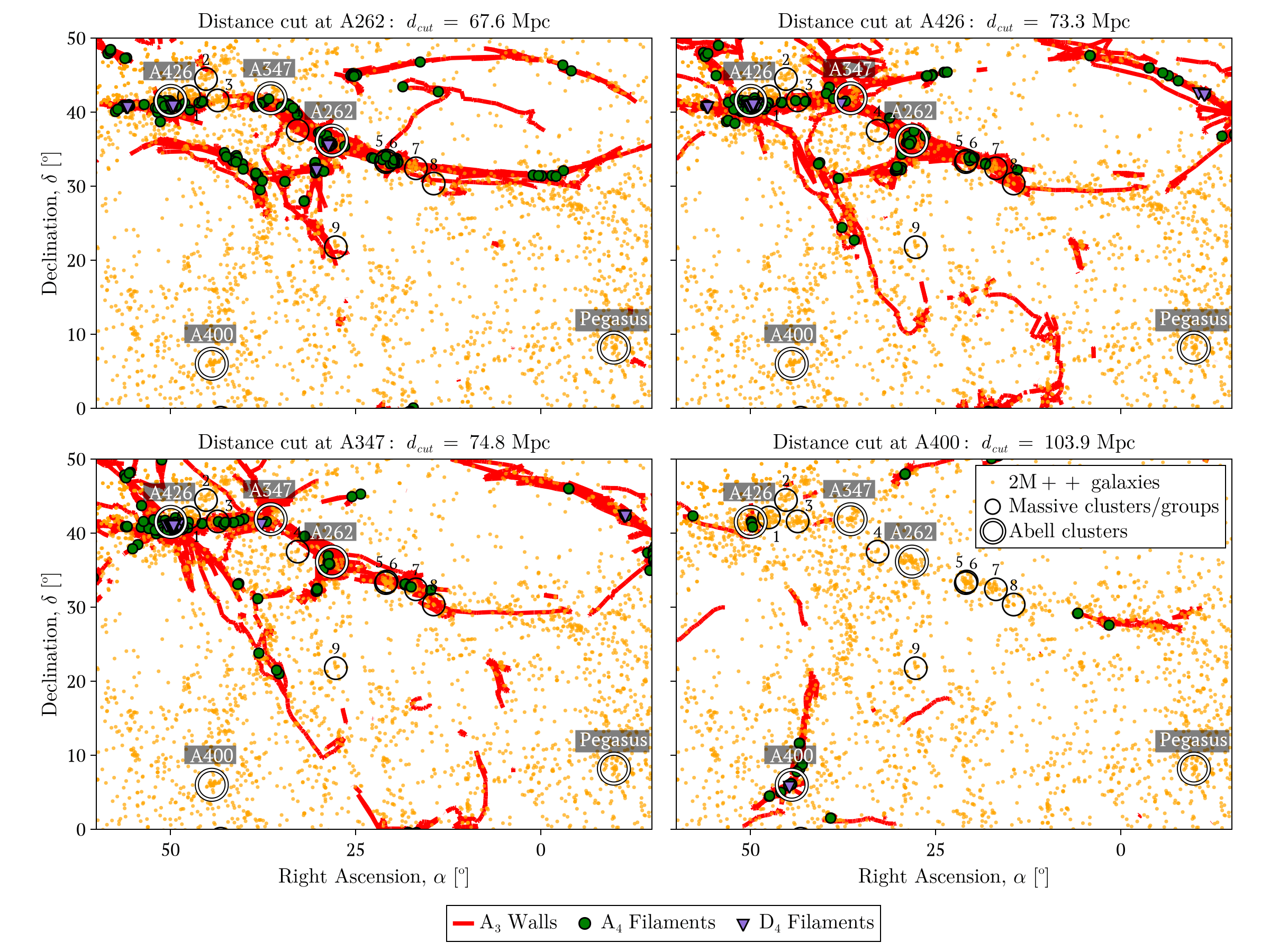}
    \caption{Slices of $\sigma = 3$ Mpc caustic skeleton from $M_1$ realisation at two different radial distances in vicinity of Pisces-Perseus supercluster viewed in equatorial coordinates. 
    Slices are taken at distances corresponding to locations of prominant Abell clusters in region: A262 at $d_{cut} = 67.6$ Mpc (top left), A426 at $d_{cut} = 73.3$ Mpc (top right), A347 at $d_{cut} = 74.8$ Mpc (bottom left) and A400 at $d_{cut} = 103.9$ Mpc (bottom right).  All 2M++ galaxies within this section of the sky are shown. A few massive clusters/groups are circled in black: (1) UGC 2562, (2) CIZAJ0300.7+4427, (3) AWM7, (4) N841, (5) NGC499, (6) NGC507, (7) NGC383, (8) N315, (9) N691.}
    \label{fig:pp_radec_slice}
\end{figure}

\Cref{fig:pp_radec_slice} provides a similar complementary view of the projected filaments around the Pisces-Perseus supercluster shown in \cref{sec:pp} \cref{fig:pp_radec}. The slices of the caustic skeleton are taken at four Abell clusters in the region, including A426, A347, A262 and A400. Although the Pegasus cluster is also present in this region, as \cref{fig:pp_radec} shows, at the smoothing scale of $\sigma = 3$ Mpc there is minimal caustic structure. The Pegasus cluster likely has a smaller characteristic scale compared to the other nearby Abell clusters. Since the A262, A426 and A437 clusters are all within roughly similar distances, the main backbone of the Pisces-Perseus supercluster is strong in those three panels.
Similar to the Coma region, the strong filamentary structures shown in \cref{fig:pp_radec}, are evident here through the walls that intersect at those filaments. 
Moreover, $D_4$ filaments are found to sit at the centres of each of these four massive clusters in the corresponding slices. The $D_4$ filaments at the centre of the A426 (Perseus) cluster are present in all three panels, and even in the further away distance cut (corresponding to A400), $A_4$ filaments are still present at its centre. This indicates that the Perseus cluster has the most filaments and extent in this region.

\bibliographystyle{JHEP}
\bibliography{refs.bib}



\end{document}